\documentclass[aps, pra, twocolumn, superscriptaddress, amsmath, tightenlines, longbibliography]{revtex4-1}

\usepackage{amssymb}
\usepackage{amsmath}
\usepackage{dcolumn}
\usepackage{graphicx}
\usepackage{mathrsfs}
\usepackage{appendix}
\usepackage{graphicx}
\usepackage{booktabs}
\usepackage{colortbl}

\definecolor{Dred}{RGB}{190,0,0}

\usepackage{url}
\usepackage[colorlinks]{hyperref}
\hypersetup{%
	plainpages=true,
	breaklinks=true,       %not default in dvips mode, so we must specify
	hypertexnames=false,  %not ideal, but needed when pagenums duplicate (`i' vs. `1')
	pageanchor=true,
	colorlinks=true,
	linkcolor={blue},
	citecolor={red},
	urlcolor={blue},
	anchorcolor={black}
}

\def \hide#1{}

\begin{document}
	\title{Chiral Quantum Network with Giant Atoms}

	%	Tunable Chiral Waveguide QED based on Josephson Metamaterials for Superconducting Quantum Networks}
	
	\author{Xin Wang}
	%\email{wangxin.phy@xjtu.edu.cn}
	\affiliation{Institute of Theoretical Physics, School of Physics, Xi'an Jiaotong University, Xi'an 710049, People’s Republic of China}
	
	\affiliation{Ministry of Education Key Laboratory for Nonequilibrium Synthesis 
		and Modulation of Condensed Matter, Shaanxi Province Key Laboratory 
		of Quantum Information and Quantum Optoelectronic Devices, School of 
		Physics, Xi'an Jiaotong University, Xi'an 710049, People’s Republic of China}
	
	\author{Hong-Rong Li}
	\affiliation{Institute of Theoretical Physics, School of Physics, Xi'an Jiaotong University, Xi'an 710049, People’s Republic of China}
	\affiliation{Ministry of Education Key Laboratory for Nonequilibrium Synthesis 
		and Modulation of Condensed Matter, Shaanxi Province Key Laboratory 
		of Quantum Information and Quantum Optoelectronic Devices, School of 
		Physics, Xi'an Jiaotong University, Xi'an 710049, People’s Republic of China}
\date{\today}

\begin{abstract}
	In superconducting quantum circuits (SQCs), chiral routing quantum 
	information is often realized with the ferrite circulators, which are 
	usually bulky, lossy and require strong magnetic fields. To overcome 
	those problems, we propose a novel method to realize chiral quantum networks by exploiting giant atom effects in SQC platforms. By assuming each coupling point being modulated with time, 
	the interaction becomes 
	momentum-dependent, and giant atoms will chirally emit photons 
	due to interference effects. The chiral factor can 
	approach 1, and both the emission direction and rate can be freely 
	tuned by the modulating signals. 
	We demonstrate that a high-fidelity state transfer between remote giant atoms can be 
	realized. Our proposal can be integrated on the superconducting chip easily, and has the potential to work as a tunable toolbox for quantum information processing in future chiral quantum networks.
\end{abstract}
\maketitle

\section{introduction}
The past two decades have witnessed a great interest in using 
superconducting quantum circuits (SQCs) as platforms for large-scale 
quantum information processing (QIP) ~\cite{Gu2017,Xu2018,Arute2019,Ye2019}. The number of programmable qubits in a single integrated chip is increasing 
rapidly~\cite{Gong2021}. In complex QIP tasks, multiple nodes separated remotely might take part in QIP simultaneously.
To mediate remote nodes as well as preserve a high-fidelity 
coherence and entanglement, the quantum network~\cite{Reiserer2015,Brekenfeld2020,Daiss2021,Awschalom2021}, which can route the quantum information encoded in 
flying qubits, becomes necessary and important~\cite{Kimble2008,Ritter2012,vanLoo2013,Humphreys2018}. 
The studies in Refs.~\cite{Vermersch2017,Xiang2017} indicate that, with an auxiliary linear oscillator, high-fidelity state transfer between remote nodes is achievable 
even when the number of thermal microwave photons is 
large. Those results show the possibilities to build all-microwave networks for SQC platform 
free of frequency transducers in the near future.

In a quantum network, chiral (or nonreciprocal) routing photons 
without information back flow is essential for deterministic quantum 
communications~\cite{Cirac1997,Stannigel2011,Ramos2016}. Chiral 
networks not only enable cascaded quantum 
circuits~\cite{Carmichael1993,Gardiner1993,Berman2020,Du2021}, but 
also can be employed for special QIP tasks which are beyond the approach of 
the conventional bidirectional networks~\cite{Cirac1997}. For 
example, due to chiral destructive interference, multiple 
quantum nodes in a chiral network can be steered in stationary 
entangled states via dissipation-driven processes~\cite{Stannigel2012}.
However, in most of current studies, chiral routing microwave photons in an SQC network still requires classical ferrite circulators~\cite{Hogan1953,Allen1956,Caloz2018}, which are 
usually bulky, lossy, and hard to be integrated on chips. 
To find better SQC nonreciprocal devices, some integrable chiral interfaces are 
proposed~\cite{Estep2014,Metelmann2015,Sounas2017,Chapman2017,Guimond2020}. 
However, those methods might lead to additional 
experimental overheads, and are usually lack of the tunability required by various kinds of QIP operations.

Given that all the nodes can emit and absorb photons 
unidirectionally, chiral networks are naturally formed without any additional
nonreciprocial 
device~\cite{Lodahl2017}. This scenario is referred as chiral quantum 
optics~\cite{Mitsch2014,Petersen2014,Pichler2015,Young2015,Bliokh2015a,leFeber2015,Grankin2018,Calaj2019}.
Most of previous studies on chiral quantum optics are 
discussed in nanophotonic systems, where the 
mechanisms are based on such as spin-momentum locking~\cite{Bliokh2015a,leFeber2015}, 
spatiotemporal acousto-optic 
modulating~\cite{Lira2012,Trainiti2016,Calaj2019,Chen2019}, etc. 
However, those methods can not be applied for SQC chips of 2D distribution. Up to now, realizing chiral emission of superconducting atoms is still in its fancy and 
rarely studied. 

In SQC platforms, giant atoms, which sizes are comparable to the 
wavelength of coupled  photons~\cite{Kockum2014,Guo2017,Kockum2018,Guo2019,Zhao2020,Wang2021,cheng2021boundary,du2021single,Soro2021}, can be realized by considering 
multiple coupling points with a photonic (or phononic) 
waveguide~\cite{Kannan2020}. 
The interference effects between different points will lead to exotic 
quantum phenomena such as frequency-dependent emissions and 
dipole-dipole interactions free of 
decoherence~\cite{Kockum2014,Kockum2018}. In 
Ref.~\cite{Guimond2020}, by considering two remote coupled atoms working as a 
composite emitter, the authors showed that chiral transports can be realized without breaking the Lorentz reciprocity. 
However, the quantum information is required to be encoded into the entangled states of two remote atoms, rather than a single giant atom.

In this work, we propose a novel method to realize chiral quantum 
networks by exploiting the quantum interference effects in a single giant atom. In our 
study, the Lorentz reciprocity is broken, and the mechanism is totally different 
from the proposal in Ref.~\cite{Guimond2020}. Compared with encoding information  
into the fragile entangled states of two small atoms, our method does not require entanglement resource, and is much more robust to decoherence noises.
The chiral emission is due to the opposite interference relation between two directions in a waveguide. By choosing suitable modulating parameters, the chiral factor can approach 1. We also demonstrate that high-fidelity state transfers between two remote nodes is achievable in our proposal. Compared with classical ferrite circulators, our approach is tunable, and can be integrated on chips easily.

\section{Momentum-dependent coupling between giant atoms and PCW}
\begin{figure}[tbph]
	\centering \includegraphics[width=8.6cm]{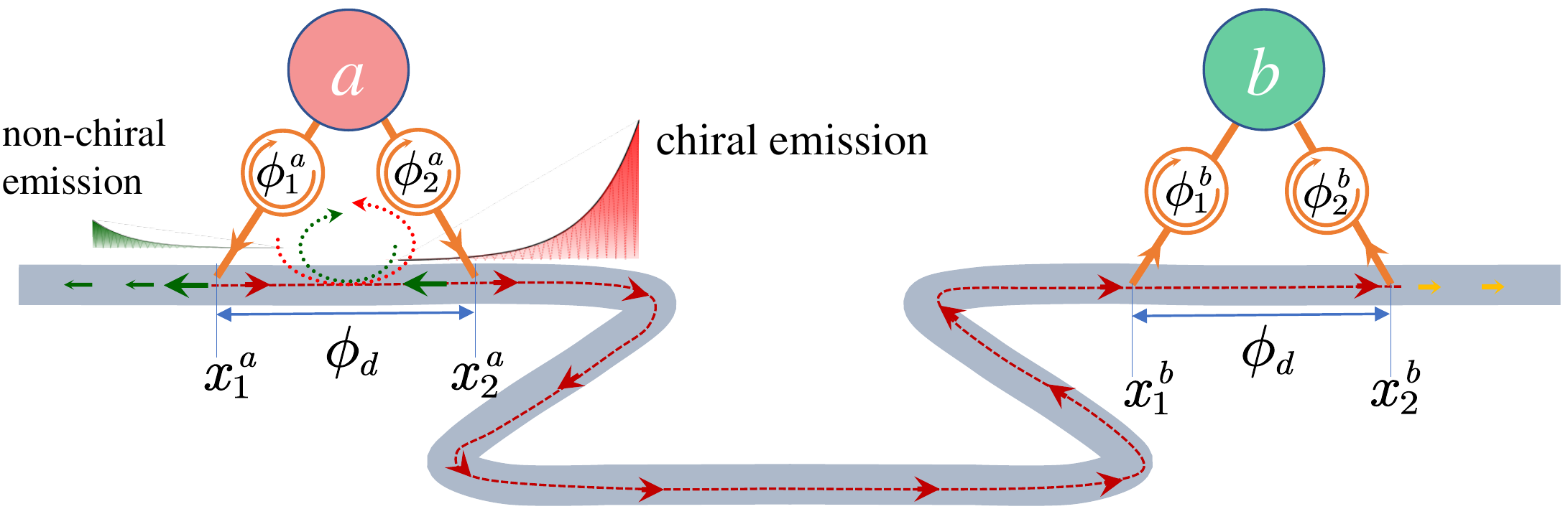}
	\caption{Sketch of our proposal with two superconducting 
		giant atoms interacting with a common photonic crystal waveguide (PCW). Atom $a$ and $b$ are both of giant atom 
		form. Their coupling points $x_{1,2}^{a,b}$ are encoded with different phases 
		$\phi_{1,2}^{a,b}$, respectively. For an itinerant photon, the distance between two coupling points in atom $a$ 
		($b$) leads to a propagating phase $\phi_d$. Those phases differences will generate an artificial gauge field in the
		coupling loops (see clockwise and counter-clockwise arrows). Due to asymmetric interference relations, the photon will be emitted/absorbed unidirectionally, and the whole system works as a chiral quantum network.}
	\label{fig1m}
\end{figure}

Our proposal is schematically depicted in Fig.~\ref{fig1m}, where giant atom $a$ ($b$) couples to the waveguide at two points which are spatially separated with a distance $x_d=x_{2}^{a(b)}-x_{1}^{a(b)}$. There will be a propagating phase for the photon emitted/absorbed by each giant atom. Besides propagating phases, each coupling point is encoded with local phases $\phi_{1,2}^{a(b)}$. For a giant atom, the interaction with right (left) propagating photons is related to the phase 
difference between $\phi_{2}^{a(b)}-\phi_{1}^{a(b)}$ and $\phi_d$ ($-\phi_d$) (see clockwise/counter-clockwise in Fig.~\ref{fig1m}). Therefore, the atom-waveguide coupling becomes 
asymmetric in momentum space, and the emission will show chiral 
preference. 

%%%%%%%%%%%%%%
\begin{table*}[tbp]
	{\normalsize \renewcommand\arraystretch{1.5}
		\begin{tabular}{>{\hfil}p{0.9in}<{\hfil}>{\hfil}p{0.85in}
				<{\hfil}>{\hfil}p{0.3in}<{\hfil}>{\hfil}p{1in}<{\hfil}>{\hfil}p{1.1in}<{\hfil}>{\hfil}p{1.3in}<{\hfil}>{\hfil}p{1.2in}<{\hfil}}
			\hline\hline  $c_{g}$ & $l_0$ & $\delta\alpha$ & $\Delta_g \; (\Delta_d)$ & $\phi_c\rightarrow \beta_{\pm}$ & $\Omega_d\rightarrow\text{switch on/off}$ & $A_1\rightarrow\Gamma_{\pm}$ \\
			\hline $2\times10^{-10}\text{F/m}$ & $5\times 10^{-6}\text{H/m}$ & 0.3 & $0.75\;(0.1)~\text{GHz}$ & $(-\pi,\pi]\rightarrow(0,1)$ & $0\sim 0.55~\text{GHz}$ & $[0,0.5]\rightarrow[0,3]~\text{MHz}$\\
			\hline\hline
	\end{tabular}}
	\caption{The system's parameters adopted for numerical simulations. The PCW parameters are set according to Refs.~\cite{Frunzio2005,Goppl2008,Clem2013}. The arrows correspond to the tuning relations between modulating signals and chiral emission parameters.} \label{table1}
\end{table*}
%%%%%%%%%%%%%%

\begin{figure*}[tbph]
	\centering \includegraphics[width=17.6cm]{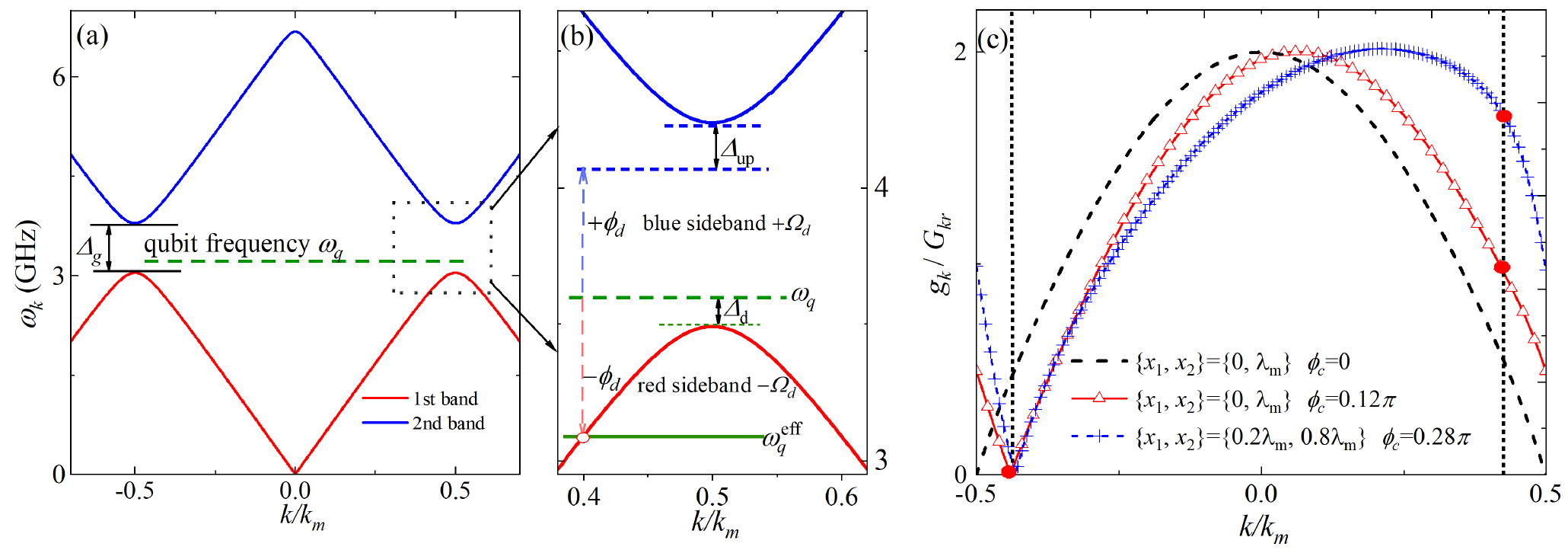}
	\caption{(a) The dispersion relation of PCW changes with wavevector $k$ for two lowest energy bands. (b) The atomic frequency $\omega_q$ is assumed to be inside the band gap with $\Delta_d\gg0$ (green dashed line), and much closer to the first band. The time-dependent modulations in giant-atom coupling lead to both red and blue sideband transitions. The blue sideband terms is far away from the 2nd band, i.e., $\Delta_{\text{up}}\gg0$, while the red sideband is resonant with the modes in the 1st band. The atomic effective frequency is shifted as $\omega_q^{\text{eff}}=\omega_q-\Omega_d.$
		(c) For different phases $\phi_c$ and 
		different coupling points $\{x_1, x_2\}$, the momentum-dependent interaction strength $g_k$ changes with $k$ for the modes in the first energy band. Parameters are adopted from Table~\ref{table1}.}
	\label{fig2m}
\end{figure*}

Not that $\phi_{1,2}^{a,b}$ are unconventional, and should be 
generated via artificial 
methods~\cite{Dalibard11,Schmidt2015,Fang2017}. 
To encode those phases into different coupling points,
we assume that the waveguide is engineered with a band 
gap. In this study, we take the photonic crystal waveguide (PCW) for example~\cite{John1991,Hung2013,Goban2014,Douglas2016,Liu2017}. In SQC platform, each unit cell in PCW is made by the 
transmission line with inductance being periodically modulated~\cite{Liu2017}.
The inductance (capacitance) per unit length is denoted as $l(x)$
($c_g$).
Consequently, the dynamics of PCW field is described by the following
wave equation~\cite{Liu2017,Wang2021}
\begin{equation}
	c_g \frac{\partial^2 \phi(x,t)}{\partial t^2} = \frac{\partial}{\partial x}\left[ \frac{1}{l(x)} \frac{\partial \phi(x,t)}{\partial x} \right],
	\label{waveeq}
\end{equation}
where $\phi(x,t)$ is the node flux at position $x$. For simplicity the impedance is assumed to be modulated with a square wave
\begin{equation}
	\frac{1}{l(x)}=\frac{1}{l_0} \Big\{1 + \delta\alpha \ \text{sgn}[\cos(k_m x) ]\Big\},
\end{equation}
where $l_0$ is the static inductance, $\delta \alpha$ is the 
modulating amplitude, and $k_m= 2\pi /\lambda_m$ is wave vector with
$\lambda_m$ being the periodic length. 

The Bloch eigen-function for 
Eq.~(\ref{waveeq}) can be derived by using Fourier series
representations~\cite{Trainiti2016,Liu2017}. Detailed methods are presented in Ref.~\cite{Wang2021}. Finally the field eigenfunction $\phi_{lk}$ is derived as
\begin{equation}
	\phi_{lk}= e^{i(\omega_l(k) t + k x)} u_{lk} (x), \quad u_{lk} (x) = \sum_{n = -\infty}^{n = \infty} c^{(l)}_{nk} e^{i n k_m x},
	\label{phifs0}
\end{equation}
where $\omega_l(k)$ is the eigenfrequency of mode $k$ in the $l$th energy band, $u_{lk} (x)$ is the Bloch  eigen-function 
satisfying $u_{lk}(x)=u_{lk} (x+\lambda_m)$, and $c^{(l)}_{nk}$ 
is the amplitude of the $n$th Fourier order. 
Consequently, the current operator of the quantized PCW is written as~\cite{Wang2021} 
\begin{equation}
	I_{w}=i\sum_{l,k}\sqrt{\frac{\hbar\omega_{l}(k)}{2L_{\text{tot}}}}\left[a_{lk} e^{-ikx} u_{lk}^{*} (x)-a_{lk}^{\dagger} e^{ikx} u_{lk} (x)\right],
	\label{Iw}
\end{equation}
where $a_{lk}$ ($a^{\dagger}_{lk}$) is the annihilation (creation) operator for the mode $k$ in the $l$th energy band, and $L_{\text{tot}}=Ll_0$ is the total inductance with $L$ being the PCW total length. By adopting parameters in Table~I, we calculate the dispersion relation in Fig.~\ref{fig2m}(a). In the first Brillouin
zone (BZ) $k\in(-0.5k_m,0.5k_m]$, there is a band gap with no propagating mode around $k\simeq \pm 0.5k_m$ [see Fig.~\ref{fig2m}(a)]. In this paper, the atomic transition frequency $\omega_q$ is set to be in this gap.
Given that the coupling between the giant atom and the PCW is weak and time-independent, the atomic emission is
significantly suppressed due to no resonant mode in the band 
gap~\cite{Goban2014,Douglas2016,Ramos2016}. 
In this case, the system will be trapped in a bound state which most energy is localized around the coupling points and cannot 
propagate~\cite{GonzlezTudela2017}. 

To realize chiral emission, we should find a method to effectively shift $\omega_q$ to be resonant with the propagating modes, as well as encoding $\phi_{1,2}^{a,b}$ into coupling points. For those purposes, each coupling points of giant atoms can be
mediated with a Josephson loop (see Appendix A). In the following we first focus on the single atom case by neglecting the index $a,b$ for 
the considered parameters (for example, 
$\phi_{i}^{a,b}=\phi_{i}$ and $x^{a,b}_i=x_i$). The coupling junctions work as tunable 
inductances, which are time-dependently modulated via external flux $\Phi^{(i)}_{\text{ext}}$ through the 
coupling 
loops~\cite{Peropadre2013,Yin2013,Geller2015,Wulschner2016,
Kounalakis2018}. As discussed in Appendix A, the system interaction Hamiltonian reads 
\begin{equation}
	H_{\text{int}}=\hbar\sum_{l}\sum_{k} 
	\left[e^{-i[\omega_q-\omega_{l}(k)]t}g_{lk}(t) 
	a_{lk}^{\dagger}\sigma_{-}+\text{H.c.}\right],
	\label{Htime}
\end{equation}
where the interaction strengths $g_{lk}(t)$ are expressed in Eq.~(\ref{tHam}), which can be modulated time-dependently via external flux
$\Phi^{(1,2)}_{\text{ext}}(t)$~\cite{Geller2015,Wulschner2016}. Given 
that the modulating amplitude is small, $g_{lk}(t)$ is written as
\begin{gather}
	g_{lk}(t)=G_{k} \sum_{\pm}e^{\pm i(\Omega_d 
	t+\phi_{1})}\left[u_{lk} (x_{1})+e^{i(kx_d\pm \phi_c)}u_{lk} 
	(x_{2})\right], \notag \\
	G_{k}=\frac{A_1}{2}\frac{L_{0}^2}{L_T}\sqrt{\frac{  
	\omega_q\omega_{l}(k)}{L_{\text{tot}}L_Q}}e^{ikx_{1}},
	\label{GKL}
\end{gather}
where $x_d=x_2-x_1$, $\phi_c=\phi_{2}-\phi_{1}$ is the relative phase difference between two 
modulating signals, and $A_{1}$ is the first order amplitude of the time-dependent mutual inductance. The higher order terms $A_{n}$ ($n\geq 2$) are of extremely low amplitudes, which are neglected in our discussion. Note that $G_{k}$ is the coupling amplitude, with $\omega_q$ ($L_Q$) being transmon frequency (inductance), and $L_0$ ($L_T$) being the share branch (Josephson) inductance in the coupling loop. Detailed discussions can be found in Appendix A.

From Eq.~(\ref{GKL}), one finds that there are two sidebands induced by the time-dependent couplings [see Fig.~\ref{fig2m}(b)]. 
The atomic frequency $\omega_q$ is initially in the band gap and 
much closer to the first energy band ($l=1$). By choosing suitable 
$\Omega_d$, we require the parameters satisfying following conditions: 
First, the blue sideband $\omega_q+\Omega_d$ is still in the 
band gap and of large detuning with the second band, i.e., 
$\Delta_{\text{up}}\gg0$. Second, the red sideband 
$\omega_q-\Omega_d$ is resonant with the first band, and also far away 
from the band edge. Under these conditions, the interactions with higher energy bands ($l\geq2$) 
are all fast oscillating terms which can be neglected. There will be plenty of resonant modes in 1st band around $\omega_{l}(k_r)=\omega_q^{\text{eff}}$ [see Fig.~\ref{fig2m}(b)]. Consequently, the effective atomic frequency is now shifted as 
\begin{equation}
	\omega_q^{\text{eff}}=\omega_q-\Omega_d.
\end{equation}

Under the rotating wave approximation, only the red sideband term
will be involved in the evolution, and therefore, 
the index $l=1$ can be neglected, i.e., $lk\rightarrow k$. 
The Hamiltonian in Eq.~(\ref{Htime}) is reduced as
\begin{equation}
	H_{\text{int}}=\hbar\sum_{k} \left[g_{k} e^{-i\Delta_k t} a_{k}^{\dagger}\sigma_{-}+\text{H.c.}\right],
	\label{Hint1}
\end{equation}
where $\Delta_k=\omega_q^{\text{eff}}-\omega_{1}(k)$.
Note that the interaction becomes momentum-dependent due to the phase difference $\phi_c$, i.e.,
\begin{equation}
	g_{k}=G_{k}\left[u_{1k} (x_{1})+e^{i(kx_d-\phi_c)}u_{1k} (x_{2}) \right],
	\label{gk_m}
\end{equation}
where we set $\phi_{1}=0$ for simplicity. Hereafter, without loss of generality, we use $x_1 = 0$ since only relative distances
matter. The time-reversal symmetry of Maxwell equations requires
$u_{-1k}(x)=u_{1k}^{*}(x)$. Therefore, given 
that $\phi_c\neq 0$, the coupling strength $g_{k}$ is asymmetric
for the right ($k>0$) and left ($k<0$) propagating modes, i.e.,
\begin{equation}
|g_{k}|\neq |g_{-k}|, \quad \phi_c\neq 0.
\end{equation}

For example, by setting $x_d=N\lambda_m$ (N is an integer), 
$u_{1k}(x_{1})=u_{1k}(x_{2})$ is valid according to Bloch 
theory. The relation between $g_k$ and $k$ is simply derived as a
cosine form 
\begin{equation}
	|g_{k}|=G_{k}|u_{1k} (x_{1})| \cos\left(\frac{kx_d-\phi_c}{2}\right).
\end{equation}
Under the condition 
$k_rd-\phi_c=(2N+1)\pi$, the coupling strength to the right (left) propagating modes around $k_r$ ($-k_r$) is zero (non-zero). Since the intensity of emission spectrum is centered around the atomic frequency $\omega_q^{\text{eff}}$, $G_k\simeq G_{k_r}$ is approximately a constant. We plot $g_{k}$ versus $k$ for different $\phi_c$ in Fig.~\ref{fig2m}(c), which shows that the symmetry 
$|g_{k}|=|g_{-k}|$ is broken when $\phi_c\neq0$. Given that $x_d\neq N\lambda_m$, $u_{1k}(x_{1})\neq 
u_{1k}(x_{2})$, and the phase difference between $u_{1k}(x_{1})$ and 
$u_{1k}(x_{2})$ also affects the asymmetric behavior of $g_k$. 
The minima coupling strength for mode $k_r$ can be derived from the following transcendental equation 
\begin{equation}
	\arg\left[\frac{u_{1k_r} (x_{2})}{u_{1k_r} (x_{1})} e^{ik_rx_d}\right]-\phi_c=(2N+1)\pi.
	\label{phaserelation}
\end{equation}
In Fig.~\ref{fig2m}(c), $g_k$ for a nonperiodic distance $\{x_1, 
x_2\}=\{0.2\lambda_m,0.8\lambda_m\}$ is plotted. The numerical result
indicates that the minima coupling point $|g_{-k_r}|=0$ is at $\phi_c=0.28\pi$. 

%Now we explain the role of the band gap in our proposal. Without this band gap, the blue sideband coupling strength proportional to $u_{lk} (x_{1})+e^{i(kx_d+\phi_c)}u_{lk} (x_{2})$ will also contribute the evolution [see Eq.~(\ref{GKL})]. From Eq.~(\ref{gk_m}) we find that the flip sign before $\phi_c$ indicates a reversed interference relation for the left and right propagation modes, which will destroy the unidirectional emission produced by the red sideband transition. Therefore, due to the existence of the band gap, only one interference effect take apparent effect, while the opposite interference effect is significantly suppressed by the large detuning to the propagating modes. Therefore, our above discussions can also be applied to other kinds of waveguides with band gaps.

The decoupling mechanism between the giant atom and modes in one propagating direction is similar to realize chiral quantum phenomena via 
generating synthetic gauge fields in discretized lattice model~\cite{Koch2011a,Vermersch2016,Ramos2016,Roushan2017,Wang2020X}. Different from those studies, our proposal is based on giant atom effects in SQC platforms, and especially feasible for conventional continuous waveguides which are more robust to disorder noise than spin-chain channels. Next we discuss how to realize tunable chiral emission of photons by 
exploiting the momentum-dependent interaction induced by giant atom effects.

\section{Chiral emission of giant atoms}
\subsection{Non-Markovian dynamics}
\begin{figure}[tbph]
	\centering \includegraphics[width=8.6cm]{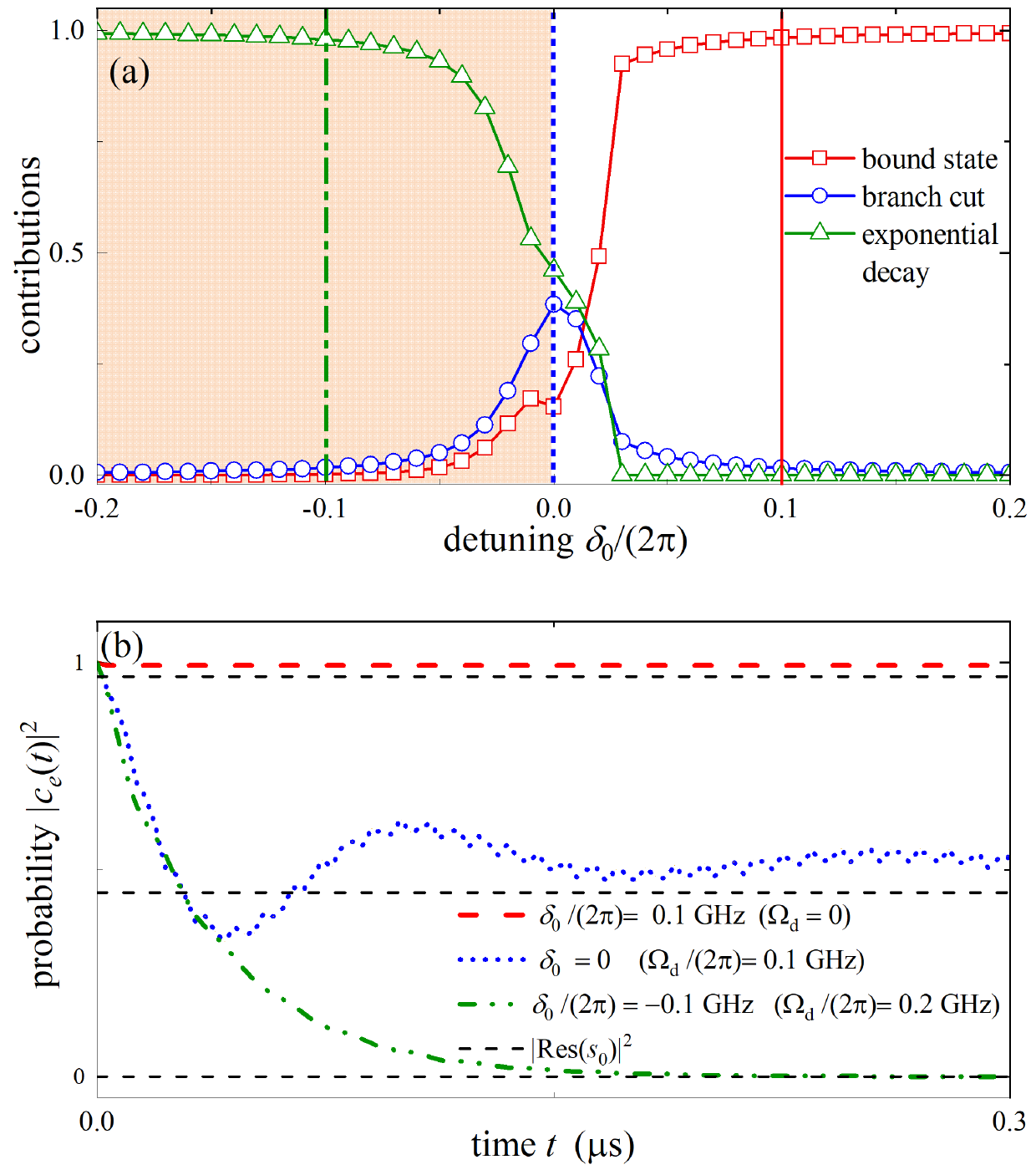}
	\caption{(a) The contributions to atomic dynamics at $t = 0$ of the bound state, branch cut and exponential decay change with $\delta_0$, respectively. (b) By tuning the drive frequency $\Omega_d$, the probability $|c_e(t)|^2$ of the atom being in its excited state changes with time for $\delta_0/(2\pi)=0.1~\text{GHz}$ (inside the band gap), $\delta_0=0$ (touching band edge) and $\delta_0/(2\pi)=-0.1~\text{GHz}$ (resonant with the 1st band). The dashed lines correspond to the steady state population calculated via the Residue theorem $|c_e(t=\infty)|^2=|\text{Res}(s_{0})|^2$. Parameters are the same with those in Fig.~\ref{fig2m}(c).}
	\label{fig3m}
\end{figure} 

As shown in Fig.~\ref{fig2m}(a), around the PCW band edge, the group velocity is $v_g\simeq0$, indicating that the wavepacket cannot propagate outside. Given that $\omega_q^{\text{eff}}$ is not far away from the band edge, both the non-decay bound state and sub-exponential decay (branch cut) will contribute significantly to the evolution due to extremely large density of states~\cite{Goban2014,Douglas2016,Liu2017}.
Specially, partial excitation in the giant atom will be trapped around the coupling points and cannot propagate to other nodes.
The emission process is highly non-Markovian. 
To exactly simulate non-Markovian dynamics and obtain the field 
distribution properties, we numerically calculate the unitary evolution 
governed by time-dependent Hamiltonian in Eq.~(\ref{Htime}) by adopting 
parameters listed in 
Table~\ref{table1}. The numerical methods are discussed in Appendix B.

\begin{figure*}[tbph]
	\centering \includegraphics[width=17.8cm]{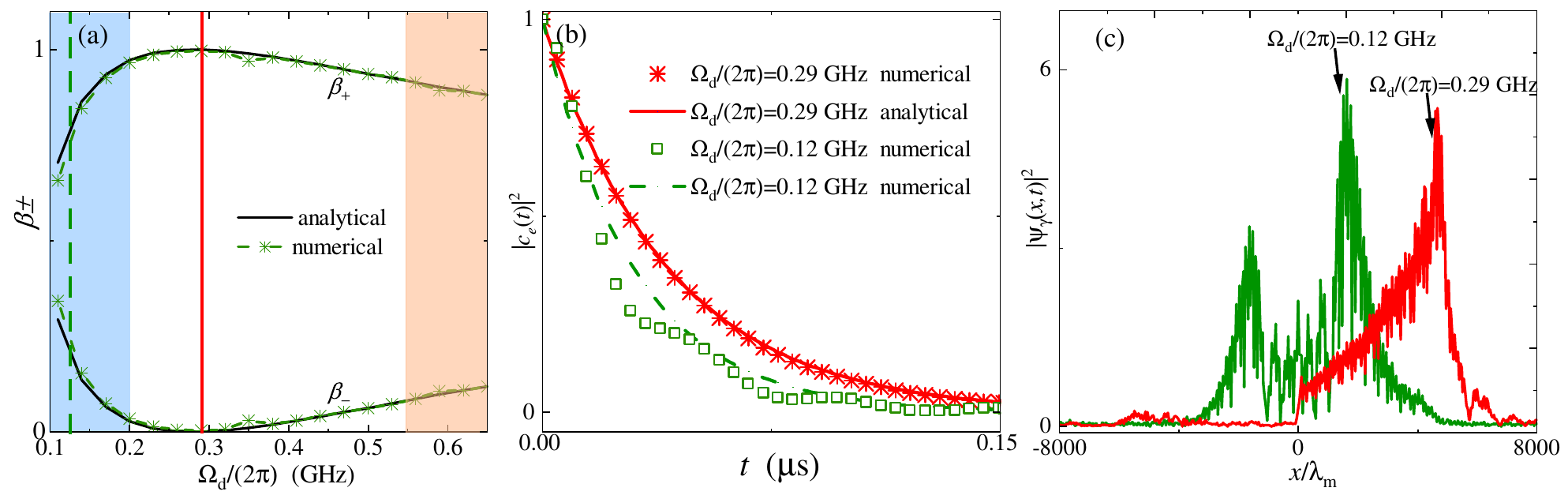}
	\caption{(a) The chiral factor $\beta_{\pm}$ change with the modulating frequency $\Omega_d$. The symbol (solid) curves are plotted according to the numerical (analytical) results given by Eq.~(\ref{Grate}) [Eq.~(\ref{beta_ana})]. In the blue area ($\Omega_d/(2\pi)<0.2~\text{GHz}$), the Markovian approximation cannot describe the decay process, and partial energy will be localized in the atom. In the orange region $\Delta_{\text{up}}/(2\pi)<0.1~\text{GHz}$ [see Fig.~\ref{fig2m}(b)], the blue sideband is too close to the 2nd band, which also destroys the chiral emission process. (b) The atom population $|c_{e}(t)|^2$ versus $t$ for $\Omega_d/(2\pi)=0.12~\text{GHz}$ and $\Omega_d/(2\pi)<0.29~\text{GHz}$. The field distributions at $t=0.15~\mu\text{s}$ are plotted in (c).}
	\label{fig4m}
\end{figure*}

Using complex analysis methods in Refs.
~\cite{cohen1998atom,GonzlezTudela2017}, non-Markovian contributions are evaluated via the resolvent operator techniques. Detailed discussions are 
presented in Appendix B. the normalized contribution weights
$w_i=W_{i}/(\sum_{i}W_i)$ changing with $\delta_0$ are shown in Fig.~\ref{fig3m}(a).
The bound state will dominate only when $\omega_q^{\text{eff}}$ is in the band gap area, i.e., $\delta_0\gg 0$. Given that $\omega_q^{\text{eff}}$ approaches the band top, these three contributions are of similar weight, as depicted in Fig.~\ref{fig3m}. 
When $\omega_q^{\text{eff}}$ is resonant with the continuous modes and far away from the band edge ($\delta_0\ll 0$), the exponential decay will dominate the evolution. 

By employing the parameters in 
Table~\ref{table1}, we numerically plot the dynamical evolution of $|c_e(t)|^2$ 
for different $\Omega_{d}$ in Fig.~\ref{fig3m}(b). The dashed horizon lines correspond to the atomic steady-state populations $|\text{Res}(s_{0})|^2$, which are obtained from complex analysis methods. In the long-time limit, the numerical evolutions asymptotically approach $|\text{Res}(s_{0})|^2$. Therefore, our numerical methods can well describe the non-Markovian dynamics due to band edge effects.
Given that $\Omega_{d}=0$, the atom hardly decays due to the large 
detuning ($\delta_0/(2\pi)=0.1~\text{GHz}$). The steady 
population is $|c_e(t=\infty)|^2\simeq 1$. 
The spontaneous radiation is strongly suppressed, and most energy is trapped in the atom [red dashed curve in Fig.~\ref{fig3m}(a)]. When gradually 
increasing the drive frequency $\Omega_{d}$, $\omega_q^{\text{eff}}$ approaches 
the band edge. Given that $\delta_0=0$, the atomic excitation only 
partially decays, while the rest part is trapped without decaying
[blue dashed line in Fig.~\ref{fig3m}(a)].
When $\Omega_{d}/(2\pi)=0.2~\text{GHz}$ ($\delta_0/(2\pi)=-0.1~\text{GHz}$), the evolution enters into the Markovian regime, where all the atomic energy decays into the PCW exponentially. Due to the mechanisms described above, the interaction between the giant atom and the PCW can be tailored freely by choosing different detuning $\delta_0$.

\subsection{Chiral emission in Markovian regime}
In a chiral quantum network, to release the information encoded in each node entirely, the excitation localized by the band edge effects should be avoided. In our proposal the effective atomic frequency $\omega_q^{\text{eff}}$ should be shifted far below the 1st band's top by adopting a large $\Omega_d$ [see Fig.~\ref{fig2m}(b)]. In this case, 
there will be plenty of modes coupling to the giant atom, which allows the atom emitting photons exponentially. Since $k_r$ is far away from the band top, the dispersion relation is approximately linear, i.e., $\Delta_k=\pm v_{g}\delta k_{\pm}$, where $\delta k_{\pm}\simeq k\pm k_r$ with $v_g$ being the group velocity at $k_r$. 
In the single-excitation subspace, the emission process is derived via Green function methods. The self-energy of the giant atom is written as (see Appendix B)
\begin{eqnarray}
\Sigma_{e}(s)&=&\sum_{i=\pm}\sum_{k}\frac{|g_{k}|^2}{s\mp iv_{g}\delta k_{\pm}} \notag \\
&\simeq& \sum_{i=\pm}|g'_{\pm k_{r}}|^2 \int_{\text{BZ}} d\delta k \frac{1}{s\mp iv_{g}\delta k_{\pm}}, \label{sigE}
\end{eqnarray}
where $\pm$ represent the right and left propagating modes, respectively.
Note that in Eq.~(\ref{sigE}) there will be a factor $L/(2\pi)$ when replacing the summation over $k$ by an integral~\cite{Scully1997}. Consequently, $g'_{\pm k_{r}}$ is expressed as
\begin{equation}
g'_{\pm k_{r}}=g_{\pm k_{r}}\sqrt{\frac{L}{2\pi}}. \label{g_renormal}
\end{equation}
As shown in Eq.~(\ref{GKL}), since $g_{\pm k_{r}}$ is proportional to $1/\sqrt{L}$ ($L_{\text{tot}}=Ll_0$), $g'_{\pm k_{r}}$ is the interacting strength independent of the waveguide length $L$, which is consistent with the Markovian spontaneous decay dynamics in the environment with an infinite length $L\rightarrow\infty$.
Both the decay rate and the energy shift can be derived from the transcendental
equation $s+\Sigma_{e}(s)=0$. By assuming the coupling strength varying slowly around the mode $\pm k_r$, we derive $\Sigma_{e}(s)$ as 
\begin{gather}
\Sigma_{e}(s)\simeq \sum_{i=\pm}i\Delta_{\pm}(s)+\Gamma_{\pm}(s), \\ 
\Delta_{\pm}(s)=\pm |g'_{\pm k_{r}}|^2\int_{\text{BZ}} d\delta k\frac{v_g\delta k_{\pm}}{s^2+(v_g\delta k_{\pm})^2}, \label{Lamb} \\
\Gamma_{\pm}(s)=|g'_{\pm k_{r}}|^2 \int_{\text{BZ}} d\delta 
 k\frac{s}{s^2+(v_g\delta k_{\pm})^2}.
\end{gather}
In the weak coupling regime, the integral bound is extended to be infinite. Moreover, the transcendental
equation $s+\Sigma_{e}(s)=0$ can be derived via the first-order iteration, i.e., by substituting  $s\rightarrow 0$ into $\Sigma_{e}(s)$~\cite{Wang2021}. 
Note that $\Delta_{\pm}(s)$ are the lamb shifts of the giant atom due to coupling with the PCW modes.
Finally we obtain the decay rate
\begin{equation}
\Gamma=\Gamma_{+}+\Gamma_{-}, \quad  \Gamma_{\pm}=\frac{\pi|g'_{\pm k_{r}}|^2}{v_g},
\label{Grate}
\end{equation}
where $\Gamma_{+}$ ($\Gamma_{-}$) is the decay rate into the right (left) propagating modes.

Note that the local decoherence rates are neglected in our discussions.
By fixing $x_d$ and $\phi_c$, there is an optimal point where the momentum-dependent coupling 
satisfies $g_{kr}\gg g_{-kr}\simeq 0$ [see Fig.~\ref{fig2m}(c)], indicating that the spontaneous emission is chiral with $\Gamma_{+}\gg \Gamma_{-}$. To show this, we first define the photonic wavefunction in real space~\cite{Scully1997} 
\begin{eqnarray}
\psi_{\gamma}(x,t)=\sum_{k}c_k(t)\sqrt{\frac{\hbar\omega_{k'}}{2L_{\text{tot}}}}e^{-ikx}
 u_{1k}^{*} (x).
\label{field_D}
\end{eqnarray}
As shown in Fig.~\ref{fig1m}, the distance between two coupling points is of order $\lambda_m$, which is 
much shorter than the photonic wavepacket, Therefore, the photonic flux 
emitted into the left (right) hand side of giant atom is defined as 
\begin{equation}
\Phi_{R/L} = \left| \int_{0}^{\pm \infty} |\psi_{\gamma}(x',t)|^2 dx' \right|, \quad t\rightarrow\infty.
\label{field_dis}
\end{equation}
The chiral factor $\beta$ is defined as~\cite{Lodahl2017} 
\begin{equation}
	\beta_{\pm}=\frac{\Gamma_{\pm}}{\Gamma_{+}+\Gamma_{-}}=\frac{\Phi_{R(L)}}{\Phi_{R}+\Phi_{R}},
	\label{beta_ana}
\end{equation}
where $\beta_{\pm}$ can be analytically (numerically) calculated 
according to Eq.~(\ref{Grate}) [Eq.~(\ref{beta_ana})]. 
In experiments, a transmon exposed in the noisy environment will experience both dissipation and dephasing simultaneously (i.e., finite lifetime $T_{1,2}$). Given that $T_{1,2}$ is short, the chiral factor in Eq.~(\ref{beta_ana}) should should be modified (see Ref.~\cite{Lodahl2017}). As discussed in Ref.~\cite{Wang2022}, current fabrication technology can increase transmon's lifetime as long as $T_{1,2}\sim 0.1~\text{ms}$. In chiral quantum networks, the information encoded in each node is often released into the quantum channel rapidly. In our discussions, the decay rate into the chiral PCW channel is set around $\Gamma_{\pm}/(2\pi)\sim 3~\text{MHz}$ (see Table~\ref{table1}). As depicted in Fig.~4(b) and Fig.~6(b), most energy of the photon will be chirally emitted within $0.1~\mu\text{s}$. Due to $T_{1,2}\gg \Gamma^{-1}_{\pm}$, it is reasonable to neglect the local decoherence in our discussions.

In Fig.~\ref{fig4m}(a), we plot $\beta_{\pm}$ versus
$\Omega_d$ by setting the coupling points at $\{x_1, 
x_2\}=\{0,\lambda_m\}$. Given that 
$\Omega_{d}/(2\pi)=0.29~\text{GHz}$, the giant atom dissipates almost 
all its energy into the right direction. In the Markovian 
regime, both the chiral factor and decaying dynamics based Eq.~(\ref{Grate}) 
and Eq.~(\ref{beta_ana}) [solid curves in Fig.~\ref{fig4m}(a, b)] match well the numerical results (curves with symbols). 

Given that $\Omega_{d}/(2\pi)=0.12~\text{GHz}$, the detuning to the 
band top is $\delta_0/(2\pi)=0.02~\text{GHz}$, indicating that the 
effective atomic frequency $\Omega_q^{\text{eff}}$ is too close to 
the band edge, and the giant atom's evolution is of sub-exponential 
decay [see 
the green curves in Fig.~\ref{fig4m}(b)]. Both the bound state and branch cut will lead to non-Markovian 
dynamics (see Appendix B).
Moreover, compared with $\Omega_{d}/(2\pi)=0.29~\text{GHz}$, the 
emission becomes bidirectional, as well as the photonic field propagates at a 
lower group velocity due to the band edge effects [see 
Fig.~\ref{fig4m}(c)]. Those unwanted non-Markovian dynamics take apparent effects within 
the blue area $\Omega_d/(2\pi)<0.2~\text{GHz}$ in
Fig.~\ref{fig4m}(a). To avoid this, one should employ a large 
modulating frequency $\Omega_{d}$ to shift $\Omega_q^{\text{eff}}$ 
far away from the band top. 
\begin{figure}[tbph]
	\centering \includegraphics[width=8.6cm]{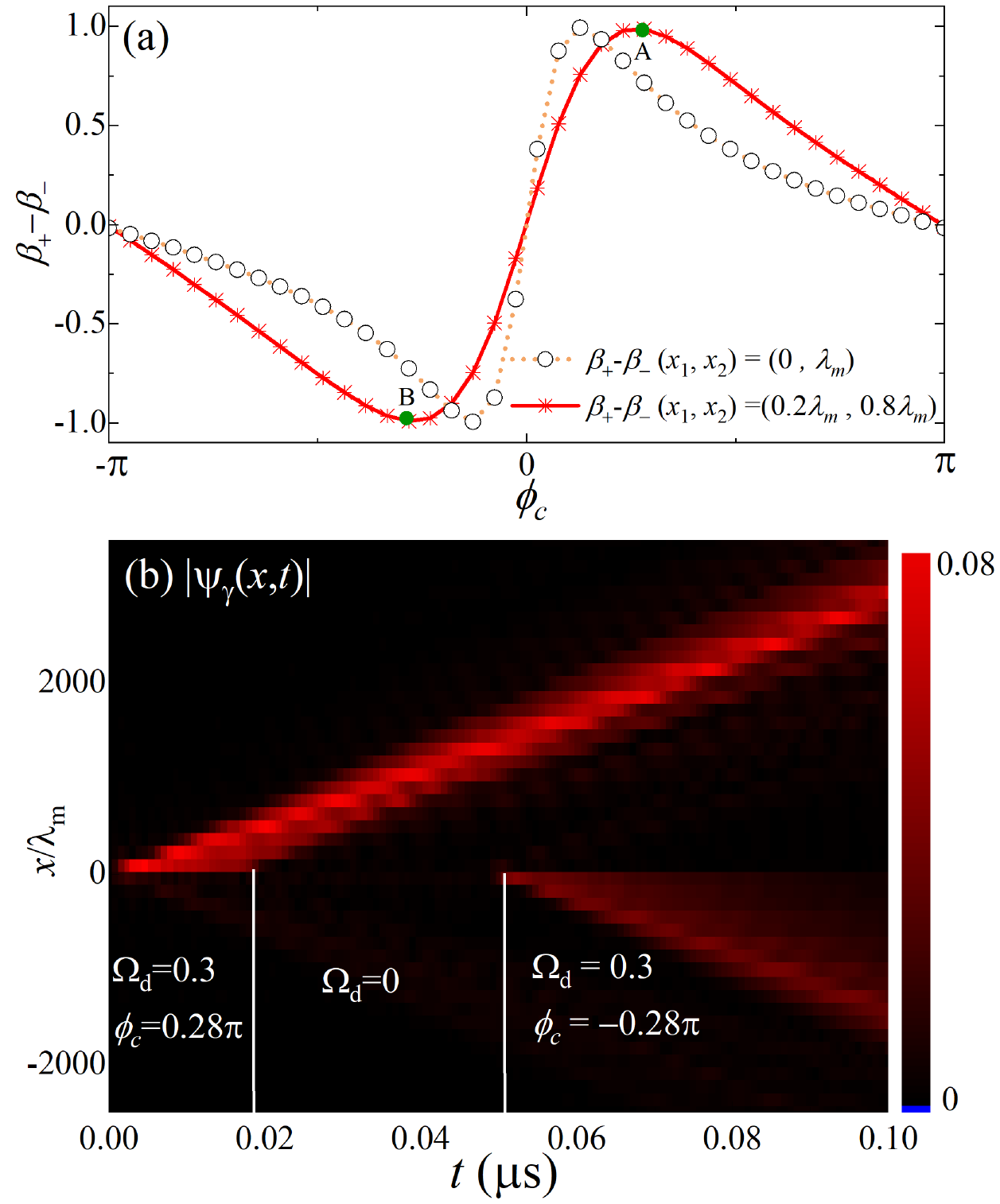}
	\caption{(a) For
		$(x_1,x_2)=(0,\lambda_m)$ and 
		$(x_1,x_2)=(0.2\lambda_m,0.8\lambda_m)$, $\beta_{+}-\beta_-$ 
		changes with $\phi_c$. (b) The field 
		distribution evolution under the following control sequence: 
		$0<t<0.016~\mu\text{s}$ (right chiral emission $\beta_{+}\simeq 
		1$); 
		$0.016~\mu\text{s}<t<0.048~\mu\text{s}$ [the atom is protected by 
		the band gap without decay ($A_1=0$ and $\Gamma=0$)]; 
		$0.048~\mu\text{s}<t<0.1~\mu\text{s}$ (left chiral emission 
		$\beta_{-}\simeq 1$). These operations split a single excitation 
		into 
		two parts propagating in the opposite directions. The parameters 
		for the right (left) chiral emission is adopted the same as 
		point A (B) in (a).}
	\label{fig5m}
\end{figure}

However, $\Omega_{d}$ cannot increase without any limitations. There is an upper bound determined by the detuning $\Delta_{\text{up}}$ to the 2nd band. In Fig.~\ref{fig4m}(a), the orange region $\Omega_d/(2\pi)>0.55~\text{GHz}$ corresponds to $\Delta_{\text{up}}/(2\pi)<0.1~\text{GHz}$, where the 2nd band might be involved in the evolution, and should be avoided in experiments. 
Due to these limitations,  the modulating frequency is limited in 
the range $\Omega_{d}/(2\pi)\in (0.2\simeq 0.55)~\text{GHz}$. Consequently, the emission rate in our numerical calculations is about $\Gamma_{\pm}/(2\pi)\in (1\simeq 
3)~\text{MHz}$  (see Table~\ref{table1}).

By tuning the phase difference $\phi_c$ between two coupling points, 
the chiral direction can be reversed [see 
Fig.~\ref{fig5m}(a)]. Given that two coupling points are at 
$(x_1,x_2)=(0,\lambda_m)$, the maximum right/left chirality is achieved 
when $\phi_c=\pm0.12\pi$. The chiral direction can be continuously tuned by simply shifting the relative phase difference $\phi_c$. When the coupling points are shifted as 
$(x_1,x_2)=(0.2\lambda_m,0.8\lambda_m)$, the condition for 
$\beta_{\pm}\simeq1$ becomes $\phi_c=\pm 0.28\pi$ [green dotted curve in 
Fig.~\ref{fig5m}(a)]. Therefore, with a shorter coupling distance the 
phase separation between maximum right and left chiral emissions is 
larger.

Due to the band gap, the information in a giant atom 
can be protected without emission or be released into the PCW by 
changing the modulating signals (as summarized in Table~I).
We take an interesting process for example to demonstrate our 
proposal's flexibility. In Fig.~\ref{fig5m}(b), by considering 
different modulating signals applied in a three-step process, we 
plot the real-space field distribution $\phi(x,t)$ changing with time.
In the first step, the modulating signal is of the maximum right emission, and the giant atom begins to dissipate its energy into the right direction. In the second step, the modulating amplitude $A_1$ 
is switched to zero, and the giant atom is prevented from decaying with no photonic flux in the PCW.
In the last step, the modulating signal with left chirality is switched on, and the field 
is released to the left direction.
Such a control sequence can split a single atomic excitation into two 
parts propagating in the opposite directions. 

\section{Cascaded quantum system and state transfer process}
Our proposal in Fig.~\ref{fig1m} can be extended as a chiral quantum 
network by considering multiple giant atoms interacting with a PCW bus. 
In this scenario, there are three distinct 
topologies, which are described as the 
separated, nested and braided giant atoms~\cite{Kockum2018}. In a long-distance quantum network, the giant atoms are usually of the conventional separated form. When 
the separation distances are comparable to the giant atom size,
considering nested and braided giant atoms becomes necessary~\cite{Soro2021}, which will be addressed in our future research.

As discussed in Appendix C, the SLH formalism can be employed to derive the master equation for two separated giant atoms chirally interacting with a common waveguide. Given that each giant atom is tuned with the maximum right chirality, i.e., $\beta_{+}=1$, the cascaded master equation is derived as
\begin{equation}
\dot{\rho}=-i(H_{\text{eff}}\rho-\rho H^{\dagger}_{\text{eff}})+L_R\rho L^{\dagger}_R. 
\label{cas_me}
\end{equation}
The non-Hermitian Hamiltonian $H_{\text{eff}}$ and the jump operator are respectively expressed as
\begin{eqnarray}
H_{\text{eff}}=\sum_{i=a,b}\frac{\omega_i}{2}\sigma_i^z -i\frac{1}{2}(S_a^{\dagger}S_a+S_b^{\dagger}S_b+2S_b^{\dagger}S_a), \label{cas_H}  \\
	L_R=S_{a}+S_{b}, \quad S_{a,b}=2i\sin \left( \phi_{a,b} \right) \sqrt{\frac{\gamma _{a,b}}{2}}\sigma _{-}^{a,b},
\end{eqnarray}
where $\sqrt{\gamma_i}$ is the interacting strength between giant atom $i$ 
and the PCW, which are assumed to be identical for each coupling 
point, and $\phi_{i}$ is the propagating phase between two coupling 
points for giant atom $i$ (see Appendix C). 
The non-Hermitian Hamiltonian $H_{\text{eff}}$ contains the 
nonreciprocal term $S_b^{\dagger}S_a$, which describes the chiral 
transport from atom $a$ to $b$ without information back flow. The last term in Eq.~(\ref{cas_me}) represents the quantum jump process by decaying a photon into the PCW irreversibly. Note that the master 
equation does not contain the retardation effects describing the wavepacket propagating between $a$ and $b$. Therefore, 
Eq.~(\ref{cas_me}) is valid when the distance between two atoms 
$L_{ab}=x_1^b-x_2^a$ is much shorter than wavepacket length. To describe the time-delay effect, we still use the numerical method in Appendix B to calculate the time-delay effects.

Given that decay rate $\Gamma_{\pm}$ changes with time, the 
chiral wavepacket can be tailed to the desired shape. In 
Refs.~\cite{Korotkov2011,Stannigel2011}, it was demonstrated that the perfect re-absorption is possible when the chiral emitted field is of time-reversal symmetry. Based on this mechanism, we discuss how to 
realize a high-fidelity state 
transfer between giant atom $a$ and $b$ in our proposal.
\begin{figure}[tbph]
	\centering \includegraphics[width=8.6cm]{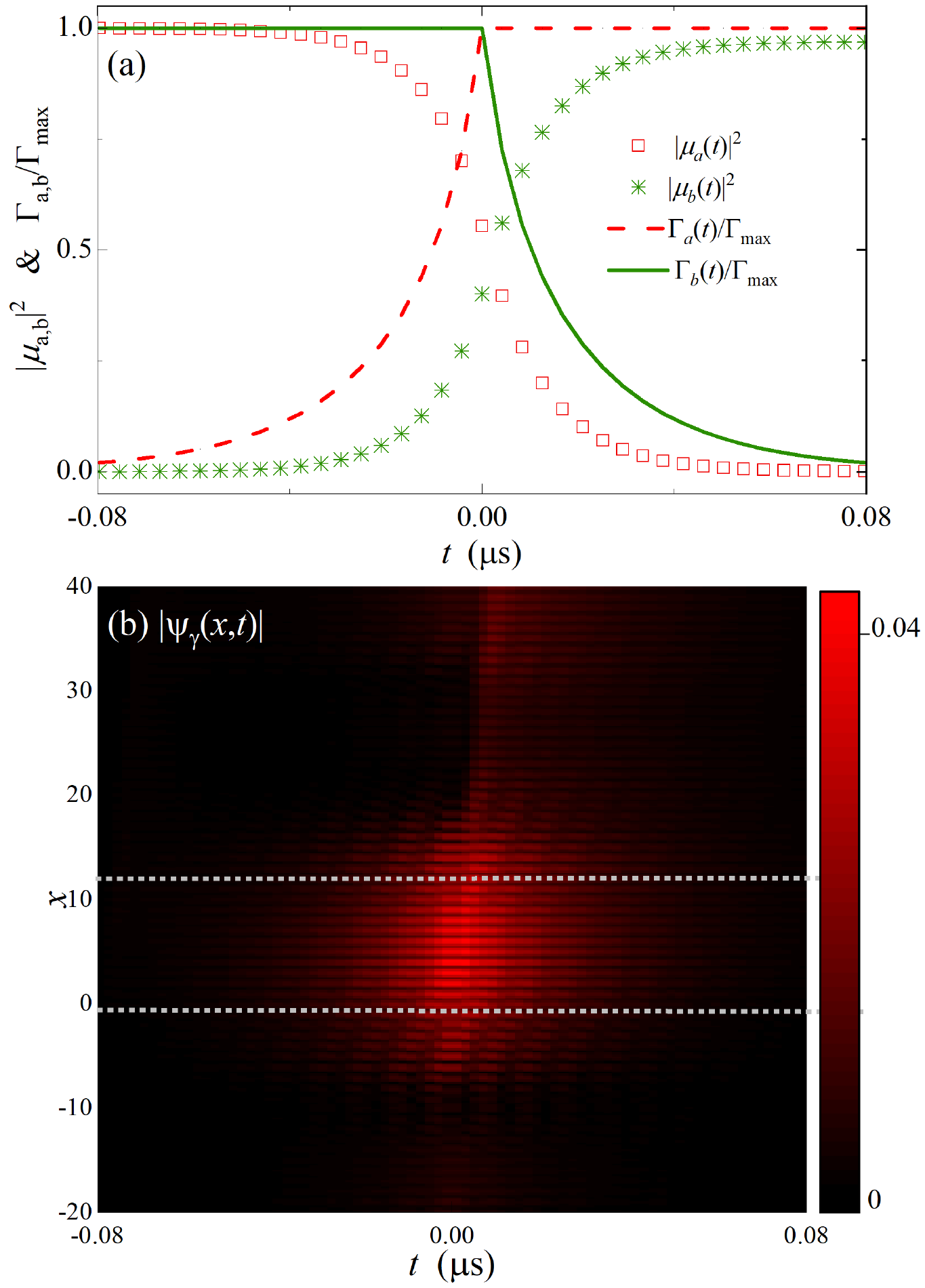}
	\caption{(a) The transfer process of a single-excitation from giant atom $a$ to $b$ in the cascaded quantum network based on our proposal. The separation distance is set as $L_{ab}=8\lambda_m$. The decay rates $\Gamma_{a,b}(t)$ are given by Eq.~(\ref{gsequence}), which are controlled by the modulating amplitude $A_1(t)$. (b) During the transfer process, the field distribution along the PCW versus time $t$ and position $x$. The dashed lines correspond to the positions of atom $a$ and $b$. Parameters are adopted as the same with those for point A in Fig.~\ref{fig5m}(a).}
	\label{fig6m}
\end{figure}

In a state transfer process, the initial state of two nodes is assumed to be
\begin{equation}
|\psi(t_i)\rangle=(c_e|e_1\rangle+c_g|g_1\rangle)\otimes|0_{\text{ch}}\rangle\otimes|g_2\rangle,
\end{equation}
where atom $a$ ($b$) is in an arbitrary superposition (ground) 
state, and
$|0_{\text{ch}}\rangle$ represents the PCW in its vacuum state. 
Given that the transfer process finishes at $t_f$ without any infidelity, the 
final state should be written as 
\begin{equation}
|\psi(t_f)\rangle=(|g_1\rangle)\otimes|0_{\text{ch}}\rangle\otimes 
(c_e|e_1\rangle+c_g|g_2\rangle).
\end{equation}
In experiments, the incoherent process [for example, the 
single-photon loss in Eq.~(\ref{cas_me})] will destroy the 
transfer fidelity. 
Therefore, the state of the system at $t$ is expressed as~\cite{Cirac1997} 
\begin{eqnarray}
|\psi_t\rangle&=&\mu_g(t)e^{i\Phi_+(t)}|gg0_{\text{ch}}\rangle+
\big[\mu_a(t)e^{-i\Phi_-(t)}|eg0_{\text{ch}}\rangle \notag \\
&&+\mu_b(t)e^{i\Phi_-(t)}|ge0_{\text{ch}}\rangle+\sum_{k} \alpha_k(t) |gg1_k\rangle \big],
\end{eqnarray}
where $$\Phi_{\pm}(t)=\frac{1}{2}\int_{t_i}^{t}\omega_{qa}^{\text{eff}}(t')dt'\pm\frac{1}{2} \int_{t_i}^{t}\omega_{qb}^{\text{eff}}(t')dt'$$
are the dynamical phases. The Lamb shift in Eq.~(\ref{Lamb}) should also be considered, i.e., $\omega_{qi}^{\text{eff}}(t)=\omega_{qi}^{\text{eff}}+\Delta_{+i}(t)+\Delta_{-i}(t)$. Similar discussion can be found in Ref.~\cite{Stannigel2012}. The Lamb shifts $\Delta_{\pm i}(t)$ for two giant atoms are derived from the transcendental
equation $s+\Sigma_{e}(s)=0$. Additionally $\Delta_{\pm i}(t)$ will be time-dependent in the following discussion, which analytical formula is hard to obtain.
In the following discussion, the numerical simulation is restricted in single excitation subspace, i.e., $\mu_g(t)=0$, which can simplify the problems led by this unknown dynamical phase.

Another reason for working in single excitation subspace is that we want to focus on both the retardation and nonlinear dispersion effects which are beyond Markovian approximation. By setting $\mu_g(t)=0$ the method in Appendix B can be employed. Exploring the dynamics of an arbitrary superposition state or multiple excitations will be intriguing questions, which will be addressed in our future studies.
Note that $\alpha_k(t)$ denotes the probability of excitation leaking
into the PCW mode $k$. To minimize this, one can control system's 
evolution to satisfy the following dark-state 
condition~\cite{Vermersch2017}
\begin{equation}
[S_{a}(t)+S_{b}(t)]|\psi_t\rangle=0,
\label{dark_con}
\end{equation}
which restricts the evolution and time-dependent decay rates satisfying the following relation
\begin{equation}
	\sqrt{\frac{\Gamma _{a}(t)}{2}}\mu_a(t)+ \sqrt{\frac{\Gamma 
	_{b}(t)}{2}}\mu_b(t)=0,	
\label{darkc}
\end{equation}
where $\Gamma _{a,b}(t)$ are defined as
\begin{equation}
\sqrt{\frac{\Gamma _{i}(t)}{2}}=2\sin \left( \phi _{i} 
\right)\sqrt{\frac{\gamma _{i}(t)}{2}}, \quad i=a,b.
\end{equation}
By combining Eq.~\ref{cas_me}) and Eq.~(\ref{darkc}), one can obtain the evolution functions $\mu_{a,b}(t)$ as
\begin{gather}
\dot{\mu}_{a}(t)=-\frac{\Gamma _{a}(t)}{2}\mu_{a}(t), \\
\dot{\mu}_{b}(t)=-\frac{\Gamma _{b}(t)}{2}\mu_{b}(t)-\sqrt{\Gamma 
_{a}(t)\Gamma _{b}(t)}\mu_{a}(t).
\end{gather}
The perfect state transfer requires the initial and final states satisfying the following boundary conditions
\begin{equation}
	\mu_{a}(t_i)=\mu_{b}(t_f)=1, \quad \mu_{a}(t_f)=\mu_{b}(t_i)=0. \label{boundary}
\end{equation}
Therefore, a high-fidelity state transfer process should satisfy the 
dark-state requirement [Eq.~(\ref{dark_con})], as well as the 
boundary condition in Eq.~(\ref{boundary}). For convenience we set $t_i=-t_f$.
To find suitable solutions, the photonic 
wavepacket from atom $a$ can be tailed with time-reversal symmetry~\cite{Korotkov2011}. In this 
case, ideal absorbing by atom $b$ is realized by 
considering a time-reversal decay rate of atom $a$, i.e., 
$\Gamma_b(t)=\Gamma_a(-t)$. As discussed in 
Ref.~\cite{Stannigel2011}, in the limit $t_i\rightarrow \infty$, the 
following control sequences satisfy 
all the above requirements 
\begin{eqnarray}
	\Gamma_{a}(t)=\Gamma_{b}(-t)=\left\{
	\begin{array}{lr}
		\Gamma_{\text{max}}\frac{e^{\Gamma_{\text{max}}t}}{2-e^{\Gamma_{\text{max}}t}}, \quad t<0,   \\
		\Gamma_{\text{max}}  \qquad t\geq 0.
	\end{array} 
\right.
\label{gsequence}
\end{eqnarray}
From Eq.~(\ref{GKL}) and Eq.~(\ref{Grate}), one finds that the decay rates can be controlled by the Fourier amplitude $A_1$ according to the following relations
	\begin{equation}
		\Gamma_{a(b)}\propto |g'_{\pm k_{r}}|^2 \propto A_{1}^2.
	\end{equation}
Therefore, the decay sequences of two giant atoms are 
realized by designing the time-dependent
Fourier amplitude $A_1(t)$ according to Eq.~(\ref{gsequence}), which corresponds to changing the amplitude of control flux in experiments (see Fig.~\ref{fig1A} and Fig.~\ref{fig2A}).

\begin{figure}[tbph]
	\centering \includegraphics[width=8.4cm]{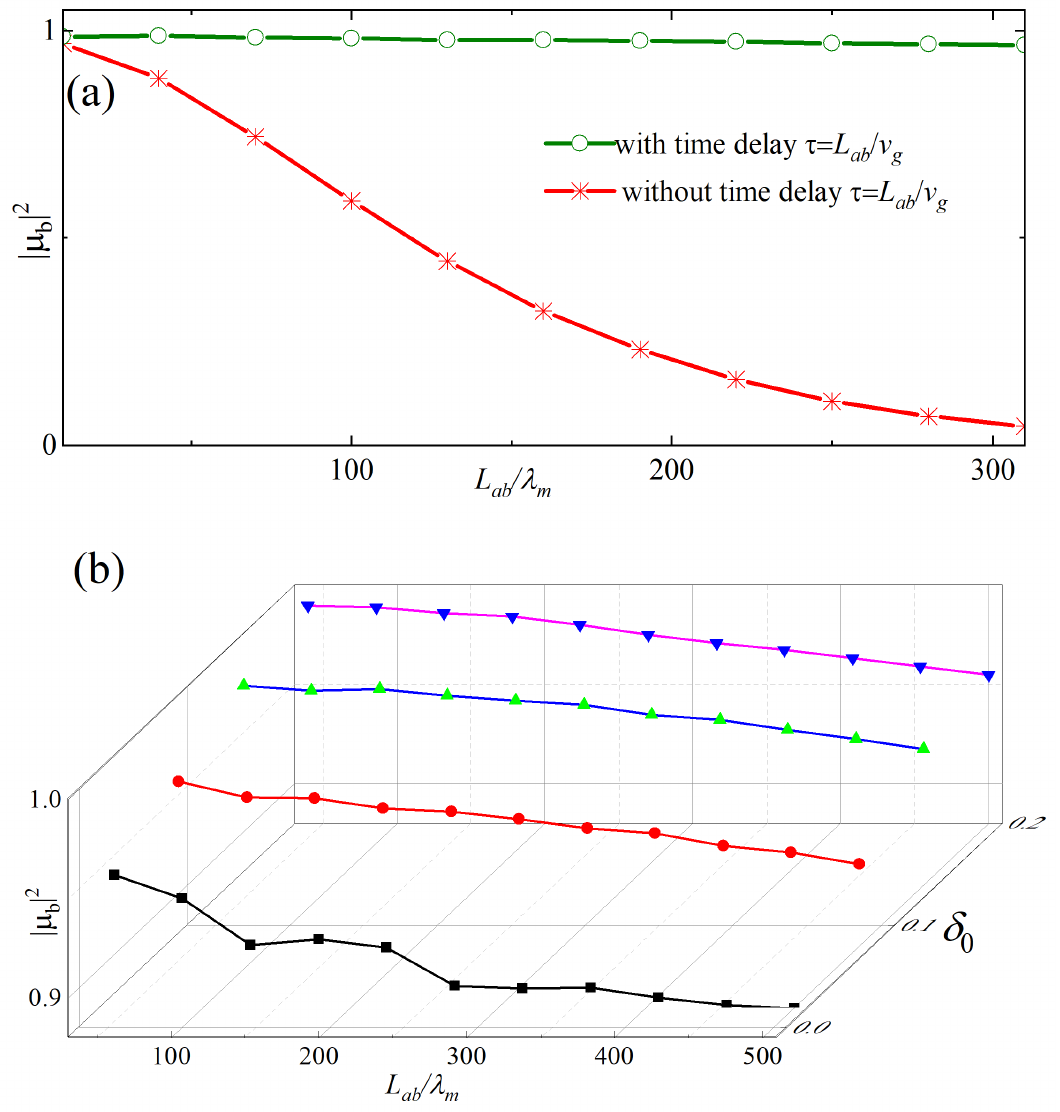}
	\caption{(a) Given that an initial excitation in atom $a$, the final transition probability $|\mu_b|^2$ changes with the separation distance $L_{ab}$ with (without) time-delay correction. (b) For different frequency detuning $\delta_0$ to the band edge, the transition probability $|\mu_b|^2$ changes with  $L_{ab}$ (with time-delay correction).}
	\label{fig7m}
\end{figure}

The chiral transfer process between $a$ and $b$ is presented in Fig.~\ref{fig6m}(a). 
At $t_f=0.08~\mu\text{s}$, the transfer 
probability is about $|\mu_{b}(t_f)|^2\simeq0.97$. The transfer 
fidelity can be enhanced by adopting a larger time period $t_f-t_i$. 
The time-dependent evolution of the field distribution in PCW during the transfer process is shown in Fig.~\ref{fig6m}(b) . One finds that the highest field intensity appears around $t\simeq 0$, which corresponds to the peak of 
time-reversal symmetric wavepacket. Due to dark-state 
conditions, the field is strongly localized between 
two atoms, with little energy leaking outside, which ensures the
high-fidelity transfer process. Therefore, both the numerical and analytical results indicate that our proposal is a well-performed chiral quantum system.

Given that $L_{ab}\gg \lambda_m$, the retardation time  
from $a$ to $b$ will take apparent effects. Since the propagating
time is approximately $\tau \simeq L_{ab}/v_g$, the modulating decay 
signals in Eq.~(\ref{gsequence}) should be modified as 
$\Gamma_{a}(t)=\Gamma_{b}(\tau-t)$.
In Fig.~\ref{fig7m}, by plotting $|\mu_b(t_f)|^2$ versus $L_{ab}$, we find that with the
time-delay signals the photon will be re-absorbed by node $b$ with high probabilities compared with the processes without time delay. However, due to PCW's nonlinear dispersing effects, the wavepacket becomes wider when increasing $\tau$ [see Fig.~\ref{fig5m}(b)]. Consequently, $|\mu_b|^2$ slightly decreases with $L_{ab}$ even with the time-delay. In Fig.~\ref{fig7m}(b), we plot $|\mu_b|^2$ versus $L_{ab}$ and detuning to the band top $\delta_0$. As discussed in Sec.~III and Appendix B, when $\omega^\text{eff}_q$ is too close to the band top, both the non-Markovian effects and nonlinear dispersion of the edge modes become apparent, which lead to a decrease of transfer $|\mu_b|^2$. Due to those effects, the fidelity decays with $L_{a,b}$ much faster when $\delta_0$ becomes smaller. To achieve a better transfer fidelity between remote nodes, one can shift the effective atomic frequency $\omega_{q}^{\text{eff}}$ far away from the band top.

\section{Conclusion and outlooks}
In this work, we discuss how to realize a chiral quantum network by 
exploiting quantum interference effects in SQC giant atoms. 
By considering time-dependent interactions with a photonic crystal waveguide, the coupling points can be encoded with different local 
phases. 
The asymmetric interference effects for the 
opposite directions will lead to chiral spontaneous 
emission of photons. We also find the parameter regimes where 
the non-Markovian dynamics led by band edge effects can be suppressed.  
The chiral factor in our 
proposal can approach 1, and both the emission direction and rate 
can be continuously
tuned by the modulating signals. Moreover, the release of 
information encoded in the giant atom can be turn on/off on demand.

Due to the tunability of our proposal, high-fidelity unidirectional 
QIP tasks, for example, the state transfer between remote nodes, can 
be realized. Compared with the classical ferrite circulators, our 
method chirally routes photons 
without strong magnetic fields, and can easily be integrated 
on the chip without 
additional overheads. 
In recent years, the interests in employing giant atoms for quantum 
information processes are increasing 
rapidly~\cite{Kannan2020,Wang2021,Soro2021}. In future, it might be 
possible to combine both small and giant atoms in superconducting 
quantum information processors to exploit their advantages and 
achieve better performance. We hope that our proposal can be a versatile quantum interface for chiral routing microwave photons in future SQC quantum networks. 

\section{Acknowledgments}
The quantum dynamical simulations are based on open source code 
QuTiP~\cite{Johansson12qutip,Johansson13qutip}. We thank Dr. Wen-Xiao Liu for helping polish the expressions of the whole manuscript.
X.W.~is supported by 
the National Natural Science
Foundation of China (NSFC) ( No.~12174303 and Grant No.~11804270), and China Postdoctoral Science Foundation No.~2018M631136.
H.R.L. is supported by the National Natural Science
Foundation of China (NSFC) (Grant No.11774284).

\emph{Note added--} We notice a similar work by Yu-Xiang Zhang et al.~\cite{Zhang2021}.

\section*{APPENDICES}
\setcounter{equation}{0}
\renewcommand{\theequation}{A\arabic{equation}}
\setcounter{figure}{0}
\renewcommand{\thefigure}{A\arabic{figure}}\

\begin{appendix}	

\section{Time-dependent coupling between superconducting atoms and PCW}
\subsection{tunable mutual inductance}
\begin{figure}[tbph]
	\centering \includegraphics[width=7.6cm]{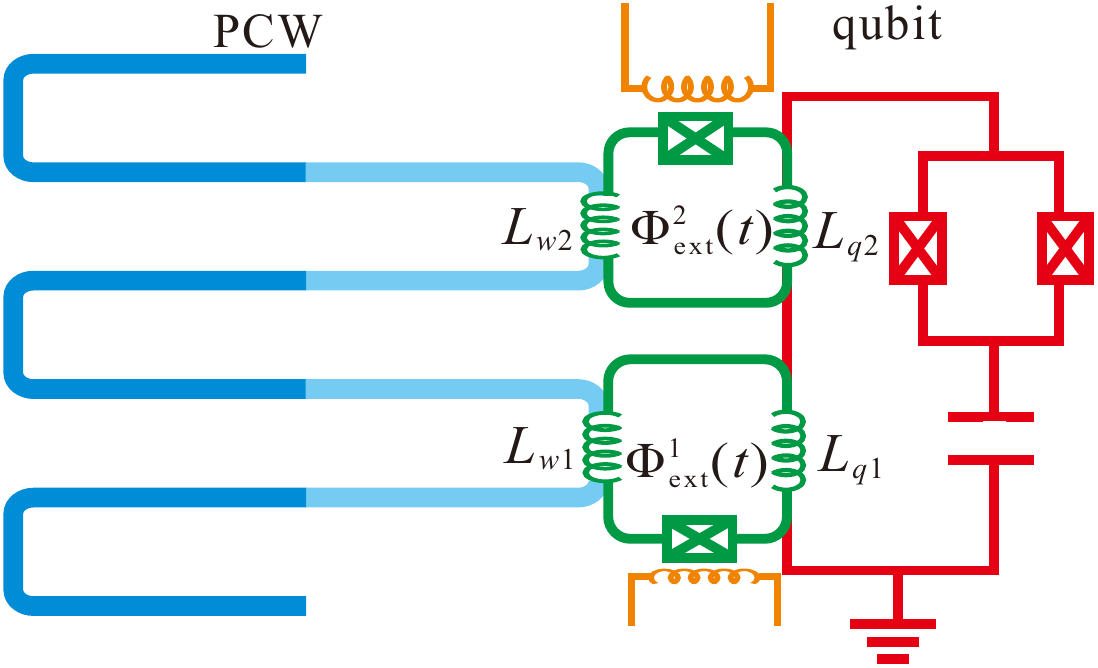}
	\caption{Tunable coupling between a photonic crystal waveguide 
		(PCW) and a superconducting giant atom of transmon form. The dark (light) blue lines 
		represent the 
		high (low) impedance positions of the PCW. The coupling points are 
		mediated via loops with Josephson junctions (green crosses). 
		At coupling point $i$, $L_{wi}$ and 
		$L_{qi}$ are the shared branch inductances in the PCW and giant atoms, 
		respectively. To realize time-dependent couplings, an external 
		time-dependent flux $\Phi_i(t)$ are applied.}
	\label{fig1A}
\end{figure}
As depicted in Fig.~\ref{fig1A}, a superconducting giant atom interacts with the PCW at two points. Each coupling point is mediated by a Josephson junction inserted in a loop. The inductance 
$L_{wi}$ and $L_{qi}$ of the $i$th loop ($i=1,2$) are the shared 
branch in the PCW and giant atoms, respectively. The 
gauge-invariant phase difference across Josephson inductance in loop 
$i$ is denoted as $\phi^{(i)}_J$. The intermediate junction can be viewed as a 
lumped inductance $L_{i}$ as
\begin{equation}
	L_{i}=\frac{L_{T}}{\cos\phi^{(i)}_J}, \quad L_{T}=\frac{\Phi_0}{2\pi I_c},
	\label{Lj0}
\end{equation}
where the critical currents of two junctions are assumed to be 
identical as $I_c$. Note that $L_{wi}$, $L_{qi}$ and $L_{j}$ forms 
the loop 
$\mathcal{C}_i$ at the $i$th coupling point, through which an 
external flux bias $\Phi^{(i)}_{\text{ext}}$ is applied. The 
inductance branch $L_{wi}$ ($L_{qi}$) is much smaller than the total inductance of the atom (PCW waveguide). The boundary relation of the 
loop $\mathcal{C}_i$ is given by~\cite{Geller2015,Wulschner2016}
\begin{equation}
	\phi^{(i)}_J=\int_{\mathcal{C}_i} \mathbf{A} d\mathbf{l} =\frac{2\pi}{\Phi_0}\left[\Phi^{(i)}_{\text{ext}}-(L_{wi}+L_{qi})I_{c}\sin\phi^{(i)}_J \right],
\end{equation}
from which one can find that $\phi^{(i)}_J$ is restricted by the 
following transcendental equation
\begin{equation}
	\phi^{(i)}_J+\beta \sin\phi^{(i)}_J=\frac{2\pi}{\Phi_0}\Phi^{(i)}_{\text{ext}}, \quad \beta= \frac{L_{wi}+L_{qi}}{L_{T}},
	\label{tranEQ}
\end{equation}
which shows that $\phi^{(i)}_J$ can be controlled by the external 
flux. Note that $\beta$ is the screening parameter and is assumed to 
be
identical for two junctions. Given that $\beta<1$, Eq.~(\ref{tranEQ}) describing the relation between 
$\phi^{(i)}_J$ and $\Phi^{(i)}_{\text{ext}}$ is single-valued. We 
assume 
$L_{wi}=L_{qi}=L_{0}$ for simplicity.
By applying the $\text{Y}-\Delta$ transformation for 
the coupling loop, the 
effective mutual inductance between PCW and giant atom  
is derived as~\cite{Wulschner2016}
\begin{equation}
	M_{gi}=\frac{L_{0}^2}{2L_0+L_{i}}=\frac{L_{0}^2}{L_T}\frac{\cos\phi^{(i)}_J}{1+\beta \cos\phi^{(i)}_J}.
	\label{mutualM}
\end{equation}
Therefore, the mutual inductance $M_{gi}$ is tunable 
by changing the external flux $\Phi^{(i)}_{\text{ext}}$. The modulating 
relation is found from the transcendental Eq.~(\ref{tranEQ}) and 
Eq.~(\ref{mutualM}). 
Moreover, the additional inductance for the 
giant atom due to the coupling loop  is 
\begin{eqnarray}
	L_s&=&\sum_{i=1,2}\left(\frac{L_{0}^2}{L_T}\frac{\cos\phi^{(i)}_J}{1+\beta \cos\phi^{(i)}_J}+\frac{L_0}{1+\beta \cos\phi^{(i)}_J}\right) \notag \\
	&=&2L_0+M_{g1}+M_{g2}.
	\label{addL}
\end{eqnarray}
%\sum_{i=1,2 \left( L_{0}-\frac{\cos\phi^{(i)}_J}{1+\beta \cos\phi^{(i)}_J}\frac{L_{0}^2}{L_T}\right)}.

\begin{figure*}[tbph]
	\centering \includegraphics[width=17.8cm]{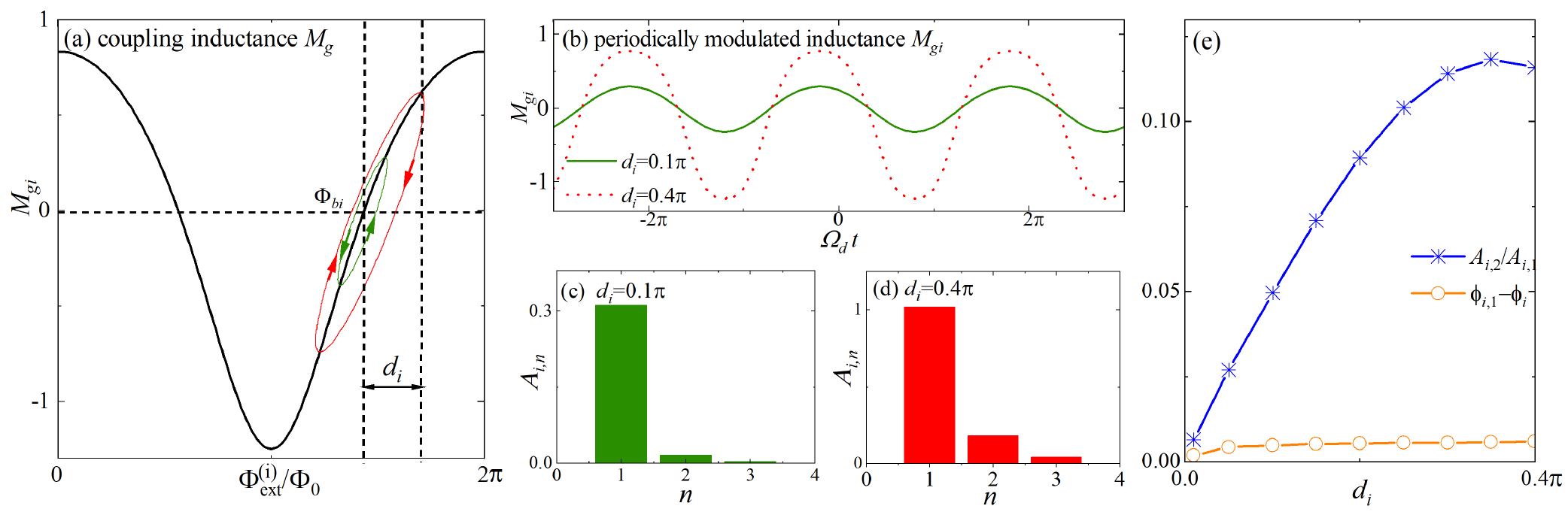}
	\caption{(a) By setting $\beta=0.2$, the effective mutual 
	inductance $M_{gi}$ (in the unit $L^2_{0}/L_{T}$) changes with 
	the external flux $\Phi^{(i)}_{\text{ext}}$. By applying a 
	time-dependent $\Phi^{(i)}_{\text{ext}}$, $M_{gi}$ is modulated 
	along a loop. (b) The inductance $M_{gi}$ changes with time given 
	that $d_i=0.1\pi$ and $d_i=0.4\pi$ respectively. (c) and (d) 
	correspond to the Fourier amplitudes $A_{i,n}$ of two signals in 
	(b), respectively. (e) The amplitude ratio $A_{n,2}/A_{n,1}$ and the phase 
	difference $\phi_{i,1}-\phi_{i}$ change with $d_i$.}
	\label{fig2A}
\end{figure*}

By assuming $\beta\ll 1$ (i.e., $L_0\ll L_{T}$), we obtain
$\phi^i_J\simeq \frac{2\pi}{\Phi_0}\Phi^{(i)}_{\text{ext}}$, and derive the
effective mutual inductance as
\begin{equation}	
M_{gi}=\frac{L_{0}^2}{L_T}\cos\left( \frac{2\pi}{\Phi_0}\Phi^{(i)}_{\text{ext}}\right),
\end{equation}
which shows that the mutual inductance $M_{gi}$ can be modulated by 
$\Phi^{(i)}_{\text{ext}}$ in a cosine form. Given that $\beta$ is comparable to 1, the modulation becomes nonlinear. We numerically plot $M_{gi}$ 
changing $\Phi^{(i)}_{\text{ext}}$ in Fig.~\ref{fig2A}(a). For example,  $\Phi^{(i)}_{\text{ext}}(t)$ can be periodically modulated as
\begin{equation}
	\Phi^{(i)}_{\text{ext}}=\Phi_{bi}+\frac{\Phi_0}{2\pi}d_i \cos(\Omega_d t+\phi_i).
\end{equation}
where $\Phi_{bi}$ is the dc part, $d_i$ and ($\phi_i$) is the modulating amplitude (phase) of the ac part at frequency $\Omega_d$. Figure~\ref{fig2A}(b) shows the mutual inductance $M_{gi}$ changing with time periodically given that $d_i=0.1\pi$ and $d_i=0.4\pi $, respectively~\cite{Roushan2017}. We analyze the frequency components of $M_{gi}(t)$ by expanding
it in the Fourier form 
\begin{equation}
	M_{gi}(t)=\frac{L_{0}^2}{L_T}\sum_{n=0}^{\infty}A_{i,n}\cos(n\Omega_d t+\phi_{i,n}).
	\label{Mgt}
\end{equation}

By numerically optimizing $\Phi_{bi}$, the dc component $A_{i,0}$ 
representing the time-independent coupling inductance can be 
eliminated. The amplitudes for each Fourier 
order are plotted in Fig.~\ref{fig2A}(c, d), which show that the contributions of the higher 
order terms ($n\geq2$) also increase with $d_i$. The ratio $A_{i,2}/A_{i,1}$ versus 
$d_{i}$ is plotted in 
Fig~\ref{fig2A}(e). It is found that
$A_{i,2}/A_{i,1}\ll 1$ is valid even when $d_i=0.4\pi$. Although
the first-order phase $\phi_{i,1}$ differs from $\phi_i$, their difference is 
very small, i.e., $\phi_{i,1}\simeq\phi_{i}$ even when $d_i$ is large. \emph{This is very 
important for our following discussions, since 
$\phi_{i,1}$ will directly determine the chiral direction of 
the giant atom.}

\subsection{Time-dependent interactions}
When inlcuding the additional inductance 
in Eq.~(\ref{addL}), the total 
inductance of the transmon is approximately expressed 
as~\cite{Koch2017}
\begin{equation}
	L_Q=L_q+L_s, \quad L_q=\left(\frac{\Phi_0}{2\pi}\right)^2 \frac{1}{E_J},
\end{equation}
where $E_J$ is the Josephson energy of the transmon, and $L_q$ 
is its Josephson inductance. Given that the transmon is 
of weak Kerr nonlinearity,
we can approximately view it as a Duffing oscillator, which quantization Hamiltonian reads 
\begin{equation}
	H_q=\hbar \Omega_q b^{\dagger}b-\frac{E_C}{12}(b+b^\dagger)^4, \quad \Omega_q=\frac{1}{\sqrt{L_QC_q}},
\end{equation}
where $C_q$ is the total transmon capacitance, $E_C=e^2/(2C_q)$ is the 
charge energy, and $b$ ($b^{\dagger}$) is the annihilation (creation) 
operator. According to Josephson relation,
the current operator of the transmon is approximately 
written as~\cite{Geller2015} 
\begin{equation}
	I_{q}\simeq \sqrt{\frac{\hbar \Omega_q}{2L_{Q}}}(b+b^\dagger). 
\end{equation} 
By considering two lowest energy levels, we write the Hamiltonian $H_q$ in the Pauli operators, i.e., by replacing $b^{\dagger}(b)\rightarrow \sigma_{+}(\sigma_{-})$
\begin{gather}
	H_q\simeq \frac{1}{2}\hbar \omega_{q}\sigma_{z}, \quad \omega_q=\Omega_q-\frac{E_C}{\hbar},   \\
	I_{q}=\sqrt{\frac{\hbar \omega_q}{2L_{Q}}}(\sigma_{-}+\sigma_{+}).
\end{gather}
In our discussion, the giant atom weakly couples to the PCW, i.e., 
$M_{gi}(t)<L_0\ll L_{q}$. Therefore, the transmon 
inductance is approximate as a constant with $L_Q\simeq L_q+2L_{0}$, which indicates the atomic 
frequency $\omega_q$ becomes time-independent.

The current operator is derived in Eq.~(\ref{Iw}), and the interaction is mediated by $M_{gi}(t)$. Therefore, the interaction Hamiltonian is
\begin{gather}
	H_{c}=\sum_{i=1,2} M_{gi}(t)I_{q}I_{w}\simeq\hbar\sum_{l}\sum_{k} \left[g_{lk}(t)a_{lk}^{\dagger}\sigma_{-}+\text{H.c.}\right], \notag \\
	g_{lk}(t)=\frac{1}{2}\sqrt{\frac{  
			\omega_q\omega_{l}(k)}{L_{\text{tot}}L_{Q}}} 
	\sum_{i=1,2}M_{gi}(t)e^{ikx_{i}} u_{lk} (x_{i}).
	\label{tHam}
\end{gather}
In our discussions, both $\Phi^{(1)}_{\text{ext}}(t)$ and $\Phi^{(2)}_{\text{ext}}(t)$ are monochromatic with identical frequency $\Omega_d$, \emph{but with different phases $\phi_{1,2}$.}
Assuming that $d_i$ is small, we neglect the higher Fourier orders ($n\geq2$) of $M_{gi}(t)$, and only consider the fundamental frequency component $A_1$. Since the phase of the first order satisfies $\phi_{1,i}\simeq\phi_{i}$ [see Fig.~\ref{fig2A}(e)], $M_{gi}(t)$ can be simplified as 
\begin{equation}
	M_{gi}(t)\simeq A_1\frac{L_{0}^2}{L_T}\cos(\Omega_d t+\phi_i),
	\
\end{equation}
where we assume $A_{i,1}=A_{1}$.
Consequently, the time-dependent interaction Hamiltonian 
in Eq.~(\ref{Htime}) is obtained.
 
\setcounter{equation}{0}
\renewcommand{\theequation}{B\arabic{equation}}

\section{Chiral emission process of giant atoms}
\subsection{Numerical methods for simulating chiral emission of giant atoms}
Although the cascaded master equation can describe the chiral photon flow between different nodes, the information such as field distribution and Non-Markovian dynamics led by band edge effects are all discarded. Those information is essential for our discussions in the main text. Therefore, we choose to numerically simulate the unitary evolution governed by the time-dependent Hamiltonian in Eq.~(\ref{Htime}), where both the atom and the photonic field information are kept. For simplicity, during the spontaneous emission process only one single excitation 
	is considered in the system. The state of the whole system is written as 
	$|\psi(t)\rangle=\sum_{k} c_k(t)|g,1_k\rangle+c_{e}(t)|e,0\rangle$. The steps of the numerical calculations are summarized as below:
	
	i) By adopting the circuit parameter in Table I, both eigen-frequencies and wavefunctions of the PCW are obtained according to Eq.~(\ref{waveeq}-\ref{phifs0}). Detailed methods can be found in Ref.~\cite{Wang2021}. In our 
    simulation, the mode number in the first BZ $k\in (-0.5k_m, 0.5k_m]$ is discretized as $N=10^4$, which is equal to consider a finite PCW with length $L=10^4\lambda_m$ in the real space. Such a large $L$ guarantees the propagating wavepacket never touching the boundary during the simulation.
	
	ii) In the single-excitation subspace, the Hamiltonian in Eq.~(\ref{Htime}) can be mapped into a matrix with dimension $N+Q$, where $Q$ is the atoms' number.
	Taking two giant atom ($Q=2$) for example, the matrix for the time-dependent Hamiltonian is 
	\begin{widetext}
		\begin{eqnarray}
		H_{\text{int}}=\left[ \begin{matrix}
				w_{lk_1}&		0&		...&		0&		g_{lk_1}\left( x_{1,2}^{a}, t \right) &		g_{lk_1}\left(  x_{1,2}^{b}, t \right)\\
				0&		w_{lk_2}&		\ddots&		...&		g_{lk_2}\left(  x_{1,2}^{a},t \right) &		g_{lk_1}\left(   x_{1,2}^{b},t \right)\\
				\vdots&		\ddots&		...&		0&		\vdots &		\vdots \\
				0&		...&		0&		w_{lk_N}&		g_{lk_N}\left(   x_{1,2}^{a},t \right) &		g_{lk_N}\left(   x_{1,2}^{b},t \right)\\
				g_{lk_1}^{*}\left( x_{1,2}^{a}, t \right)&		g_{lk_2}^{*}\left(  x_{1,2}^{a},t \right)&		...&		g_{lk_N}^{*}\left( x_{1,2}^{a}, t \right)&		w_{qa} &	0\\
				g_{lk_1}^{*}\left(  x_{1,2}^{b},t \right)&		g_{lk_2}^{*}\left(  x_{1,2}^{b},t \right)&		...&		g_{lk_N}^{*}\left(  x_{1,2}^{b},t \right)&		0 &		w_{qb}\\
			\end{matrix} \right],
		\label{Hintmat}
		\end{eqnarray}
	\end{widetext}
where $l$ and $k_i$ denote the energy band and wave number index, respectively. The coupling strength between atom $a$ ($b$) with mode $lk_i$ is $g_{lk_i}\left(  x_{1,2}^{a(b)}, t \right)$, which is numerically obtained from Eq.~(\ref{GKL}). Since the coupling positions' information is included into $g_{lk_i}\left(  x_{1,2}^{b}, t \right)$, the cascaded properties of noise and retardation effects due to field propagation are already considered in our calculations.
	
	iii) One can numerically solve the evolution governed by $H_{\text{int}}$ in Eq.~(\ref{Hintmat}). Note that the step of the discretized time should be much smaller than the modulating coupling period $T=1/\Omega_d$. In the simulation, the information of each step's state will be recorded. During the state transfer process, the time-dependent 
	decay rates of two giant atoms are controlled by the amplitude of $g_{lk_i}\left(x_{1,2}^{a(b)}, t \right)$.  The controlling sequence is encoded in Fourier amplitude $A_{1}(t)$ according to Eq.~(\ref{GKL}) and Eq.~(\ref{Grate}).
	
	iv) By extracting the amplitudes of all the modes $c_{lk}(t_i)$ for different $t=t_i$, one can recover $\psi_{\gamma}(x,t_i)$ via 
	Eq.~(\ref{field_D}), which describes the field distribution versus $x$. By plotting $\psi_{\gamma}(x,t_i)$ for 
	different time $t_i$, the spatiotemporal propagating processes of the photonic field in Fig.~\ref{fig5m} and Fig.~\ref{fig6m}(b) are
	obtained. Note that the nonlinear dispersion relation is already included into the diagonal terms $w_{lk_i}$.
	
	Compared with the cascaded master equations, the above numerical 
	method allows to observe both the field propagating effects and non-Markovian dynamics. Since the spatial distributing relations among multiple coupling points are described by $g_{lk_i}\left(  x_{1,2}^{b}, t \right)$, the cascaded properties (downstream/upstream relations of the chiral 
	noise) are also kept. Due to this, we can observe the interference due to propagating phases and retardation effects, which will be discussed in the main text.

\begin{figure}[tbph]
	\centering \includegraphics[width=8.6cm]{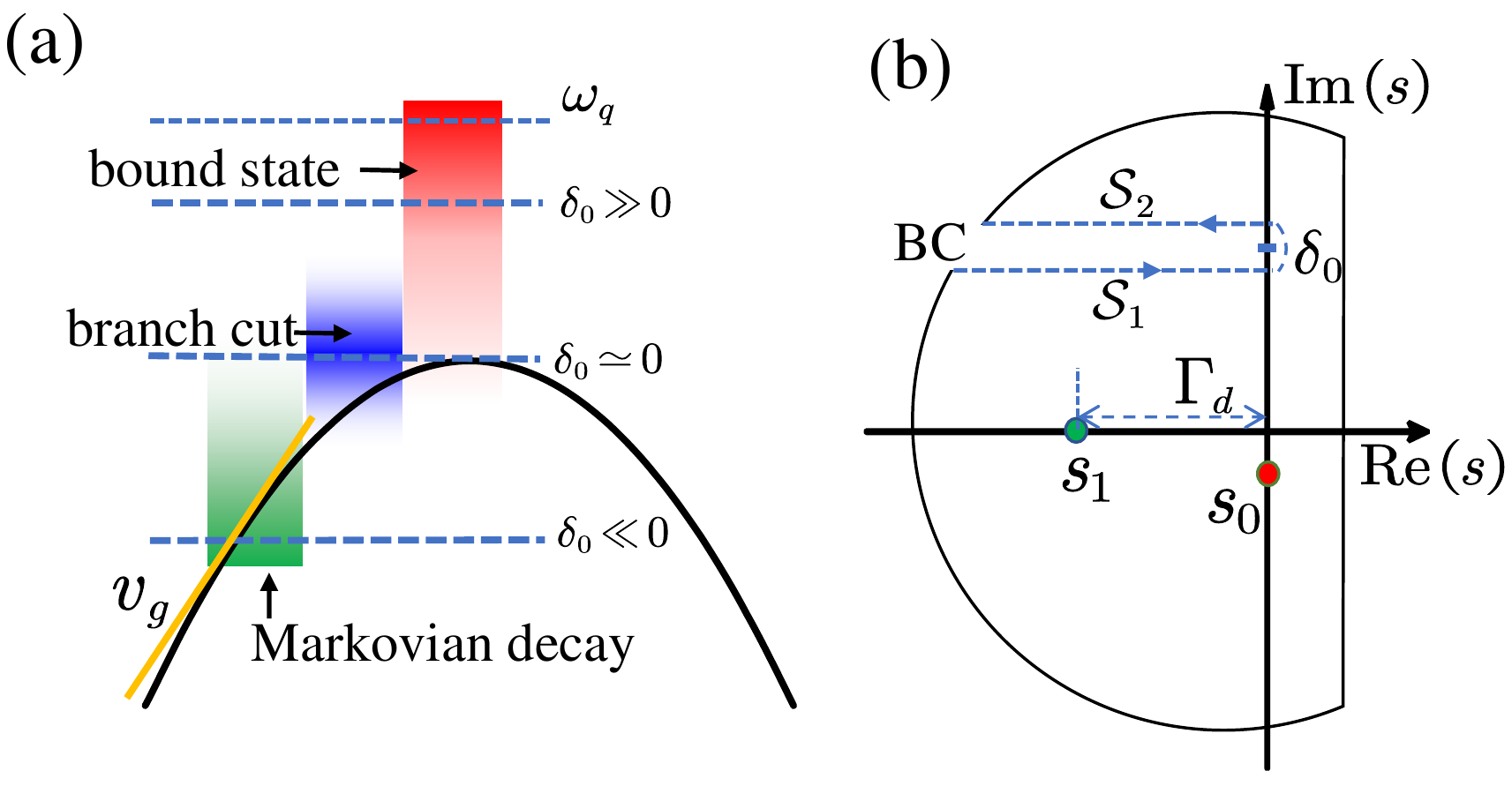}
	\caption{(a) By changing the modulating frequency $\Omega_d$, the 
	effective atomic frequency $\omega_q^{\text{eff}}$ can be tuned 
	to different regimes. In the limit $\delta_0\gg0$ 
	($\delta_0\ll0$), the atom evolution is dominated by the bound 
	state (Markovian decay). Around $\delta_0\simeq 0$, the branch 
	cut contribution will dominate. (b) Contour 
	integral used for calculating the atom decaying into PCW. The 
	poles $s_{0}$ and $s_{1}$ correspond to the contributions of 
	bound state and Markovian decay, respectively. The branch cut 
	(BC) will lead to a non-exponential decay.}
	\label{fig3A}
\end{figure}

\subsection{Analyzing the non-Markovian dynamics}
The atomic frequency $\omega_q^{\text{eff}}$ is around the top of the 1st band (see Fig.~\ref{fig2m}). Additionally, 
the blue sideband is of large detuning to the 2nd energy band 
(i.e., $\Delta_{\text{up}}\gg0$). Under those conditions, the giant 
atom approximately only interacts with the 1st energy band. Since 
$\Omega_{d}$ can not be very large, $\omega_q^{\text{eff}}$ is near the band top. As discussed in 
Refs.~\cite{cohen1998atom,Ramos2016,GonzlezTudela2017}, the band edge 
effects might 
lead to non-Markovian dynamics during the spontaneous decay.
In the following the band edge effects will be addressed, and find the 
parameter regimes where the Markovian decay will dominate the chiral 
emission process.

By considering the lowest energy level $l=1$, the interaction Hamiltonian is 
given in Eq.~(\ref{Htime}), and 
the evolution is derived from the following differential equations
\begin{gather}
	\dot{c}_{e}(t)=-i\sum_k g_{k} e^{i\Delta_k t}c_k(t), \label{cet}  \\
	\dot{c}_k(t)=-ig_k^*e^{-i\Delta_k t}c_{e}(t).
	\label{cakt}
\end{gather}
By defining $e^{-i\Delta _kt}C_k\left( t \right) =c_k\left( t \right) 
$, Eq.~(\ref{cet}-\ref{cakt}) are derived in Laplace space as
\begin{gather}
	\tilde{c}_e\left( s \right) =\frac{1}{s+\Sigma_{e}(s)}, \quad \Sigma_{e}(s)=\sum_k{\frac{|g_k|^2}{s-i\Delta _k}}, \\
	\tilde{C}_k\left( s \right) =\frac{ig_{k}^{*}\tilde{c}_a\left( s \right)}{i\Delta _k-s},
\end{gather}
where $\Sigma_{e}(s)$ is the self-energy, and the time-dependent evolution is recovered by the inverse Laplace transformation~\cite{Ramos2016}
\begin{equation}
	c_e\left( t \right)=\frac{1}{2\pi i}\lim_{E\rightarrow\infty}\int_{\epsilon-iE}^{\epsilon+iE} \tilde{c}_a\left( s \right)e^{st}ds, \quad \epsilon>0.
\end{equation}
Given that $\omega_q^{\text{eff}}$ is around the band top, the linear dispersion is not valid. We should approximate the dispersion
relation as quadratic, i.e., $\Delta_k \simeq \delta_0+\alpha _0\left( k\pm k_0 \right) ^2$, with $\delta _0$ being the detuning with the band top and $\alpha _0$ being the curvature of the dispersion relation [see Fig.~\ref{fig3A}(a)]. Finally
the self-energy term in $\tilde{c}_a\left( s \right)$ is derived as
\begin{equation}
	\Sigma_{e}(s)\simeq\sum_{\pm}\int_{\pm k_m}^{0} dk \frac{|g_k|^2}{s-i[\delta _0+\alpha _0\left( k\mp k_0 \right) ^2]}.
\end{equation}

\textit{We now analyze the band edge effects during the spontaneous decay process}. Since $\omega_q^{\text{eff}}$ is around the band top $\omega_{k0}$, the coupling  strength around $k\simeq \pm k_0$ can be viewed as a constant , i.e., $|g_k|\simeq|g_{k0}|$. Therefore, the self-energy is derived as~\cite{Wang2021} 
\begin{equation}
	\Sigma_{e}(s)=
	-\frac{\pi |g'_{k0}|^2}{\sqrt{-\alpha _0\left( \delta _0+is \right)}}.
\end{equation}
where we approximately extend the integral bound to be infinite, i.e., $k_m\rightarrow \infty$~\cite{Wang2021}. Similar to Eq.~(\ref{g_renormal}), we define $g'_{k0}=g_{k0}\sqrt{L/(2\pi)}$.
The inverse Laplace transform shows that the time-dependent evolution is dominated by the contour integral of $\tilde{c}_a\left(s\right)$ depicted in Fig.~\ref{fig3A}(b). The isolated poles inside the counter are derived from the transcendental equation
\begin{equation}
	s+\Sigma_{e}(s)=0,
\end{equation}
which are marked with the solid dots in Fig.~\ref{fig3A}(b). One pole $s_0$ 
is on the imaginary axes, which describes the bound state 
which does not decay. Another complex pole $s_{1}$ with 
$\text{Re}(s_{1})=\Gamma_d<0$ represents the exponential 
decaying process.

Additionally, since $\sqrt{-\alpha _0\left( \delta _0+is \right)}$ is 
a multi-valued function, we have to take a detour [dashed arrows in 
Fig.~\ref{fig3A}(b)] to avoid point $s=i\delta_0$ being enclosed by 
the contour loop. Consequently, there is a branch cut 
$\mathcal{S}_1\rightarrow\mathcal{S}_2$ at $s=i\delta_0$, which 
analytically continues to the second Riemann sheet. One can simply 
replace $\sqrt{...}\rightarrow-\sqrt{...}$ in $\mathcal{S}_2$. 
Setting $s=y+i\delta_0$ in this branch cut, their contributions to 
the evolution are written as~\cite{Bello2019} 
\begin{eqnarray}
	\sum_{i=1,2}\mathcal{S}_i(t)&=&\frac{1}{2\pi i}\int_{-\infty}^{0}dy 
	\Big[ \frac{1}{ y+i\delta_0 -\frac{\pi |g_{k0}|^2}{\sqrt{-i\alpha _0y}} } \notag \\
	&&-\frac{1}{ y+i\delta_0+\frac{\pi |g_{k0}|^2}{\sqrt{-i\alpha _0y}} }\Big]e^{(y-i\delta_0)t}.
\end{eqnarray} 
 
Together with the contributions from the isolated poles,
the time-dependent evolution is now obtained via the residue theorem
\begin{equation}
	c_e\left( t \right)=\sum_{i=0,1} \text{Res}(s_{i})e^{s_i t}+\sum_{i=1,2}\mathcal{S}_i(t),
	\label{cet}
\end{equation}
where $\text{Res}(s_{i})$ is the residue of the pole $s_i$ and given by the following relation
\begin{equation}
	\text{Res}(s_{i})=\frac{1}{1+\partial_s \Sigma_{e}(s)}\Big|_{s=s_{i}}.
\end{equation}
It is hard to derive $\mathcal{S}_i(t)$ analytically. However, we can infer its behavior around the band edge, where its contribution reaches maximum. Given that $\delta_0=0$ and $y\simeq 0$, the branch cut contribution is derived as~\cite{Bello2019}  
\begin{equation}
	\sum_{i=1,2}\mathcal{S}_i(t)\simeq\frac{\sqrt{i\alpha _0}}{2i|g_{k0}|^2}\frac{e^{i\delta _0t}}{\left( \pi t \right) ^{3/2}}+\mathcal{O}( t  ^{-5/2}),
\end{equation}
which indicates that branch cut describes a sub-exponential decay with a power-law behavior. For the evolution described in Eq.~(\ref{cet}), both the branch cut and decay term will vanish in the limit $t\rightarrow\infty$. Therefore, the steady state population is only determined by the bound state
\begin{equation}
	|c_e(t=\infty)|^2=|\text{Res}(s_{0})|^2,
\end{equation}
which shows that the excitation in the atom cannot totally decay into the PCW, unless the bound state contribution is extremely low. In Fig.~\ref{fig3m}(b), one finds that the steady state population $|c_e(t)|^2$ asymptotically approaches $|\text{Res}(s_{0})|^2$ for different $\delta_0$. Therefore, our analysis can well describe the system's non-Markovian dynamics.

At $t=0$, Eq.~(\ref{cet}) describes the contribution weights of the bound states and decay process which are evaluated by $W_{0,1}=|\text{Res}(s_{0,1})|$. The contribution weight of the branch cut can be obtained from the normalization condition $$W_{2}=|1-\sum_{i=0,1}\text{Res}(s_{i})|.$$ According to above discussions, the normalized contribution weights
$w_i=W_{i}/(\sum_{i}W_i)$ changing with detuning $\delta_0$ is plotted in Fig.~\ref{fig3m}(a). Their contributions are marked with color bars in Fig.~\ref{fig3A}(a). Detailed discussions can be found in the main text.

\setcounter{equation}{0}
\renewcommand{\theequation}{C\arabic{equation}}
\section{SLH formula for multiple giant atoms in a chiral quantum network}
By employing the SLH formalism~\cite{Gough2009,Combes2017,Kockum2018}, we will derive the cascaded master equation for multiple giant atoms chirally interacting with a PCW. For an open quantum system with $n$ input-output channels, the general form for an SLH triplets is $G=(\mathbf{S},\mathbf{L},H)$, where $\mathbf{S}$ is an $n\times n$ scattering matrix, $\mathbf{L}$ is the $n\times1$ vector representing the jump operators to the coupled channels, and $H$ is the system's Hamiltonian. 
Detailed discussions can be found in Ref.~\cite{Kockum2018,Soro2021}.
For the simplest network with two separated giant atoms in Fig.~\ref{fig1m}, the SLH triplet of each coupling point is
\begin{widetext}
	\begin{eqnarray}
		G^{a(b)}_{R,1}&=&\left( \text{1},\sqrt{\frac{\gamma_{a(b),1}}{2}}\sigma _{-}^{a(b)}, \frac{\omega_{a(b)}}{2}\sigma _{z}^{a(b)} \right),  \quad G^{a(b)}_{L,1}=\left( \text{1,}\sqrt{\frac{\gamma_{a(b),1}}{2}}\sigma _{-}^{a(b)}, 0 \right), \notag
		\\
		G^{a(b)}_{R,2}&=&\left( \text{1,}e^{-i\phi^{a(b)}_{c}}\sqrt{\frac{\gamma _{a(b),2}}{2}}\sigma _{-}^{a(b)},0 \right), \quad G^{a(b)}_{L,2}=\left( \text{1,}e^{-i\phi^{a(b)}_{c}}\sqrt{\frac{\gamma _{a(b),2}}{2}}\sigma _{-}^{a(b)},0 \right), \label{gab}
	\end{eqnarray}
\end{widetext}
where $L$ ($R$) represents the left (right) propagating channels, and $\sqrt{\gamma_{a(b),i}}$ is interacting strength between atom $a$ ($b$) and point $i$. Since the right and left channels are expressed independently, $\mathbf{S}$ and $\mathbf{L}$ are simplified as one component by setting $n=1$. Different from conventional giant atoms interacting with a bidirectional 1D waveguide, there are relative phase differences $\phi_c^{a,b}$ in $\mathbf{L}$, which are encoded by the time-dependent modulation [see Eq.~(\ref{gk_m})]. In Eq.~(\ref{gab}), since only relative phase difference matters, the phases at points $x_{1}^{a,b}$ is set as zero, and the phase differences $\phi_c^{a,b}$ are encoded in the jump operators at point $x_{2}^{a,b}$.

In chiral quantum networks, since the distance between giant atoms $ L_{ab}=x_2^a-x_1^b$ is much larger than their size,
we mainly focus on the separation case. The central modes of the fields emitted by the giant atoms is at $k_r\simeq \omega_q^{\text{eff}}/v_g$. 
Denoting $\phi_{L}=k_r L_{ab}$ as the propagating phase between $a$ and $b$, the SLH triplet for the interaction with the right propagating field is derived from the series product relation 
\begin{equation}
	G_R=G^{b}_{R,2}\lhd G_{\phi_{b}}\lhd G^{b}_{R,1}\lhd G_{\phi_L}\lhd G^{a}_{R,2}\lhd G_{\phi_{a}}\lhd G^{a}_{R,1},
\end{equation}
where $\lhd$ represents the series product between two SLH triplets,  $G_{\phi_{i}}=(e^{i\phi_{i}},0,0)$, $G_{\phi_{L}}=(e^{i\phi_{L}},0,0)$,
and $\phi_i$ is the phase difference of the PCW wavefunction between two coupling points of giant atom $i$, i.e., 
\begin{equation}
\phi_i\simeq \arg\left[\frac{u_{1k_r} (x^i_{2})}{u_{1k_r} (x^i_{1})} e^{ik_r (x^i_2-x^i_1)}\right], i=a,b.
\end{equation}
Finally, the SLH formula of $G_R$ is derived as
\begin{widetext}
	\begin{eqnarray}
		S_R&=&e^{i\phi _a}e^{i\phi _L}e^{i\phi _b},
		\\
		L_R&=&e^{i\phi_L}e^{i\phi_b}\left[ e^{i\phi _a}\sqrt{\frac{\gamma _{a,1}}{2}}+e^{-i\phi^a _{c}}\sqrt{\frac{\gamma _{a,2}}{2}} \right] \sigma _{-}^{a}+\left[ e^{i\phi _b}\sqrt{\frac{\gamma _{b,1}}{2}}+e^{-i\phi^b _{c}}\sqrt{\frac{\gamma _{b,2}}{2}} \right] \sigma _{-}^{b},
		\\
		H_R&=&\left[\frac{\omega _a}{2}+\sin \left( \phi _a+\phi^a _{c} \right) \sqrt{\frac{\gamma _{a,1}}{2}}\sqrt{\frac{\gamma _{a,2}}{2}}\right]\sigma _{z}^{a}+\left[\frac{\omega _b}{2}+\sin \left( \phi _b+\phi^b_{c} \right) \sqrt{\frac{\gamma _{b,1}}{2}}\sqrt{\frac{\gamma _{b,2}}{2}}\right]\sigma _{z}^{b} \notag\\
		&&+\frac{1}{2i}\Big[ \big( e^{i\phi _a}e^{i\phi_b}\sqrt{\frac{\gamma _{a,1}}{2}}\sqrt{\frac{\gamma _{b,1}}{2}}+e^{-i\phi^a _{c}}e^{i\phi _L}\sqrt{\frac{\gamma _{a,2}}{2}}\sqrt{\frac{\gamma _{b,1}}{2}} \notag\\
		&&+e^{i\phi _a}e^{i\phi _L}e^{i\phi _b}e^{i\phi^b_{c}}\sqrt{\frac{\gamma  _{a,1}}{2}}\sqrt{\frac{\gamma _{b,2}}{2}}+e^{-i\phi^a _{c}}e^{i\phi _L}e^{i\phi _b}e^{i\phi^b _{c}}\sqrt{\frac{\gamma _{a,2}}{2}}\sqrt{\frac{\gamma _{b,2}}{2}} \big) \sigma _{+}^{b}\sigma _{-}^{a}-\text{H.c.} \Big],
	\end{eqnarray}
\end{widetext}
where $L_R$ is the jump operator to the right propagating modes. The terms proportional to $\sin (\phi _{a,b}+\phi^{a,b} _{c})$ in $H_R$ are the Lamb shifts.
Similarly, the LSH triplet describing coupling with the left propagating modes is eppressed as
\begin{widetext}
	\begin{eqnarray}
		S_L&=& e^{i\phi _a}e^{i\phi _L}e^{i\phi _b},
		\\
		L_L&=&e^{i\phi _L}e^{i\phi _a}\left[ e^{-i\phi^b_{c}}e^{i\phi _b}\sqrt{\frac{\gamma _{b,2}}{2}}+\sqrt{\frac{\gamma _{b,1}}{2}} \right] \sigma _{-}^{b}+\left[ e^{i\phi_a}e^{-i\phi^a_{c}}\sqrt{\frac{\gamma _{a,2}}{2}}+\sqrt{\frac{\gamma _{a,1}}{2}} \right] \sigma _{-}^{a},
		\\
		H_L&=&\sin \left( \phi _a-\phi^a_{c} \right) \sqrt{\frac{\gamma _{a,1}}{2}}\sqrt{\frac{\gamma _{a,2}}{2}}\sigma _{z}^{a}+\sin \left( \phi _b-\phi^b_{c} \right) \sqrt{\frac{\gamma _{b,1}}{2}}\sqrt{\frac{\gamma _{b,2}}{2}}\sigma _{z}^{b} \notag \\
		&&+\frac{1}{2i}\Big[ \big( e^{-i\phi _L}e^{-i\phi^a_{c}}e^{i\phi^b_{c}}e^{-i\phi_b}\sqrt{\frac{\gamma _{a,2}}{2}}\sqrt{\frac{\gamma _{b,2}}{2}}+e^{-i\phi _L}e^{-i\phi^a_{c}}\sqrt{\frac{\gamma _{a,2}}{2}}\sqrt{\frac{\gamma _{b,1}}{2}}  \notag \\
		&&+e^{-i\phi _a}e^{-i\phi _L}e^{-i\phi _b}e^{i\phi^b_{c}}\sqrt{\frac{\gamma _{a,1}}{2}}\sqrt{\frac{\gamma _{b,2}}{2}}+e^{-i\phi_b}e^{-i\phi _a}\sqrt{\frac{\gamma _{a,1}}{2}}\sqrt{\frac{\gamma _{b,1}}{2}} \big) \sigma _{+}^{b}\sigma _{-}^{a}-\text{H.c.} \Big].
	\end{eqnarray}
\end{widetext}
In an ideal chiral quantum network, all the nodes only emit photons into one direction. 
To realize right chiral emission, the jump operator to the left channel is required to be zero, i.e., $\mathbf{L}_L=0$, which results in
\begin{eqnarray}
	e^{i\phi _a}e^{-i\phi^a_{c}}\sqrt{\frac{\gamma _{a,2}}{2}}+\sqrt{\frac{\gamma _{a,1}}{2}}=0, \label{phase_r}  \notag \\
	e^{-i\phi^b_{c}}e^{i\phi _b}\sqrt{\frac{\gamma _{b,2}}{2}}+\sqrt{\frac{\gamma _{b,1}}{2}}=\text{0}.
\end{eqnarray}
Given that the coupling strengths to each point are identical 
$\gamma _{a(b),i}=\gamma_{a(b)}$, the phase relation in Eq.~(\ref{phase_r}) should satisfy 
\begin{equation}
	\phi_{a(b)}-\phi^{a(b)}_{c}=(2N+1)\pi,
\end{equation}
which is the same as the condition in Eq.~(\ref{phaserelation}).
In this case, the SLH triplet of the left propagating modes is reduced as $L_{L}=0$ and $H_{L}=0$, i.e., the giant atoms decouple with the left modes due to the destructive interference. The jump operators $L_R$ representing emitting photons to the right propagating modes are written as 
\begin{gather}
	L_R=S_{a}+S_{b}, \notag \\
	S_i=2i\sin \left( \phi_i \right) \sqrt{\frac{\gamma _{i}}{2}}\sigma _{-}^{i}, \quad  i=a,b,
\end{gather}
where $S_{a(b)}$ corresponds to the individual jump operator for atom $a$ ($b$). Note that the phase term $e^{i(\phi _L+\phi _b)}$ is neglected when
$L_{ab}=(x_2^a-x_1^b)$ is much shorter than wavepacket length of the photons. That is, we neglect the retardation effects and assume the density matrix of the system as $\rho(t-\tau)=\rho(t)$, where $\tau \simeq L_{ab}/v_g$ is propagating time between $a$ and $b$. When $\tau$ is comparable to the decay time, the retardation effects should be considered (see Fig.~\ref{fig7m} and related discussions).

By defining the renormalized atomic frequency as
\begin{equation}
	\omega_i=\omega _i-\sin \left(2\phi _{i} \right)\gamma _{i}, \quad i=a,b,
\end{equation}
we express $H_R$ as 
\begin{equation}
	H_{R}=\frac{\omega_a}{2}\sigma_z^a+\frac{\omega_b}{2}\sigma_z^b+\frac{i}{2} \left(S^{\dagger}_{a}S_{b}-\text{H.c.}\right).
\end{equation}
Therefore, the master equation of the system is 
\begin{equation}
	\dot{\rho}=-i[H_R,\rho]+L_R\rho L^{\dagger}_R-\frac{1}{2} L^{\dagger}_R L_R\rho
	-\frac{1}{2} \rho  L^{\dagger}_R L_R.
\end{equation}
By recombining the non-Hermitian terms into $H_R$ and defining the effective Hamiltonian $H_{\text{eff}}$, one can derive the cascaded master Equation~(\ref{cas_H}).
\end{appendix}

%\bibliography{chiral_nw_ref}

\begin{thebibliography}{85}%
	\makeatletter
	\providecommand \@ifxundefined [1]{%
		\@ifx{#1\undefined}
	}%
	\providecommand \@ifnum [1]{%
		\ifnum #1\expandafter \@firstoftwo
		\else \expandafter \@secondoftwo
		\fi
	}%
	\providecommand \@ifx [1]{%
		\ifx #1\expandafter \@firstoftwo
		\else \expandafter \@secondoftwo
		\fi
	}%
	\providecommand \natexlab [1]{#1}%
	\providecommand \enquote  [1]{``#1''}%
	\providecommand \bibnamefont  [1]{#1}%
	\providecommand \bibfnamefont [1]{#1}%
	\providecommand \citenamefont [1]{#1}%
	\providecommand \href@noop [0]{\@secondoftwo}%
	\providecommand \href [0]{\begingroup \@sanitize@url \@href}%
	\providecommand \@href[1]{\@@startlink{#1}\@@href}%
	\providecommand \@@href[1]{\endgroup#1\@@endlink}%
	\providecommand \@sanitize@url [0]{\catcode `\\12\catcode `\$12\catcode
		`\&12\catcode `\#12\catcode `\^12\catcode `\_12\catcode `\%12\relax}%
	\providecommand \@@startlink[1]{}%
	\providecommand \@@endlink[0]{}%
	\providecommand \url  [0]{\begingroup\@sanitize@url \@url }%
	\providecommand \@url [1]{\endgroup\@href {#1}{\urlprefix }}%
	\providecommand \urlprefix  [0]{URL }%
	\providecommand \Eprint [0]{\href }%
	\providecommand \doibase [0]{http://dx.doi.org/}%
	\providecommand \selectlanguage [0]{\@gobble}%
	\providecommand \bibinfo  [0]{\@secondoftwo}%
	\providecommand \bibfield  [0]{\@secondoftwo}%
	\providecommand \translation [1]{[#1]}%
	\providecommand \BibitemOpen [0]{}%
	\providecommand \bibitemStop [0]{}%
	\providecommand \bibitemNoStop [0]{.\EOS\space}%
	\providecommand \EOS [0]{\spacefactor3000\relax}%
	\providecommand \BibitemShut  [1]{\csname bibitem#1\endcsname}%
	\let\auto@bib@innerbib\@empty
	%</preamble>
	\bibitem [{\citenamefont {Gu}\ \emph {et~al.}(2017)\citenamefont {Gu},
		\citenamefont {Kockum}, \citenamefont {Miranowicz}, \citenamefont {Liu},\
		and\ \citenamefont {Nori}}]{Gu2017}%
	\BibitemOpen
	\bibfield  {author} {\bibinfo {author} {\bibfnamefont {X.}~\bibnamefont
			{Gu}}, \bibinfo {author} {\bibfnamefont {A.~F.}\ \bibnamefont {Kockum}},
		\bibinfo {author} {\bibfnamefont {A.}~\bibnamefont {Miranowicz}}, \bibinfo
		{author} {\bibfnamefont {Y.-X.}\ \bibnamefont {Liu}}, \ and\ \bibinfo
		{author} {\bibfnamefont {F.}~\bibnamefont {Nori}},\ }\bibfield  {title}
	{\enquote {\bibinfo {title} {Microwave photonics with superconducting quantum
				circuits},}\ }\href {https://doi.org/10.1016/j.physrep.2017.10.002}
	{\bibfield  {journal} {\bibinfo  {journal} {Phys. Rep.}\ }\textbf {\bibinfo
			{volume} {718-719}},\ \bibinfo {pages} {1} (\bibinfo {year}
		{2017})}\BibitemShut {NoStop}%
	\bibitem [{\citenamefont {Xu}\ \emph {et~al.}(2018)\citenamefont {Xu},
		\citenamefont {Cai}, \citenamefont {Ma}, \citenamefont {Mu}, \citenamefont
		{Hu}, \citenamefont {Chen}, \citenamefont {Wang}, \citenamefont {Song},
		\citenamefont {Xue}, \citenamefont {Yin},\ and\ \citenamefont
		{Sun}}]{Xu2018}%
	\BibitemOpen
	\bibfield  {author} {\bibinfo {author} {\bibfnamefont {Y.}~\bibnamefont
			{Xu}}, \bibinfo {author} {\bibfnamefont {W.}~\bibnamefont {Cai}}, \bibinfo
		{author} {\bibfnamefont {Y.}~\bibnamefont {Ma}}, \bibinfo {author}
		{\bibfnamefont {X.}~\bibnamefont {Mu}}, \bibinfo {author} {\bibfnamefont
			{L.}~\bibnamefont {Hu}}, \bibinfo {author} {\bibfnamefont {Tao}\ \bibnamefont
			{Chen}}, \bibinfo {author} {\bibfnamefont {H.}~\bibnamefont {Wang}}, \bibinfo
		{author} {\bibfnamefont {Y.~P.}\ \bibnamefont {Song}}, \bibinfo {author}
		{\bibfnamefont {Zheng-Yuan}\ \bibnamefont {Xue}}, \bibinfo {author}
		{\bibfnamefont {Zhang-qi}\ \bibnamefont {Yin}}, \ and\ \bibinfo {author}
		{\bibfnamefont {L.}~\bibnamefont {Sun}},\ }\bibfield  {title} {\enquote
		{\bibinfo {title} {Single-loop realization of arbitrary nonadiabatic
				holonomic single-qubit quantum gates in a superconducting circuit},}\ }\href
	{\doibase 10.1103/PhysRevLett.121.110501} {\bibfield  {journal} {\bibinfo
			{journal} {Phys. Rev. Lett.}\ }\textbf {\bibinfo {volume} {121}},\ \bibinfo
		{pages} {110501} (\bibinfo {year} {2018})}\BibitemShut {NoStop}%
	\bibitem [{\citenamefont {et~al.}(2019{\natexlab{a}})}]{Arute2019}%
	\BibitemOpen
	\bibfield  {author} {\bibinfo {author} {\bibfnamefont {F.~Arute}\
			\bibnamefont {et~al.}},\ }\bibfield  {title} {\enquote {\bibinfo {title}
			{Quantum supremacy using a programmable superconducting processor},}\ }\href
	{\doibase 10.1038/s41586-019-1666-5} {\bibfield  {journal} {\bibinfo
			{journal} {Nature (London)}\ }\textbf {\bibinfo {volume} {574}},\ \bibinfo
		{pages} {505} (\bibinfo {year} {2019}{\natexlab{a}})}\BibitemShut {NoStop}%
	\bibitem [{\citenamefont {et~al.}(2019{\natexlab{b}})}]{Ye2019}%
	\BibitemOpen
	\bibfield  {author} {\bibinfo {author} {\bibfnamefont {Y.-S.~Ye}\
			\bibnamefont {et~al.}},\ }\bibfield  {title} {\enquote {\bibinfo {title}
			{Propagation and localization of collective excitations on a 24-qubit
				superconducting processor},}\ }\href {\doibase
		10.1103/PhysRevLett.123.050502} {\bibfield  {journal} {\bibinfo  {journal}
			{Phys. Rev. Lett.}\ }\textbf {\bibinfo {volume} {123}},\ \bibinfo {pages}
		{050502} (\bibinfo {year} {2019}{\natexlab{b}})}\BibitemShut {NoStop}%
	\bibitem [{\citenamefont {et~al.}(2021)}]{Gong2021}%
	\BibitemOpen
	\bibfield  {author} {\bibinfo {author} {\bibfnamefont {M.~Gong}\ \bibnamefont
			{et~al.}},\ }\bibfield  {title} {\enquote {\bibinfo {title} {Quantum walks on
				a programmable two-dimensional 62-qubit superconducting processor},}\ }\href
	{\doibase 10.1126/science.abg7812} {\bibfield  {journal} {\bibinfo  {journal}
			{Science}\ }\textbf {\bibinfo {volume} {372}},\ \bibinfo {pages} {948}
		(\bibinfo {year} {2021})}\BibitemShut {NoStop}%
	\bibitem [{\citenamefont {Reiserer}\ and\ \citenamefont
		{Rempe}(2015)}]{Reiserer2015}%
	\BibitemOpen
	\bibfield  {author} {\bibinfo {author} {\bibfnamefont {A.}~\bibnamefont
			{Reiserer}}\ and\ \bibinfo {author} {\bibfnamefont {G.}~\bibnamefont
			{Rempe}},\ }\bibfield  {title} {\enquote {\bibinfo {title} {Cavity-based
				quantum networks with single atoms and optical photons},}\ }\href {\doibase
		10.1103/RevModPhys.87.1379} {\bibfield  {journal} {\bibinfo  {journal} {Rev.
				Mod. Phys.}\ }\textbf {\bibinfo {volume} {87}},\ \bibinfo {pages}
		{1379--1418} (\bibinfo {year} {2015})}\BibitemShut {NoStop}%
	\bibitem [{\citenamefont {Brekenfeld}\ \emph {et~al.}(2020)\citenamefont
		{Brekenfeld}, \citenamefont {Niemietz}, \citenamefont {Christesen},\ and\
		\citenamefont {Rempe}}]{Brekenfeld2020}%
	\BibitemOpen
	\bibfield  {author} {\bibinfo {author} {\bibfnamefont {M.}~\bibnamefont
			{Brekenfeld}}, \bibinfo {author} {\bibfnamefont {D.}~\bibnamefont
			{Niemietz}}, \bibinfo {author} {\bibfnamefont {J.~D.}\ \bibnamefont
			{Christesen}}, \ and\ \bibinfo {author} {\bibfnamefont {G.}~\bibnamefont
			{Rempe}},\ }\bibfield  {title} {\enquote {\bibinfo {title} {A quantum network
				node with crossed optical fibre cavities},}\ }\href {\doibase
		10.1038/s41567-020-0855-3} {\bibfield  {journal} {\bibinfo  {journal} {Nat.
				Phys.}\ }\textbf {\bibinfo {volume} {16}},\ \bibinfo {pages} {647} (\bibinfo
		{year} {2020})}\BibitemShut {NoStop}%
	\bibitem [{\citenamefont {Daiss}\ \emph {et~al.}(2021)\citenamefont {Daiss},
		\citenamefont {Langenfeld}, \citenamefont {Welte}, \citenamefont {Distante},
		\citenamefont {Thomas}, \citenamefont {Hartung}, \citenamefont {Morin},\ and\
		\citenamefont {Rempe}}]{Daiss2021}%
	\BibitemOpen
	\bibfield  {author} {\bibinfo {author} {\bibfnamefont {S.}~\bibnamefont
			{Daiss}}, \bibinfo {author} {\bibfnamefont {S.}~\bibnamefont {Langenfeld}},
		\bibinfo {author} {\bibfnamefont {S.}~\bibnamefont {Welte}}, \bibinfo
		{author} {\bibfnamefont {E.}~\bibnamefont {Distante}}, \bibinfo {author}
		{\bibfnamefont {P.}~\bibnamefont {Thomas}}, \bibinfo {author} {\bibfnamefont
			{L.}~\bibnamefont {Hartung}}, \bibinfo {author} {\bibfnamefont
			{O.}~\bibnamefont {Morin}}, \ and\ \bibinfo {author} {\bibfnamefont
			{G.}~\bibnamefont {Rempe}},\ }\bibfield  {title} {\enquote {\bibinfo {title}
			{A quantum-logic gate between distant quantum-network modules},}\ }\href
	{\doibase 10.1126/science.abe3150} {\bibfield  {journal} {\bibinfo  {journal}
			{Science}\ }\textbf {\bibinfo {volume} {371}},\ \bibinfo {pages} {614}
		(\bibinfo {year} {2021})}\BibitemShut {NoStop}%
	\bibitem [{\citenamefont {et~al}(2021)}]{Awschalom2021}%
	\BibitemOpen
	\bibfield  {author} {\bibinfo {author} {\bibfnamefont {D.~Awschalom}\
			\bibnamefont {et~al}},\ }\bibfield  {title} {\enquote {\bibinfo {title}
			{Development of quantum interconnects ({QuICs}) for next-generation
				information technologies},}\ }\href
	{https://doi.org/10.1103/prxquantum.2.017002} {\bibfield  {journal} {\bibinfo
			{journal} {{PRX} Quantum}\ }\textbf {\bibinfo {volume} {2}} (\bibinfo {year}
		{2021})}\BibitemShut {NoStop}%
	\bibitem [{\citenamefont {Kimble}(2008)}]{Kimble2008}%
	\BibitemOpen
	\bibfield  {author} {\bibinfo {author} {\bibfnamefont {H.~J.}\ \bibnamefont
			{Kimble}},\ }\bibfield  {title} {\enquote {\bibinfo {title} {The quantum
				internet},}\ }\href {\doibase 10.1038/nature07127} {\bibfield  {journal}
		{\bibinfo  {journal} {Nature (London)}\ }\textbf {\bibinfo {volume} {453}},\
		\bibinfo {pages} {1023} (\bibinfo {year} {2008})}\BibitemShut {NoStop}%
	\bibitem [{\citenamefont {Ritter}\ \emph {et~al.}(2012)\citenamefont {Ritter},
		\citenamefont {N\"{o}lleke}, \citenamefont {Hahn}, \citenamefont {Reiserer},
		\citenamefont {Neuzner}, \citenamefont {Uphoff}, \citenamefont {M\"{u}cke},
		\citenamefont {Figueroa}, \citenamefont {Bochmann},\ and\ \citenamefont
		{Rempe}}]{Ritter2012}%
	\BibitemOpen
	\bibfield  {author} {\bibinfo {author} {\bibfnamefont {S.}~\bibnamefont
			{Ritter}}, \bibinfo {author} {\bibfnamefont {C.}~\bibnamefont {N\"{o}lleke}},
		\bibinfo {author} {\bibfnamefont {C.}~\bibnamefont {Hahn}}, \bibinfo {author}
		{\bibfnamefont {A.}~\bibnamefont {Reiserer}}, \bibinfo {author}
		{\bibfnamefont {A.}~\bibnamefont {Neuzner}}, \bibinfo {author} {\bibfnamefont
			{M.}~\bibnamefont {Uphoff}}, \bibinfo {author} {\bibfnamefont
			{M.}~\bibnamefont {M\"{u}cke}}, \bibinfo {author} {\bibfnamefont
			{E.}~\bibnamefont {Figueroa}}, \bibinfo {author} {\bibfnamefont
			{J.}~\bibnamefont {Bochmann}}, \ and\ \bibinfo {author} {\bibfnamefont
			{G.}~\bibnamefont {Rempe}},\ }\bibfield  {title} {\enquote {\bibinfo {title}
			{An elementary quantum network of single atoms in optical cavities},}\ }\href
	{\doibase 10.1038/nature11023} {\bibfield  {journal} {\bibinfo  {journal}
			{Nature (London)}\ }\textbf {\bibinfo {volume} {484}},\ \bibinfo {pages}
		{195} (\bibinfo {year} {2012})}\BibitemShut {NoStop}%
	\bibitem [{\citenamefont {van Loo}\ \emph {et~al.}(2013)\citenamefont {van
			Loo}, \citenamefont {Fedorov}, \citenamefont {Lalumiere}, \citenamefont
		{Sanders}, \citenamefont {Blais},\ and\ \citenamefont
		{Wallraff}}]{vanLoo2013}%
	\BibitemOpen
	\bibfield  {author} {\bibinfo {author} {\bibfnamefont {A.~F.}\ \bibnamefont
			{van Loo}}, \bibinfo {author} {\bibfnamefont {A.}~\bibnamefont {Fedorov}},
		\bibinfo {author} {\bibfnamefont {K.}~\bibnamefont {Lalumiere}}, \bibinfo
		{author} {\bibfnamefont {B.~C.}\ \bibnamefont {Sanders}}, \bibinfo {author}
		{\bibfnamefont {A.}~\bibnamefont {Blais}}, \ and\ \bibinfo {author}
		{\bibfnamefont {A.}~\bibnamefont {Wallraff}},\ }\bibfield  {title} {\enquote
		{\bibinfo {title} {Photon-mediated interactions between distant artificial
				atoms},}\ }\href {\doibase 10.1126/science.1244324} {\bibfield  {journal}
		{\bibinfo  {journal} {Science}\ }\textbf {\bibinfo {volume} {342}},\ \bibinfo
		{pages} {1494} (\bibinfo {year} {2013})}\BibitemShut {NoStop}%
	\bibitem [{\citenamefont {Humphreys}\ \emph {et~al.}(2018)\citenamefont
		{Humphreys}, \citenamefont {Kalb}, \citenamefont {Morits}, \citenamefont
		{Schouten}, \citenamefont {Vermeulen}, \citenamefont {Twitchen},
		\citenamefont {Markham},\ and\ \citenamefont {Hanson}}]{Humphreys2018}%
	\BibitemOpen
	\bibfield  {author} {\bibinfo {author} {\bibfnamefont {P.~C.}\ \bibnamefont
			{Humphreys}}, \bibinfo {author} {\bibfnamefont {N.}~\bibnamefont {Kalb}},
		\bibinfo {author} {\bibfnamefont {J.~P.~J.}\ \bibnamefont {Morits}}, \bibinfo
		{author} {\bibfnamefont {R.~N.}\ \bibnamefont {Schouten}}, \bibinfo {author}
		{\bibfnamefont {R.~F.~L.}\ \bibnamefont {Vermeulen}}, \bibinfo {author}
		{\bibfnamefont {D.~J.}\ \bibnamefont {Twitchen}}, \bibinfo {author}
		{\bibfnamefont {M.}~\bibnamefont {Markham}}, \ and\ \bibinfo {author}
		{\bibfnamefont {R.}~\bibnamefont {Hanson}},\ }\bibfield  {title} {\enquote
		{\bibinfo {title} {Deterministic delivery of remote entanglement on a quantum
				network},}\ }\href {\doibase 10.1038/s41586-018-0200-5} {\bibfield  {journal}
		{\bibinfo  {journal} {Nature (London)}\ }\textbf {\bibinfo {volume} {558}},\
		\bibinfo {pages} {268} (\bibinfo {year} {2018})}\BibitemShut {NoStop}%
	\bibitem [{\citenamefont {Vermersch}\ \emph {et~al.}(2017)\citenamefont
		{Vermersch}, \citenamefont {Guimond}, \citenamefont {Pichler},\ and\
		\citenamefont {Zoller}}]{Vermersch2017}%
	\BibitemOpen
	\bibfield  {author} {\bibinfo {author} {\bibfnamefont {B.}~\bibnamefont
			{Vermersch}}, \bibinfo {author} {\bibfnamefont {P.-O.}\ \bibnamefont
			{Guimond}}, \bibinfo {author} {\bibfnamefont {H.}~\bibnamefont {Pichler}}, \
		and\ \bibinfo {author} {\bibfnamefont {P.}~\bibnamefont {Zoller}},\
	}\bibfield  {title} {\enquote {\bibinfo {title} {Quantum state transfer via
				noisy photonic and phononic waveguides},}\ }\href
	{https://doi.org/10.1103/physrevlett.118.133601} {\bibfield  {journal}
		{\bibinfo  {journal} {Phys. Rev. Lett.}\ }\textbf {\bibinfo {volume} {118}},\
		\bibinfo {pages} {133601} (\bibinfo {year} {2017})}\BibitemShut {NoStop}%
	\bibitem [{\citenamefont {Xiang}\ \emph {et~al.}(2017)\citenamefont {Xiang},
		\citenamefont {Zhang}, \citenamefont {Jiang},\ and\ \citenamefont
		{Rabl}}]{Xiang2017}%
	\BibitemOpen
	\bibfield  {author} {\bibinfo {author} {\bibfnamefont {Z.-L.}\ \bibnamefont
			{Xiang}}, \bibinfo {author} {\bibfnamefont {M.-Z.}\ \bibnamefont {Zhang}},
		\bibinfo {author} {\bibfnamefont {L.}~\bibnamefont {Jiang}}, \ and\ \bibinfo
		{author} {\bibfnamefont {P.}~\bibnamefont {Rabl}},\ }\bibfield  {title}
	{\enquote {\bibinfo {title} {Intracity quantum communication via thermal
				microwave networks},}\ }\href {https://doi.org/10.1103/physrevx.7.011035}
	{\bibfield  {journal} {\bibinfo  {journal} {Phys. Rev. X}\ }\textbf {\bibinfo
			{volume} {7}},\ \bibinfo {pages} {011035} (\bibinfo {year}
		{2017})}\BibitemShut {NoStop}%
	\bibitem [{\citenamefont {Cirac}\ \emph {et~al.}(1997)\citenamefont {Cirac},
		\citenamefont {Zoller}, \citenamefont {Kimble},\ and\ \citenamefont
		{Mabuchi}}]{Cirac1997}%
	\BibitemOpen
	\bibfield  {author} {\bibinfo {author} {\bibfnamefont {J.~I.}\ \bibnamefont
			{Cirac}}, \bibinfo {author} {\bibfnamefont {P.}~\bibnamefont {Zoller}},
		\bibinfo {author} {\bibfnamefont {H.~J.}\ \bibnamefont {Kimble}}, \ and\
		\bibinfo {author} {\bibfnamefont {H.}~\bibnamefont {Mabuchi}},\ }\bibfield
	{title} {\enquote {\bibinfo {title} {Quantum state transfer and entanglement
				distribution among distant nodes in a quantum network},}\ }\href {\doibase
		10.1103/PhysRevLett.78.3221} {\bibfield  {journal} {\bibinfo  {journal}
			{Phys. Rev. Lett.}\ }\textbf {\bibinfo {volume} {78}},\ \bibinfo {pages}
		{3221} (\bibinfo {year} {1997})}\BibitemShut {NoStop}%
	\bibitem [{\citenamefont {Stannigel}\ \emph
		{et~al.}(2011{\natexlab{a}})\citenamefont {Stannigel}, \citenamefont {Rabl},
		\citenamefont {S{\o}rensen}, \citenamefont {Lukin},\ and\ \citenamefont
		{Zoller}}]{Stannigel2011}%
	\BibitemOpen
	\bibfield  {author} {\bibinfo {author} {\bibfnamefont {K.}~\bibnamefont
			{Stannigel}}, \bibinfo {author} {\bibfnamefont {P.}~\bibnamefont {Rabl}},
		\bibinfo {author} {\bibfnamefont {A.~S.}\ \bibnamefont {S{\o}rensen}},
		\bibinfo {author} {\bibfnamefont {M.~D.}\ \bibnamefont {Lukin}}, \ and\
		\bibinfo {author} {\bibfnamefont {P.}~\bibnamefont {Zoller}},\ }\bibfield
	{title} {\enquote {\bibinfo {title} {Optomechanical transducers for
				quantum-information processing},}\ }\href
	{https://doi.org/10.1103/physreva.84.042341} {\bibfield  {journal} {\bibinfo
			{journal} {Phys. Rev. A}\ }\textbf {\bibinfo {volume} {84}},\ \bibinfo
		{pages} {042341} (\bibinfo {year} {2011}{\natexlab{a}})}\BibitemShut
	{NoStop}%
	\bibitem [{\citenamefont {Ramos}\ \emph {et~al.}(2016)\citenamefont {Ramos},
		\citenamefont {Vermersch}, \citenamefont {Hauke}, \citenamefont {Pichler},\
		and\ \citenamefont {Zoller}}]{Ramos2016}%
	\BibitemOpen
	\bibfield  {author} {\bibinfo {author} {\bibfnamefont {T.}~\bibnamefont
			{Ramos}}, \bibinfo {author} {\bibfnamefont {B.}~\bibnamefont {Vermersch}},
		\bibinfo {author} {\bibfnamefont {P.}~\bibnamefont {Hauke}}, \bibinfo
		{author} {\bibfnamefont {H.}~\bibnamefont {Pichler}}, \ and\ \bibinfo
		{author} {\bibfnamefont {P.}~\bibnamefont {Zoller}},\ }\bibfield  {title}
	{\enquote {\bibinfo {title} {Non-markovian dynamics in chiral quantum
				networks with spins and photons},}\ }\href
	{https://doi.org/10.1103/physreva.93.062104} {\bibfield  {journal} {\bibinfo
			{journal} {Phys. Rev. A}\ }\textbf {\bibinfo {volume} {93}},\ \bibinfo
		{pages} {062104} (\bibinfo {year} {2016})}\BibitemShut {NoStop}%
	\bibitem [{\citenamefont {Carmichael}(1993)}]{Carmichael1993}%
	\BibitemOpen
	\bibfield  {author} {\bibinfo {author} {\bibfnamefont {H.~J.}\ \bibnamefont
			{Carmichael}},\ }\bibfield  {title} {\enquote {\bibinfo {title} {Quantum
				trajectory theory for cascaded open systems},}\ }\href {\doibase
		10.1103/PhysRevLett.70.2273} {\bibfield  {journal} {\bibinfo  {journal}
			{Phys. Rev. Lett.}\ }\textbf {\bibinfo {volume} {70}},\ \bibinfo {pages}
		{2273} (\bibinfo {year} {1993})}\BibitemShut {NoStop}%
	\bibitem [{\citenamefont {Gardiner}(1993)}]{Gardiner1993}%
	\BibitemOpen
	\bibfield  {author} {\bibinfo {author} {\bibfnamefont {C.~W.}\ \bibnamefont
			{Gardiner}},\ }\bibfield  {title} {\enquote {\bibinfo {title} {Driving a
				quantum system with the output field from another driven quantum system},}\
	}\href {\doibase 10.1103/PhysRevLett.70.2269} {\bibfield  {journal} {\bibinfo
			{journal} {Phys. Rev. Lett.}\ }\textbf {\bibinfo {volume} {70}},\ \bibinfo
		{pages} {2269} (\bibinfo {year} {1993})}\BibitemShut {NoStop}%
	\bibitem [{\citenamefont {Berman}(2020)}]{Berman2020}%
	\BibitemOpen
	\bibfield  {author} {\bibinfo {author} {\bibfnamefont {P.~R.}\ \bibnamefont
			{Berman}},\ }\bibfield  {title} {\enquote {\bibinfo {title} {Theory of two
				atoms in a chiral waveguide},}\ }\href
	{https://doi.org/10.1103/physreva.101.013830} {\bibfield  {journal} {\bibinfo
			{journal} {Phys. Rev. A}\ }\textbf {\bibinfo {volume} {101}},\ \bibinfo
		{pages} {013830} (\bibinfo {year} {2020})}\BibitemShut {NoStop}%
	\bibitem [{\citenamefont {Du}\ \emph {et~al.}(2021{\natexlab{a}})\citenamefont
		{Du}, \citenamefont {Cai}, \citenamefont {Wu}, \citenamefont {Wang},\ and\
		\citenamefont {Li}}]{Du2021}%
	\BibitemOpen
	\bibfield  {author} {\bibinfo {author} {\bibfnamefont {L.}~\bibnamefont
			{Du}}, \bibinfo {author} {\bibfnamefont {M.-R.}\ \bibnamefont {Cai}},
		\bibinfo {author} {\bibfnamefont {J.-H.}\ \bibnamefont {Wu}}, \bibinfo
		{author} {\bibfnamefont {Z.-H.}\ \bibnamefont {Wang}}, \ and\ \bibinfo
		{author} {\bibfnamefont {Y.}~\bibnamefont {Li}},\ }\bibfield  {title}
	{\enquote {\bibinfo {title} {Single-photon nonreciprocal excitation transfer
				with non-markovian retarded effects},}\ }\href
	{https://doi.org/10.1103/physreva.103.053701} {\bibfield  {journal} {\bibinfo
			{journal} {Phys. Rev. A}\ }\textbf {\bibinfo {volume} {103}},\ \bibinfo
		{pages} {053701} (\bibinfo {year} {2021}{\natexlab{a}})}\BibitemShut
	{NoStop}%
	\bibitem [{\citenamefont {Stannigel}\ \emph
		{et~al.}(2011{\natexlab{b}})\citenamefont {Stannigel}, \citenamefont {Rabl},
		\citenamefont {S\o{}rensen}, \citenamefont {Lukin},\ and\ \citenamefont
		{Zoller}}]{Stannigel2012}%
	\BibitemOpen
	\bibfield  {author} {\bibinfo {author} {\bibfnamefont {K.}~\bibnamefont
			{Stannigel}}, \bibinfo {author} {\bibfnamefont {P.}~\bibnamefont {Rabl}},
		\bibinfo {author} {\bibfnamefont {A.~S.}\ \bibnamefont {S\o{}rensen}},
		\bibinfo {author} {\bibfnamefont {M.~D.}\ \bibnamefont {Lukin}}, \ and\
		\bibinfo {author} {\bibfnamefont {P.}~\bibnamefont {Zoller}},\ }\bibfield
	{title} {\enquote {\bibinfo {title} {Optomechanical transducers for
				quantum-information processing},}\ }\href {\doibase
		10.1103/PhysRevA.84.042341} {\bibfield  {journal} {\bibinfo  {journal} {Phys.
				Rev. A}\ }\textbf {\bibinfo {volume} {84}},\ \bibinfo {pages} {042341}
		(\bibinfo {year} {2011}{\natexlab{b}})}\BibitemShut {NoStop}%
	\bibitem [{\citenamefont {Hogan}(1953)}]{Hogan1953}%
	\BibitemOpen
	\bibfield  {author} {\bibinfo {author} {\bibfnamefont {C.~L.}\ \bibnamefont
			{Hogan}},\ }\bibfield  {title} {\enquote {\bibinfo {title} {The ferromagnetic
				faraday effect at microwave frequencies and its applications},}\ }\href
	{\doibase 10.1103/revmodphys.25.253} {\bibfield  {journal} {\bibinfo
			{journal} {Rev. Mod. Phys.}\ }\textbf {\bibinfo {volume} {25}},\ \bibinfo
		{pages} {253} (\bibinfo {year} {1953})}\BibitemShut {NoStop}%
	\bibitem [{\citenamefont {Allen}(1956)}]{Allen1956}%
	\BibitemOpen
	\bibfield  {author} {\bibinfo {author} {\bibfnamefont {P.J.}\ \bibnamefont
			{Allen}},\ }\bibfield  {title} {\enquote {\bibinfo {title} {The turnstile
				circulator},}\ }\href {\doibase 10.1109/tmtt.1956.1125066} {\bibfield
		{journal} {\bibinfo  {journal} {{IEEE} Transactions on Microwave Theory and
				Techniques}\ }\textbf {\bibinfo {volume} {4}},\ \bibinfo {pages} {223}
		(\bibinfo {year} {1956})}\BibitemShut {NoStop}%
	\bibitem [{\citenamefont {Caloz}\ \emph {et~al.}(2018)\citenamefont {Caloz},
		\citenamefont {Al{\`{u}}}, \citenamefont {Tretyakov}, \citenamefont {Sounas},
		\citenamefont {Achouri},\ and\ \citenamefont {Deck-L{\'{e}}ger}}]{Caloz2018}%
	\BibitemOpen
	\bibfield  {author} {\bibinfo {author} {\bibfnamefont {C.}~\bibnamefont
			{Caloz}}, \bibinfo {author} {\bibfnamefont {A.}~\bibnamefont {Al{\`{u}}}},
		\bibinfo {author} {\bibfnamefont {S.}~\bibnamefont {Tretyakov}}, \bibinfo
		{author} {\bibfnamefont {D.}~\bibnamefont {Sounas}}, \bibinfo {author}
		{\bibfnamefont {K.}~\bibnamefont {Achouri}}, \ and\ \bibinfo {author}
		{\bibfnamefont {Z.}~\bibnamefont {Deck-L{\'{e}}ger}},\ }\bibfield  {title}
	{\enquote {\bibinfo {title} {Electromagnetic nonreciprocity},}\ }\href
	{https://doi.org/10.1103/physrevapplied.10.047001} {\bibfield  {journal}
		{\bibinfo  {journal} {Phys. Rev. Applied}\ }\textbf {\bibinfo {volume}
			{10}},\ \bibinfo {pages} {047001} (\bibinfo {year} {2018})}\BibitemShut
	{NoStop}%
	\bibitem [{\citenamefont {Estep}\ \emph {et~al.}(2014)\citenamefont {Estep},
		\citenamefont {Sounas}, \citenamefont {Soric},\ and\ \citenamefont
		{Al{\`{u}}}}]{Estep2014}%
	\BibitemOpen
	\bibfield  {author} {\bibinfo {author} {\bibfnamefont {N.~A.}\ \bibnamefont
			{Estep}}, \bibinfo {author} {\bibfnamefont {D.~L.}\ \bibnamefont {Sounas}},
		\bibinfo {author} {\bibfnamefont {J.}~\bibnamefont {Soric}}, \ and\ \bibinfo
		{author} {\bibfnamefont {A.}~\bibnamefont {Al{\`{u}}}},\ }\bibfield  {title}
	{\enquote {\bibinfo {title} {Magnetic-free non-reciprocity and isolation
				based on parametrically modulated coupled-resonator loops},}\ }\href
	{\doibase 10.1038/nphys3134} {\bibfield  {journal} {\bibinfo  {journal} {Nat.
				Phys.}\ }\textbf {\bibinfo {volume} {10}},\ \bibinfo {pages} {923--927}
		(\bibinfo {year} {2014})}\BibitemShut {NoStop}%
	\bibitem [{\citenamefont {Metelmann}\ and\ \citenamefont
		{Clerk}(2015)}]{Metelmann2015}%
	\BibitemOpen
	\bibfield  {author} {\bibinfo {author} {\bibfnamefont {A.}~\bibnamefont
			{Metelmann}}\ and\ \bibinfo {author} {\bibfnamefont {A.~A.}\ \bibnamefont
			{Clerk}},\ }\bibfield  {title} {\enquote {\bibinfo {title} {Nonreciprocal
				photon transmission and amplification via reservoir engineering},}\ }\href
	{https://doi.org/10.1103/physrevx.5.021025} {\bibfield  {journal} {\bibinfo
			{journal} {Phys. Rev. X}\ }\textbf {\bibinfo {volume} {5}},\ \bibinfo {pages}
		{021025} (\bibinfo {year} {2015})}\BibitemShut {NoStop}%
	\bibitem [{\citenamefont {Sounas}\ and\ \citenamefont
		{Al{\`{u}}}(2017)}]{Sounas2017}%
	\BibitemOpen
	\bibfield  {author} {\bibinfo {author} {\bibfnamefont {D.~L.}\ \bibnamefont
			{Sounas}}\ and\ \bibinfo {author} {\bibfnamefont {A.}~\bibnamefont
			{Al{\`{u}}}},\ }\bibfield  {title} {\enquote {\bibinfo {title}
			{Non-reciprocal photonics based on time modulation},}\ }\href {\doibase
		10.1038/s41566-017-0051-x} {\bibfield  {journal} {\bibinfo  {journal} {Nat.
				Photonics}\ }\textbf {\bibinfo {volume} {11}},\ \bibinfo {pages} {774}
		(\bibinfo {year} {2017})}\BibitemShut {NoStop}%
	\bibitem [{\citenamefont {Chapman}\ \emph {et~al.}(2017)\citenamefont
		{Chapman}, \citenamefont {Rosenthal}, \citenamefont {Kerckhoff},
		\citenamefont {Moores}, \citenamefont {Vale}, \citenamefont {Mates},
		\citenamefont {Hilton}, \citenamefont {Lalumi\`ere}, \citenamefont {Blais},\
		and\ \citenamefont {Lehnert}}]{Chapman2017}%
	\BibitemOpen
	\bibfield  {author} {\bibinfo {author} {\bibfnamefont {B.~J.}\ \bibnamefont
			{Chapman}}, \bibinfo {author} {\bibfnamefont {E.~I.}\ \bibnamefont
			{Rosenthal}}, \bibinfo {author} {\bibfnamefont {J.}~\bibnamefont
			{Kerckhoff}}, \bibinfo {author} {\bibfnamefont {B.~A.}\ \bibnamefont
			{Moores}}, \bibinfo {author} {\bibfnamefont {L.~R.}\ \bibnamefont {Vale}},
		\bibinfo {author} {\bibfnamefont {J.~A.~B.}\ \bibnamefont {Mates}}, \bibinfo
		{author} {\bibfnamefont {G.~C.}\ \bibnamefont {Hilton}}, \bibinfo {author}
		{\bibfnamefont {K.}~\bibnamefont {Lalumi\`ere}}, \bibinfo {author}
		{\bibfnamefont {A.}~\bibnamefont {Blais}}, \ and\ \bibinfo {author}
		{\bibfnamefont {K.~W.}\ \bibnamefont {Lehnert}},\ }\bibfield  {title}
	{\enquote {\bibinfo {title} {Widely tunable on-chip microwave circulator for
				superconducting quantum circuits},}\ }\href {\doibase
		10.1103/PhysRevX.7.041043} {\bibfield  {journal} {\bibinfo  {journal} {Phys.
				Rev. X}\ }\textbf {\bibinfo {volume} {7}},\ \bibinfo {pages} {041043}
		(\bibinfo {year} {2017})}\BibitemShut {NoStop}%
	\bibitem [{\citenamefont {Guimond}\ \emph {et~al.}(2020)\citenamefont
		{Guimond}, \citenamefont {Vermersch}, \citenamefont {Juan}, \citenamefont
		{Sharafiev}, \citenamefont {Kirchmair},\ and\ \citenamefont
		{Zoller}}]{Guimond2020}%
	\BibitemOpen
	\bibfield  {author} {\bibinfo {author} {\bibfnamefont {P.~O.}\ \bibnamefont
			{Guimond}}, \bibinfo {author} {\bibfnamefont {B.}~\bibnamefont {Vermersch}},
		\bibinfo {author} {\bibfnamefont {M.~L.}\ \bibnamefont {Juan}}, \bibinfo
		{author} {\bibfnamefont {A.}~\bibnamefont {Sharafiev}}, \bibinfo {author}
		{\bibfnamefont {G.}~\bibnamefont {Kirchmair}}, \ and\ \bibinfo {author}
		{\bibfnamefont {P.}~\bibnamefont {Zoller}},\ }\bibfield  {title} {\enquote
		{\bibinfo {title} {A unidirectional on-chip photonic interface for
				superconducting circuits},}\ }\href
	{https://www.nature.com/articles/s41534-020-0261-9.pdf} {\bibfield  {journal}
		{\bibinfo  {journal} {npj Quantum Information}\ }\textbf {\bibinfo {volume}
			{6}},\ \bibinfo {pages} {32} (\bibinfo {year} {2020})}\BibitemShut {NoStop}%
	\bibitem [{\citenamefont {Lodahl}\ \emph {et~al.}(2017)\citenamefont {Lodahl},
		\citenamefont {Mahmoodian}, \citenamefont {Stobbe}, \citenamefont
		{Rauschenbeutel}, \citenamefont {Schneeweiss}, \citenamefont {Volz},
		\citenamefont {Pichler},\ and\ \citenamefont {Zoller}}]{Lodahl2017}%
	\BibitemOpen
	\bibfield  {author} {\bibinfo {author} {\bibfnamefont {P.}~\bibnamefont
			{Lodahl}}, \bibinfo {author} {\bibfnamefont {S.}~\bibnamefont {Mahmoodian}},
		\bibinfo {author} {\bibfnamefont {S.}~\bibnamefont {Stobbe}}, \bibinfo
		{author} {\bibfnamefont {A.}~\bibnamefont {Rauschenbeutel}}, \bibinfo
		{author} {\bibfnamefont {P.}~\bibnamefont {Schneeweiss}}, \bibinfo {author}
		{\bibfnamefont {J.}~\bibnamefont {Volz}}, \bibinfo {author} {\bibfnamefont
			{H.}~\bibnamefont {Pichler}}, \ and\ \bibinfo {author} {\bibfnamefont
			{P.}~\bibnamefont {Zoller}},\ }\bibfield  {title} {\enquote {\bibinfo {title}
			{Chiral quantum optics},}\ }\href {\doibase 10.1038/nature21037} {\bibfield
		{journal} {\bibinfo  {journal} {Nature (London)}\ }\textbf {\bibinfo {volume}
			{541}},\ \bibinfo {pages} {473} (\bibinfo {year} {2017})}\BibitemShut
	{NoStop}%
	\bibitem [{\citenamefont {Mitsch}\ \emph {et~al.}(2014)\citenamefont {Mitsch},
		\citenamefont {Sayrin}, \citenamefont {Albrecht}, \citenamefont
		{Schneeweiss},\ and\ \citenamefont {Rauschenbeutel}}]{Mitsch2014}%
	\BibitemOpen
	\bibfield  {author} {\bibinfo {author} {\bibfnamefont {R.}~\bibnamefont
			{Mitsch}}, \bibinfo {author} {\bibfnamefont {C.}~\bibnamefont {Sayrin}},
		\bibinfo {author} {\bibfnamefont {B.}~\bibnamefont {Albrecht}}, \bibinfo
		{author} {\bibfnamefont {P.}~\bibnamefont {Schneeweiss}}, \ and\ \bibinfo
		{author} {\bibfnamefont {A.}~\bibnamefont {Rauschenbeutel}},\ }\bibfield
	{title} {\enquote {\bibinfo {title} {Quantum state-controlled directional
				spontaneous emission of photons into a nanophotonic waveguide},}\ }\href
	{https://doi.org/10.1038/ncomms6713} {\bibfield  {journal} {\bibinfo
			{journal} {Nat. Commun.}\ }\textbf {\bibinfo {volume} {5}},\ \bibinfo {pages}
		{6713} (\bibinfo {year} {2014})}\BibitemShut {NoStop}%
	\bibitem [{\citenamefont {Petersen}\ \emph {et~al.}(2014)\citenamefont
		{Petersen}, \citenamefont {Volz},\ and\ \citenamefont
		{Rauschenbeutel}}]{Petersen2014}%
	\BibitemOpen
	\bibfield  {author} {\bibinfo {author} {\bibfnamefont {J.}~\bibnamefont
			{Petersen}}, \bibinfo {author} {\bibfnamefont {J.}~\bibnamefont {Volz}}, \
		and\ \bibinfo {author} {\bibfnamefont {A.}~\bibnamefont {Rauschenbeutel}},\
	}\bibfield  {title} {\enquote {\bibinfo {title} {Chiral nanophotonic
				waveguide interface based on spin-orbit interaction of light},}\ }\href
	{\doibase 10.1126/science.1257671} {\bibfield  {journal} {\bibinfo  {journal}
			{Science}\ }\textbf {\bibinfo {volume} {346}},\ \bibinfo {pages} {67}
		(\bibinfo {year} {2014})}\BibitemShut {NoStop}%
	\bibitem [{\citenamefont {Pichler}\ \emph {et~al.}(2015)\citenamefont
		{Pichler}, \citenamefont {Ramos}, \citenamefont {Daley},\ and\ \citenamefont
		{Zoller}}]{Pichler2015}%
	\BibitemOpen
	\bibfield  {author} {\bibinfo {author} {\bibfnamefont {H.}~\bibnamefont
			{Pichler}}, \bibinfo {author} {\bibfnamefont {T.}~\bibnamefont {Ramos}},
		\bibinfo {author} {\bibfnamefont {A.~J.}\ \bibnamefont {Daley}}, \ and\
		\bibinfo {author} {\bibfnamefont {P.}~\bibnamefont {Zoller}},\ }\bibfield
	{title} {\enquote {\bibinfo {title} {Quantum optics of chiral spin
				networks},}\ }\href {https://doi.org/10.1103/physreva.91.042116} {\bibfield
		{journal} {\bibinfo  {journal} {Phys. Rev. A}\ }\textbf {\bibinfo {volume}
			{91}},\ \bibinfo {pages} {042116} (\bibinfo {year} {2015})}\BibitemShut
	{NoStop}%
	\bibitem [{\citenamefont {Young}\ \emph {et~al.}(2015)\citenamefont {Young},
		\citenamefont {Thijssen}, \citenamefont {Beggs}, \citenamefont
		{Androvitsaneas}, \citenamefont {Kuipers}, \citenamefont {Rarity},
		\citenamefont {Hughes},\ and\ \citenamefont {Oulton}}]{Young2015}%
	\BibitemOpen
	\bibfield  {author} {\bibinfo {author} {\bibfnamefont {A.~B.}\ \bibnamefont
			{Young}}, \bibinfo {author} {\bibfnamefont
			{A.{\hspace{0.167em}}C.{\hspace{0.167em}}T.}\ \bibnamefont {Thijssen}},
		\bibinfo {author} {\bibfnamefont {D.{\hspace{0.167em}}M.}\ \bibnamefont
			{Beggs}}, \bibinfo {author} {\bibfnamefont {P.}~\bibnamefont
			{Androvitsaneas}}, \bibinfo {author} {\bibfnamefont {L.}~\bibnamefont
			{Kuipers}}, \bibinfo {author} {\bibfnamefont {J.{\hspace{0.167em}}G.}\
			\bibnamefont {Rarity}}, \bibinfo {author} {\bibfnamefont {S.}~\bibnamefont
			{Hughes}}, \ and\ \bibinfo {author} {\bibfnamefont {R.}~\bibnamefont
			{Oulton}},\ }\bibfield  {title} {\enquote {\bibinfo {title} {Polarization
				engineering in photonic crystal waveguides for spin-photon entanglers},}\
	}\href {\doibase 10.1103/physrevlett.115.153901} {\bibfield  {journal}
		{\bibinfo  {journal} {Phys. Rev. Lett.}\ }\textbf {\bibinfo {volume} {115}},\
		\bibinfo {pages} {153901} (\bibinfo {year} {2015})}\BibitemShut {NoStop}%
	\bibitem [{\citenamefont {Bliokh}\ and\ \citenamefont
		{Nori}(2015)}]{Bliokh2015a}%
	\BibitemOpen
	\bibfield  {author} {\bibinfo {author} {\bibfnamefont {K.~Y.}\ \bibnamefont
			{Bliokh}}\ and\ \bibinfo {author} {\bibfnamefont {F.}~\bibnamefont {Nori}},\
	}\bibfield  {title} {\enquote {\bibinfo {title} {Transverse and longitudinal
				angular momenta of light},}\ }\href {\doibase 10.1016/j.physrep.2015.06.003}
	{\bibfield  {journal} {\bibinfo  {journal} {Phys. Rep.}\ }\textbf {\bibinfo
			{volume} {592}},\ \bibinfo {pages} {1} (\bibinfo {year} {2015})}\BibitemShut
	{NoStop}%
	\bibitem [{\citenamefont {le~Feber}\ \emph {et~al.}(2015)\citenamefont
		{le~Feber}, \citenamefont {Rotenberg},\ and\ \citenamefont
		{Kuipers}}]{leFeber2015}%
	\BibitemOpen
	\bibfield  {author} {\bibinfo {author} {\bibfnamefont {B.}~\bibnamefont
			{le~Feber}}, \bibinfo {author} {\bibfnamefont {N.}~\bibnamefont {Rotenberg}},
		\ and\ \bibinfo {author} {\bibfnamefont {L.}~\bibnamefont {Kuipers}},\
	}\bibfield  {title} {\enquote {\bibinfo {title} {Nanophotonic control of
				circular dipole emission},}\ }\href {https://doi.org/10.1038/ncomms7695}
	{\bibfield  {journal} {\bibinfo  {journal} {Nat. Communications}\ }\textbf
		{\bibinfo {volume} {6}},\ \bibinfo {pages} {7695} (\bibinfo {year}
		{2015})}\BibitemShut {NoStop}%
	\bibitem [{\citenamefont {Grankin}\ \emph {et~al.}(2018)\citenamefont
		{Grankin}, \citenamefont {Guimond}, \citenamefont {Vasilyev}, \citenamefont
		{Vermersch},\ and\ \citenamefont {Zoller}}]{Grankin2018}%
	\BibitemOpen
	\bibfield  {author} {\bibinfo {author} {\bibfnamefont {A.}~\bibnamefont
			{Grankin}}, \bibinfo {author} {\bibfnamefont {P.~O.}\ \bibnamefont
			{Guimond}}, \bibinfo {author} {\bibfnamefont {D.~V.}\ \bibnamefont
			{Vasilyev}}, \bibinfo {author} {\bibfnamefont {B.}~\bibnamefont {Vermersch}},
		\ and\ \bibinfo {author} {\bibfnamefont {P.}~\bibnamefont {Zoller}},\
	}\bibfield  {title} {\enquote {\bibinfo {title} {Free-space photonic quantum
				link and chiral quantum optics},}\ }\href {\doibase
		10.1103/physreva.98.043825} {\bibfield  {journal} {\bibinfo  {journal} {Phys.
				Rev. A}\ }\textbf {\bibinfo {volume} {98}},\ \bibinfo {pages} {043825}
		(\bibinfo {year} {2018})}\BibitemShut {NoStop}%
	\bibitem [{\citenamefont {Calaj{\'{o}}}\ \emph {et~al.}(2019)\citenamefont
		{Calaj{\'{o}}}, \citenamefont {Schuetz}, \citenamefont {Pichler},
		\citenamefont {Lukin}, \citenamefont {Schneeweiss}, \citenamefont {Volz},\
		and\ \citenamefont {Rabl}}]{Calaj2019}%
	\BibitemOpen
	\bibfield  {author} {\bibinfo {author} {\bibfnamefont {G.}~\bibnamefont
			{Calaj{\'{o}}}}, \bibinfo {author} {\bibfnamefont {M.~J.~A.}\ \bibnamefont
			{Schuetz}}, \bibinfo {author} {\bibfnamefont {H.}~\bibnamefont {Pichler}},
		\bibinfo {author} {\bibfnamefont {M.~D.}\ \bibnamefont {Lukin}}, \bibinfo
		{author} {\bibfnamefont {P.}~\bibnamefont {Schneeweiss}}, \bibinfo {author}
		{\bibfnamefont {J.}~\bibnamefont {Volz}}, \ and\ \bibinfo {author}
		{\bibfnamefont {P.}~\bibnamefont {Rabl}},\ }\bibfield  {title} {\enquote
		{\bibinfo {title} {Quantum acousto-optic control of light-matter interactions
				in nanophotonic networks},}\ }\href
	{https://doi.org/10.1103/physreva.99.053852} {\bibfield  {journal} {\bibinfo
			{journal} {Phys. Rev. A}\ }\textbf {\bibinfo {volume} {99}},\ \bibinfo
		{pages} {053852} (\bibinfo {year} {2019})}\BibitemShut {NoStop}%
	\bibitem [{\citenamefont {Lira}\ \emph {et~al.}(2012)\citenamefont {Lira},
		\citenamefont {Yu}, \citenamefont {Fan},\ and\ \citenamefont
		{Lipson}}]{Lira2012}%
	\BibitemOpen
	\bibfield  {author} {\bibinfo {author} {\bibfnamefont {H.}~\bibnamefont
			{Lira}}, \bibinfo {author} {\bibfnamefont {Z.-F.}\ \bibnamefont {Yu}},
		\bibinfo {author} {\bibfnamefont {S.-H.}\ \bibnamefont {Fan}}, \ and\
		\bibinfo {author} {\bibfnamefont {M.}~\bibnamefont {Lipson}},\ }\bibfield
	{title} {\enquote {\bibinfo {title} {Electrically driven nonreciprocity
				induced by interband photonic transition on a silicon chip},}\ }\href
	{\doibase 10.1103/PhysRevLett.109.033901} {\bibfield  {journal} {\bibinfo
			{journal} {Phys. Rev. Lett.}\ }\textbf {\bibinfo {volume} {109}},\ \bibinfo
		{pages} {033901} (\bibinfo {year} {2012})}\BibitemShut {NoStop}%
	\bibitem [{\citenamefont {Trainiti}\ and\ \citenamefont
		{Ruzzene}(2016)}]{Trainiti2016}%
	\BibitemOpen
	\bibfield  {author} {\bibinfo {author} {\bibfnamefont {G.}~\bibnamefont
			{Trainiti}}\ and\ \bibinfo {author} {\bibfnamefont {M.}~\bibnamefont
			{Ruzzene}},\ }\bibfield  {title} {\enquote {\bibinfo {title} {Non-reciprocal
				elastic wave propagation in spatiotemporal periodic structures},}\ }\href
	{\doibase 10.1088/1367-2630/18/8/083047} {\bibfield  {journal} {\bibinfo
			{journal} {New J. Phys.}\ }\textbf {\bibinfo {volume} {18}},\ \bibinfo
		{pages} {083047} (\bibinfo {year} {2016})}\BibitemShut {NoStop}%
	\bibitem [{\citenamefont {Chen}\ \emph {et~al.}(2019)\citenamefont {Chen},
		\citenamefont {Li}, \citenamefont {Nassar}, \citenamefont {Norris},
		\citenamefont {Daraio},\ and\ \citenamefont {Huang}}]{Chen2019}%
	\BibitemOpen
	\bibfield  {author} {\bibinfo {author} {\bibfnamefont {Y.-Y.}\ \bibnamefont
			{Chen}}, \bibinfo {author} {\bibfnamefont {X.-P.}\ \bibnamefont {Li}},
		\bibinfo {author} {\bibfnamefont {H.}~\bibnamefont {Nassar}}, \bibinfo
		{author} {\bibfnamefont {A.~N.}\ \bibnamefont {Norris}}, \bibinfo {author}
		{\bibfnamefont {C.}~\bibnamefont {Daraio}}, \ and\ \bibinfo {author}
		{\bibfnamefont {G.-L.}\ \bibnamefont {Huang}},\ }\bibfield  {title} {\enquote
		{\bibinfo {title} {Nonreciprocal wave propagation in a continuum-based
				metamaterial with space-time modulated resonators},}\ }\href
	{https://doi.org/10.1103/physrevapplied.11.064052} {\bibfield  {journal}
		{\bibinfo  {journal} {Phys. Rev. Applied}\ }\textbf {\bibinfo {volume}
			{11}},\ \bibinfo {pages} {064052} (\bibinfo {year} {2019})}\BibitemShut
	{NoStop}%
	\bibitem [{\citenamefont {Kockum}\ \emph {et~al.}(2014)\citenamefont {Kockum},
		\citenamefont {Delsing},\ and\ \citenamefont {Johansson}}]{Kockum2014}%
	\BibitemOpen
	\bibfield  {author} {\bibinfo {author} {\bibfnamefont {A.~F.}\ \bibnamefont
			{Kockum}}, \bibinfo {author} {\bibfnamefont {P.}~\bibnamefont {Delsing}}, \
		and\ \bibinfo {author} {\bibfnamefont {G.}~\bibnamefont {Johansson}},\
	}\bibfield  {title} {\enquote {\bibinfo {title} {Designing
				frequency-dependent relaxation rates and {L}amb shifts for a giant artificial
				atom},}\ }\href {\doibase 10.1103/PhysRevA.90.013837} {\bibfield  {journal}
		{\bibinfo  {journal} {Phys. Rev. A}\ }\textbf {\bibinfo {volume} {90}},\
		\bibinfo {pages} {013837} (\bibinfo {year} {2014})}\BibitemShut {NoStop}%
	\bibitem [{\citenamefont {Guo}\ \emph {et~al.}(2017)\citenamefont {Guo},
		\citenamefont {Grimsmo}, \citenamefont {Kockum}, \citenamefont {Pletyukhov},\
		and\ \citenamefont {Johansson}}]{Guo2017}%
	\BibitemOpen
	\bibfield  {author} {\bibinfo {author} {\bibfnamefont {L.-Z.}\ \bibnamefont
			{Guo}}, \bibinfo {author} {\bibfnamefont {A.}~\bibnamefont {Grimsmo}},
		\bibinfo {author} {\bibfnamefont {A.~F.}\ \bibnamefont {Kockum}}, \bibinfo
		{author} {\bibfnamefont {M.}~\bibnamefont {Pletyukhov}}, \ and\ \bibinfo
		{author} {\bibfnamefont {G.}~\bibnamefont {Johansson}},\ }\bibfield  {title}
	{\enquote {\bibinfo {title} {Giant acoustic atom: A single quantum system
				with a deterministic time delay},}\ }\href {\doibase
		10.1103/PhysRevA.95.053821} {\bibfield  {journal} {\bibinfo  {journal} {Phys.
				Rev. A}\ }\textbf {\bibinfo {volume} {95}},\ \bibinfo {pages} {053821}
		(\bibinfo {year} {2017})}\BibitemShut {NoStop}%
	\bibitem [{\citenamefont {Kockum}\ \emph {et~al.}(2018)\citenamefont {Kockum},
		\citenamefont {Johansson},\ and\ \citenamefont {Nori}}]{Kockum2018}%
	\BibitemOpen
	\bibfield  {author} {\bibinfo {author} {\bibfnamefont {A.~F.}\ \bibnamefont
			{Kockum}}, \bibinfo {author} {\bibfnamefont {G.}~\bibnamefont {Johansson}}, \
		and\ \bibinfo {author} {\bibfnamefont {F.}~\bibnamefont {Nori}},\ }\bibfield
	{title} {\enquote {\bibinfo {title} {Decoherence-free interaction between
				giant atoms in waveguide quantum electrodynamics},}\ }\href {\doibase
		10.1103/PhysRevLett.120.140404} {\bibfield  {journal} {\bibinfo  {journal}
			{Phys. Rev. Lett.}\ }\textbf {\bibinfo {volume} {120}},\ \bibinfo {pages}
		{140404} (\bibinfo {year} {2018})}\BibitemShut {NoStop}%
	\bibitem [{\citenamefont {Guo}\ \emph {et~al.}(2020)\citenamefont {Guo},
		\citenamefont {Kockum}, \citenamefont {Marquardt},\ and\ \citenamefont
		{Johansson}}]{Guo2019}%
	\BibitemOpen
	\bibfield  {author} {\bibinfo {author} {\bibfnamefont {L.-Z.}\ \bibnamefont
			{Guo}}, \bibinfo {author} {\bibfnamefont {A.~F.}\ \bibnamefont {Kockum}},
		\bibinfo {author} {\bibfnamefont {F.}~\bibnamefont {Marquardt}}, \ and\
		\bibinfo {author} {\bibfnamefont {G.}~\bibnamefont {Johansson}},\ }\bibfield
	{title} {\enquote {\bibinfo {title} {Oscillating bound states for a giant
				atom},}\ }\href {\doibase 10.1103/PhysRevResearch.2.043014} {\bibfield
		{journal} {\bibinfo  {journal} {Phys. Rev. Research}\ }\textbf {\bibinfo
			{volume} {2}},\ \bibinfo {pages} {043014} (\bibinfo {year}
		{2020})}\BibitemShut {NoStop}%
	\bibitem [{\citenamefont {Zhao}\ and\ \citenamefont {Wang}(2020)}]{Zhao2020}%
	\BibitemOpen
	\bibfield  {author} {\bibinfo {author} {\bibfnamefont {W.}~\bibnamefont
			{Zhao}}\ and\ \bibinfo {author} {\bibfnamefont {Z.}~\bibnamefont {Wang}},\
	}\bibfield  {title} {\enquote {\bibinfo {title} {{Single-photon scattering
					and bound states in an atom-waveguide system with two or multiple coupling
					points}},}\ }\href {\doibase 10.1103/PhysRevA.101.053855} {\bibfield
		{journal} {\bibinfo  {journal} {Phys. Rev. A}\ }\textbf {\bibinfo {volume}
			{101}},\ \bibinfo {pages} {053855} (\bibinfo {year} {2020})}\BibitemShut
	{NoStop}%
	\bibitem [{\citenamefont {Wang}\ \emph {et~al.}(2021)\citenamefont {Wang},
		\citenamefont {Liu}, \citenamefont {Kockum}, \citenamefont {Li},\ and\
		\citenamefont {Nori}}]{Wang2021}%
	\BibitemOpen
	\bibfield  {author} {\bibinfo {author} {\bibfnamefont {X.}~\bibnamefont
			{Wang}}, \bibinfo {author} {\bibfnamefont {T.}~\bibnamefont {Liu}}, \bibinfo
		{author} {\bibfnamefont {A.~F.}\ \bibnamefont {Kockum}}, \bibinfo {author}
		{\bibfnamefont {H.-R.}\ \bibnamefont {Li}}, \ and\ \bibinfo {author}
		{\bibfnamefont {F.}~\bibnamefont {Nori}},\ }\bibfield  {title} {\enquote
		{\bibinfo {title} {Tunable chiral bound states with giant atoms},}\ }\href
	{https://doi.org/10.1103/physrevlett.126.043602} {\bibfield  {journal}
		{\bibinfo  {journal} {Phys. Rev. Lett.}\ }\textbf {\bibinfo {volume} {126}},\
		\bibinfo {pages} {043602} (\bibinfo {year} {2021})}\BibitemShut {NoStop}%
	\bibitem [{\citenamefont {Cheng}\ \emph {et~al.}(2021)\citenamefont {Cheng},
		\citenamefont {Wang},\ and\ \citenamefont {Liu}}]{cheng2021boundary}%
	\BibitemOpen
	\bibfield  {author} {\bibinfo {author} {\bibfnamefont {W.-J}\ \bibnamefont
			{Cheng}}, \bibinfo {author} {\bibfnamefont {Z.-H.}\ \bibnamefont {Wang}}, \
		and\ \bibinfo {author} {\bibfnamefont {Y.-X.}\ \bibnamefont {Liu}},\
	}\bibfield  {title} {\enquote {\bibinfo {title} {Boundary effect and dressed
				states of a giant atom in a topological waveguide},}\ }\href
	{https://arxiv.org/abs/2103.04542} {\bibfield  {journal} {\bibinfo  {journal}
			{arXiv preprint arXiv:2103.04542}\ } (\bibinfo {year} {2021})}\BibitemShut
	{NoStop}%
	\bibitem [{\citenamefont {Du}\ \emph {et~al.}(2021{\natexlab{b}})\citenamefont
		{Du}, \citenamefont {Chen},\ and\ \citenamefont {Li}}]{du2021single}%
	\BibitemOpen
	\bibfield  {author} {\bibinfo {author} {\bibfnamefont {L.}~\bibnamefont
			{Du}}, \bibinfo {author} {\bibfnamefont {Y.-T.}\ \bibnamefont {Chen}}, \ and\
		\bibinfo {author} {\bibfnamefont {Y.}~\bibnamefont {Li}},\ }\bibfield
	{title} {\enquote {\bibinfo {title} {Nonreciprocal frequency conversion with
				chiral $\mathrm{\ensuremath{\Lambda}}$-type atoms},}\ }\href {\doibase
		10.1103/PhysRevResearch.3.043226} {\bibfield  {journal} {\bibinfo  {journal}
			{Phys. Rev. Research}\ }\textbf {\bibinfo {volume} {3}},\ \bibinfo {pages}
		{043226} (\bibinfo {year} {2021}{\natexlab{b}})}\BibitemShut {NoStop}%
	\bibitem [{\citenamefont {Soro}\ and\ \citenamefont {Kockum}(2021)}]{Soro2021}%
	\BibitemOpen
	\bibfield  {author} {\bibinfo {author} {\bibfnamefont {A.}~\bibnamefont
			{Soro}}\ and\ \bibinfo {author} {\bibfnamefont {A.~F.}\ \bibnamefont
			{Kockum}},\ }\bibfield  {title} {\enquote {\bibinfo {title} {Chiral quantum
				optics with giant atoms},}\ }\href {https://arxiv.org/pdf/2106.11946v1.pdf}
	{\bibfield  {journal} {\bibinfo  {journal} {preprint arXiv:2106.11946}\ }
		(\bibinfo {year} {2021})}\BibitemShut {NoStop}%
	\bibitem [{\citenamefont {Kannan}\ \emph {et~al.}(2020)\citenamefont {Kannan},
		\citenamefont {Ruckriegel}, \citenamefont {Campbell}, \citenamefont {Kockum},
		\citenamefont {Braum{\"{u}}ller}, \citenamefont {Kim}, \citenamefont
		{Kjaergaard}, \citenamefont {Krantz}, \citenamefont {Melville}, \citenamefont
		{Niedzielski}, \citenamefont {Veps{\"{a}}l{\"{a}}inen}, \citenamefont
		{Winik}, \citenamefont {Yoder}, \citenamefont {Nori}, \citenamefont
		{Orlando}, \citenamefont {Gustavsson},\ and\ \citenamefont
		{Oliver}}]{Kannan2020}%
	\BibitemOpen
	\bibfield  {author} {\bibinfo {author} {\bibfnamefont {B.}~\bibnamefont
			{Kannan}}, \bibinfo {author} {\bibfnamefont {M.~J.}\ \bibnamefont
			{Ruckriegel}}, \bibinfo {author} {\bibfnamefont {D.~L.}\ \bibnamefont
			{Campbell}}, \bibinfo {author} {\bibfnamefont {A.~F.}\ \bibnamefont
			{Kockum}}, \bibinfo {author} {\bibfnamefont {J.}~\bibnamefont
			{Braum{\"{u}}ller}}, \bibinfo {author} {\bibfnamefont {D.~K.}\ \bibnamefont
			{Kim}}, \bibinfo {author} {\bibfnamefont {M.}~\bibnamefont {Kjaergaard}},
		\bibinfo {author} {\bibfnamefont {P.}~\bibnamefont {Krantz}}, \bibinfo
		{author} {\bibfnamefont {A.}~\bibnamefont {Melville}}, \bibinfo {author}
		{\bibfnamefont {B.~M.}\ \bibnamefont {Niedzielski}}, \bibinfo {author}
		{\bibfnamefont {A.}~\bibnamefont {Veps{\"{a}}l{\"{a}}inen}}, \bibinfo
		{author} {\bibfnamefont {R.}~\bibnamefont {Winik}}, \bibinfo {author}
		{\bibfnamefont {J.~L.}\ \bibnamefont {Yoder}}, \bibinfo {author}
		{\bibfnamefont {F.}~\bibnamefont {Nori}}, \bibinfo {author} {\bibfnamefont
			{T.~P.}\ \bibnamefont {Orlando}}, \bibinfo {author} {\bibfnamefont
			{S.}~\bibnamefont {Gustavsson}}, \ and\ \bibinfo {author} {\bibfnamefont
			{W.~D.}\ \bibnamefont {Oliver}},\ }\bibfield  {title} {\enquote {\bibinfo
			{title} {{Waveguide quantum electrodynamics with superconducting artificial
					giant atoms}},}\ }\href {\doibase 10.1038/s41586-020-2529-9} {\bibfield
		{journal} {\bibinfo  {journal} {Nature (London)}\ }\textbf {\bibinfo {volume}
			{583}},\ \bibinfo {pages} {775} (\bibinfo {year} {2020})}\BibitemShut
	{NoStop}%
	\bibitem [{\citenamefont {Frunzio}\ \emph {et~al.}(2005)\citenamefont
		{Frunzio}, \citenamefont {Wallraff}, \citenamefont {Schuster}, \citenamefont
		{Majer},\ and\ \citenamefont {Schoelkopf}}]{Frunzio2005}%
	\BibitemOpen
	\bibfield  {author} {\bibinfo {author} {\bibfnamefont {L.}~\bibnamefont
			{Frunzio}}, \bibinfo {author} {\bibfnamefont {A.}~\bibnamefont {Wallraff}},
		\bibinfo {author} {\bibfnamefont {D.}~\bibnamefont {Schuster}}, \bibinfo
		{author} {\bibfnamefont {J.}~\bibnamefont {Majer}}, \ and\ \bibinfo {author}
		{\bibfnamefont {R.}~\bibnamefont {Schoelkopf}},\ }\bibfield  {title}
	{\enquote {\bibinfo {title} {Fabrication and characterization of
				superconducting circuit qed devices for quantum computation},}\ }\href
	{\doibase 10.1109/TASC.2005.850084} {\bibfield  {journal} {\bibinfo
			{journal} {IEEE Transactions on Applied Superconductivity}\ }\textbf
		{\bibinfo {volume} {15}},\ \bibinfo {pages} {860--863} (\bibinfo {year}
		{2005})}\BibitemShut {NoStop}%
	\bibitem [{\citenamefont {Göppl}\ \emph {et~al.}(2008)\citenamefont {Göppl},
		\citenamefont {Fragner}, \citenamefont {Baur}, \citenamefont {Bianchetti},
		\citenamefont {Filipp}, \citenamefont {Fink}, \citenamefont {Leek},
		\citenamefont {Puebla}, \citenamefont {Steffen},\ and\ \citenamefont
		{Wallraff}}]{Goppl2008}%
	\BibitemOpen
	\bibfield  {author} {\bibinfo {author} {\bibfnamefont {M.}~\bibnamefont
			{Göppl}}, \bibinfo {author} {\bibfnamefont {A.}~\bibnamefont {Fragner}},
		\bibinfo {author} {\bibfnamefont {M.}~\bibnamefont {Baur}}, \bibinfo {author}
		{\bibfnamefont {R.}~\bibnamefont {Bianchetti}}, \bibinfo {author}
		{\bibfnamefont {S.}~\bibnamefont {Filipp}}, \bibinfo {author} {\bibfnamefont
			{J.~M.}\ \bibnamefont {Fink}}, \bibinfo {author} {\bibfnamefont {P.~J.}\
			\bibnamefont {Leek}}, \bibinfo {author} {\bibfnamefont {G.}~\bibnamefont
			{Puebla}}, \bibinfo {author} {\bibfnamefont {L.}~\bibnamefont {Steffen}}, \
		and\ \bibinfo {author} {\bibfnamefont {A.}~\bibnamefont {Wallraff}},\
	}\bibfield  {title} {\enquote {\bibinfo {title} {Coplanar waveguide
				resonators for circuit quantum electrodynamics},}\ }\href {\doibase
		10.1063/1.3010859} {\bibfield  {journal} {\bibinfo  {journal} {J. Appl.
				Phys.}\ }\textbf {\bibinfo {volume} {104}},\ \bibinfo {pages} {113904}
		(\bibinfo {year} {2008})}\BibitemShut {NoStop}%
	\bibitem [{\citenamefont {Clem}(2013)}]{Clem2013}%
	\BibitemOpen
	\bibfield  {author} {\bibinfo {author} {\bibfnamefont {J.~R.}\ \bibnamefont
			{Clem}},\ }\bibfield  {title} {\enquote {\bibinfo {title} {Inductances and
				attenuation constant for a thin-film superconducting coplanar waveguide
				resonator},}\ }\href {https://aip.scitation.org/doi/full/10.1063/1.4773070}
	{\bibfield  {journal} {\bibinfo  {journal} {J. Appl. Phys.}\ }\textbf
		{\bibinfo {volume} {113}},\ \bibinfo {pages} {013910} (\bibinfo {year}
		{2013})}\BibitemShut {NoStop}%
	\bibitem [{\citenamefont {Dalibard}\ \emph {et~al.}(2011)\citenamefont
		{Dalibard}, \citenamefont {Gerbier}, \citenamefont
		{Juzeli\ifmmode~\bar{u}\else \={u}\fi{}nas},\ and\ \citenamefont
		{\"Ohberg}}]{Dalibard11}%
	\BibitemOpen
	\bibfield  {author} {\bibinfo {author} {\bibfnamefont {J.}~\bibnamefont
			{Dalibard}}, \bibinfo {author} {\bibfnamefont {F.}~\bibnamefont {Gerbier}},
		\bibinfo {author} {\bibfnamefont {G.}~\bibnamefont
			{Juzeli\ifmmode~\bar{u}\else \={u}\fi{}nas}}, \ and\ \bibinfo {author}
		{\bibfnamefont {P.}~\bibnamefont {\"Ohberg}},\ }\bibfield  {title} {\enquote
		{\bibinfo {title} {Colloquium: Artificial gauge potentials for neutral
				atoms},}\ }\href {\doibase 10.1103/RevModPhys.83.1523} {\bibfield  {journal}
		{\bibinfo  {journal} {Rev. Mod. Phys.}\ }\textbf {\bibinfo {volume} {83}},\
		\bibinfo {pages} {1523} (\bibinfo {year} {2011})}\BibitemShut {NoStop}%
	\bibitem [{\citenamefont {Schmidt}\ \emph {et~al.}(2015)\citenamefont
		{Schmidt}, \citenamefont {Kessler}, \citenamefont {Peano}, \citenamefont
		{Painter},\ and\ \citenamefont {Marquardt}}]{Schmidt2015}%
	\BibitemOpen
	\bibfield  {author} {\bibinfo {author} {\bibfnamefont {M.}~\bibnamefont
			{Schmidt}}, \bibinfo {author} {\bibfnamefont {S.}~\bibnamefont {Kessler}},
		\bibinfo {author} {\bibfnamefont {V.}~\bibnamefont {Peano}}, \bibinfo
		{author} {\bibfnamefont {O.}~\bibnamefont {Painter}}, \ and\ \bibinfo
		{author} {\bibfnamefont {F.}~\bibnamefont {Marquardt}},\ }\bibfield  {title}
	{\enquote {\bibinfo {title} {Optomechanical creation of magnetic fields for
				photons on a lattice},}\ }\href {\doibase 10.1364/optica.2.000635} {\bibfield
		{journal} {\bibinfo  {journal} {Optica}\ }\textbf {\bibinfo {volume} {2}},\
		\bibinfo {pages} {635} (\bibinfo {year} {2015})}\BibitemShut {NoStop}%
	\bibitem [{\citenamefont {Fang}\ \emph {et~al.}(2017)\citenamefont {Fang},
		\citenamefont {Luo}, \citenamefont {Metelmann}, \citenamefont {Matheny},
		\citenamefont {Marquardt}, \citenamefont {Clerk},\ and\ \citenamefont
		{Painter}}]{Fang2017}%
	\BibitemOpen
	\bibfield  {author} {\bibinfo {author} {\bibfnamefont {K.-J.}\ \bibnamefont
			{Fang}}, \bibinfo {author} {\bibfnamefont {J.}~\bibnamefont {Luo}}, \bibinfo
		{author} {\bibfnamefont {A.}~\bibnamefont {Metelmann}}, \bibinfo {author}
		{\bibfnamefont {M.~H.}\ \bibnamefont {Matheny}}, \bibinfo {author}
		{\bibfnamefont {F.}~\bibnamefont {Marquardt}}, \bibinfo {author}
		{\bibfnamefont {A.~A.}\ \bibnamefont {Clerk}}, \ and\ \bibinfo {author}
		{\bibfnamefont {O.}~\bibnamefont {Painter}},\ }\bibfield  {title} {\enquote
		{\bibinfo {title} {Generalized non-reciprocity in an optomechanical circuit
				via synthetic magnetism and reservoir engineering},}\ }\href {\doibase
		10.1038/nphys4009} {\bibfield  {journal} {\bibinfo  {journal} {Nat. Phys.}\
		}\textbf {\bibinfo {volume} {13}},\ \bibinfo {pages} {465} (\bibinfo {year}
		{2017})}\BibitemShut {NoStop}%
	\bibitem [{\citenamefont {John}\ and\ \citenamefont {Wang}(1991)}]{John1991}%
	\BibitemOpen
	\bibfield  {author} {\bibinfo {author} {\bibfnamefont {S.}~\bibnamefont
			{John}}\ and\ \bibinfo {author} {\bibfnamefont {J.}~\bibnamefont {Wang}},\
	}\bibfield  {title} {\enquote {\bibinfo {title} {Quantum optics of localized
				light in a photonic band-gap},}\ }\href {\doibase DOI
		10.1103/PhysRevB.43.12772} {\bibfield  {journal} {\bibinfo  {journal} {Phys.
				Rev. B}\ }\textbf {\bibinfo {volume} {43}},\ \bibinfo {pages} {12772}
		(\bibinfo {year} {1991})}\BibitemShut {NoStop}%
	\bibitem [{\citenamefont {Hung}\ \emph {et~al.}(2013)\citenamefont {Hung},
		\citenamefont {Meenehan}, \citenamefont {Chang}, \citenamefont {Painter},\
		and\ \citenamefont {Kimble}}]{Hung2013}%
	\BibitemOpen
	\bibfield  {author} {\bibinfo {author} {\bibfnamefont {C.~L.}\ \bibnamefont
			{Hung}}, \bibinfo {author} {\bibfnamefont {S.~M.}\ \bibnamefont {Meenehan}},
		\bibinfo {author} {\bibfnamefont {D.~E.}\ \bibnamefont {Chang}}, \bibinfo
		{author} {\bibfnamefont {O.}~\bibnamefont {Painter}}, \ and\ \bibinfo
		{author} {\bibfnamefont {H.~J.}\ \bibnamefont {Kimble}},\ }\bibfield  {title}
	{\enquote {\bibinfo {title} {Trapped atoms in one-dimensional photonic
				crystals},}\ }\href {\doibase Artn 083026 10.1088/1367-2630/15/8/083026}
	{\bibfield  {journal} {\bibinfo  {journal} {New J. Phys.}\ }\textbf {\bibinfo
			{volume} {15}},\ \bibinfo {pages} {083026} (\bibinfo {year}
		{2013})}\BibitemShut {NoStop}%
	\bibitem [{\citenamefont {Goban}\ \emph {et~al.}(2014)\citenamefont {Goban},
		\citenamefont {Hung}, \citenamefont {Yu}, \citenamefont {Hood}, \citenamefont
		{Muniz}, \citenamefont {Lee}, \citenamefont {Martin}, \citenamefont
		{McClung}, \citenamefont {Choi}, \citenamefont {Chang}, \citenamefont
		{Painter},\ and\ \citenamefont {Kimble}}]{Goban2014}%
	\BibitemOpen
	\bibfield  {author} {\bibinfo {author} {\bibfnamefont {A.}~\bibnamefont
			{Goban}}, \bibinfo {author} {\bibfnamefont {C.-L.}\ \bibnamefont {Hung}},
		\bibinfo {author} {\bibfnamefont {S.-P.}\ \bibnamefont {Yu}}, \bibinfo
		{author} {\bibfnamefont {J.D.}\ \bibnamefont {Hood}}, \bibinfo {author}
		{\bibfnamefont {J.A.}\ \bibnamefont {Muniz}}, \bibinfo {author}
		{\bibfnamefont {J.H.}\ \bibnamefont {Lee}}, \bibinfo {author} {\bibfnamefont
			{M.J.}\ \bibnamefont {Martin}}, \bibinfo {author} {\bibfnamefont {A.C.}\
			\bibnamefont {McClung}}, \bibinfo {author} {\bibfnamefont {K.S.}\
			\bibnamefont {Choi}}, \bibinfo {author} {\bibfnamefont {D.E.}\ \bibnamefont
			{Chang}}, \bibinfo {author} {\bibfnamefont {O.}~\bibnamefont {Painter}}, \
		and\ \bibinfo {author} {\bibfnamefont {H.J.}\ \bibnamefont {Kimble}},\
	}\bibfield  {title} {\enquote {\bibinfo {title} {Atom{\textendash}light
				interactions in photonic crystals},}\ }\href
	{https://doi.org/10.1038/ncomms4808} {\bibfield  {journal} {\bibinfo
			{journal} {Nat. Communication}\ }\textbf {\bibinfo {volume} {5}},\ \bibinfo
		{pages} {4808} (\bibinfo {year} {2014})}\BibitemShut {NoStop}%
	\bibitem [{\citenamefont {Douglas}\ \emph {et~al.}(2016)\citenamefont
		{Douglas}, \citenamefont {Caneva},\ and\ \citenamefont
		{Chang}}]{Douglas2016}%
	\BibitemOpen
	\bibfield  {author} {\bibinfo {author} {\bibfnamefont {J.~S.}\ \bibnamefont
			{Douglas}}, \bibinfo {author} {\bibfnamefont {T.}~\bibnamefont {Caneva}}, \
		and\ \bibinfo {author} {\bibfnamefont {D.~E.}\ \bibnamefont {Chang}},\
	}\bibfield  {title} {\enquote {\bibinfo {title} {Photon molecules in atomic
				gases trapped near photonic crystal waveguides},}\ }\href
	{https://doi.org/10.1103/physrevx.6.031017} {\bibfield  {journal} {\bibinfo
			{journal} {Phys. Rev. X}\ }\textbf {\bibinfo {volume} {6}},\ \bibinfo {pages}
		{031017} (\bibinfo {year} {2016})}\BibitemShut {NoStop}%
	\bibitem [{\citenamefont {Liu}\ and\ \citenamefont {Houck}(2017)}]{Liu2017}%
	\BibitemOpen
	\bibfield  {author} {\bibinfo {author} {\bibfnamefont {Y.~B.}\ \bibnamefont
			{Liu}}\ and\ \bibinfo {author} {\bibfnamefont {A.~A.}\ \bibnamefont
			{Houck}},\ }\bibfield  {title} {\enquote {\bibinfo {title} {Quantum
				electrodynamics near a photonic bandgap},}\ }\href {\doibase
		10.1038/Nphys3834} {\bibfield  {journal} {\bibinfo  {journal} {Nat. Phys.}\
		}\textbf {\bibinfo {volume} {13}},\ \bibinfo {pages} {48} (\bibinfo {year}
		{2017})}\BibitemShut {NoStop}%
	\bibitem [{\citenamefont {Gonz{\'{a}}lez-Tudela}\ and\ \citenamefont
		{Cirac}(2017)}]{GonzlezTudela2017}%
	\BibitemOpen
	\bibfield  {author} {\bibinfo {author} {\bibfnamefont {A.}~\bibnamefont
			{Gonz{\'{a}}lez-Tudela}}\ and\ \bibinfo {author} {\bibfnamefont {J.~I.}\
			\bibnamefont {Cirac}},\ }\bibfield  {title} {\enquote {\bibinfo {title}
			{{Markovian and non-Markovian dynamics of quantum emitters coupled to
					two-dimensional structured reservoirs}},}\ }\href
	{https://doi.org/10.1103/physreva.96.043811} {\bibfield  {journal} {\bibinfo
			{journal} {Phys. Rev. A}\ }\textbf {\bibinfo {volume} {96}},\ \bibinfo
		{pages} {043811} (\bibinfo {year} {2017})}\BibitemShut {NoStop}%
	\bibitem [{\citenamefont {Peropadre}\ \emph {et~al.}(2013)\citenamefont
		{Peropadre}, \citenamefont {Zueco}, \citenamefont {Wulschner}, \citenamefont
		{Deppe}, \citenamefont {Marx}, \citenamefont {Gross},\ and\ \citenamefont
		{García-Ripoll}}]{Peropadre2013}%
	\BibitemOpen
	\bibfield  {author} {\bibinfo {author} {\bibfnamefont {B.}~\bibnamefont
			{Peropadre}}, \bibinfo {author} {\bibfnamefont {D.}~\bibnamefont {Zueco}},
		\bibinfo {author} {\bibfnamefont {F.}~\bibnamefont {Wulschner}}, \bibinfo
		{author} {\bibfnamefont {F.}~\bibnamefont {Deppe}}, \bibinfo {author}
		{\bibfnamefont {A.}~\bibnamefont {Marx}}, \bibinfo {author} {\bibfnamefont
			{R.}~\bibnamefont {Gross}}, \ and\ \bibinfo {author} {\bibfnamefont {J.~J.}\
			\bibnamefont {García-Ripoll}},\ }\bibfield  {title} {\enquote {\bibinfo
			{title} {Tunable coupling engineering between superconducting resonators:
				From sidebands to effective gauge fields},}\ }\href {\doibase
		10.1103/PhysRevB.87.134504} {\bibfield  {journal} {\bibinfo  {journal} {Phys.
				Rev. B}\ }\textbf {\bibinfo {volume} {87}},\ \bibinfo {pages} {134504}
		(\bibinfo {year} {2013})}\BibitemShut {NoStop}%
	\bibitem [{\citenamefont {Yin}\ \emph {et~al.}(2013)\citenamefont {Yin},
		\citenamefont {Chen}, \citenamefont {Sank}, \citenamefont {O’Malley},
		\citenamefont {White}, \citenamefont {Barends}, \citenamefont {Kelly},
		\citenamefont {Lucero}, \citenamefont {Mariantoni}, \citenamefont {Megrant},
		\citenamefont {Neill}, \citenamefont {Vainsencher}, \citenamefont {Wenner},
		\citenamefont {Korotkov}, \citenamefont {Cleland},\ and\ \citenamefont
		{Martinis}}]{Yin2013}%
	\BibitemOpen
	\bibfield  {author} {\bibinfo {author} {\bibfnamefont {Y.}~\bibnamefont
			{Yin}}, \bibinfo {author} {\bibfnamefont {Y.}~\bibnamefont {Chen}}, \bibinfo
		{author} {\bibfnamefont {D.}~\bibnamefont {Sank}}, \bibinfo {author}
		{\bibfnamefont {P.~J.~J.}\ \bibnamefont {O’Malley}}, \bibinfo {author}
		{\bibfnamefont {T.~C.}\ \bibnamefont {White}}, \bibinfo {author}
		{\bibfnamefont {R.}~\bibnamefont {Barends}}, \bibinfo {author} {\bibfnamefont
			{J.}~\bibnamefont {Kelly}}, \bibinfo {author} {\bibfnamefont
			{E.}~\bibnamefont {Lucero}}, \bibinfo {author} {\bibfnamefont
			{M.}~\bibnamefont {Mariantoni}}, \bibinfo {author} {\bibfnamefont
			{A.}~\bibnamefont {Megrant}}, \bibinfo {author} {\bibfnamefont
			{C.}~\bibnamefont {Neill}}, \bibinfo {author} {\bibfnamefont
			{A.}~\bibnamefont {Vainsencher}}, \bibinfo {author} {\bibfnamefont
			{J.}~\bibnamefont {Wenner}}, \bibinfo {author} {\bibfnamefont {A.~N.}\
			\bibnamefont {Korotkov}}, \bibinfo {author} {\bibfnamefont {A.~N.}\
			\bibnamefont {Cleland}}, \ and\ \bibinfo {author} {\bibfnamefont {J.~M.}\
			\bibnamefont {Martinis}},\ }\bibfield  {title} {\enquote {\bibinfo {title}
			{Catch and release of microwave photon states},}\ }\href {\doibase
		10.1103/PhysRevLett.110.107001} {\bibfield  {journal} {\bibinfo  {journal}
			{Phys. Rev. Lett.}\ }\textbf {\bibinfo {volume} {110}},\ \bibinfo {pages}
		{107001} (\bibinfo {year} {2013})}\BibitemShut {NoStop}%
	\bibitem [{\citenamefont {Geller}\ \emph {et~al.}(2015)\citenamefont {Geller},
		\citenamefont {Donate}, \citenamefont {Chen}, \citenamefont {Fang},
		\citenamefont {Leung}, \citenamefont {Neill}, \citenamefont {Roushan},\ and\
		\citenamefont {Martinis}}]{Geller2015}%
	\BibitemOpen
	\bibfield  {author} {\bibinfo {author} {\bibfnamefont {M.~R.}\ \bibnamefont
			{Geller}}, \bibinfo {author} {\bibfnamefont {E.}~\bibnamefont {Donate}},
		\bibinfo {author} {\bibfnamefont {Y.}~\bibnamefont {Chen}}, \bibinfo {author}
		{\bibfnamefont {M.~T.}\ \bibnamefont {Fang}}, \bibinfo {author}
		{\bibfnamefont {N.}~\bibnamefont {Leung}}, \bibinfo {author} {\bibfnamefont
			{C.}~\bibnamefont {Neill}}, \bibinfo {author} {\bibfnamefont
			{P.}~\bibnamefont {Roushan}}, \ and\ \bibinfo {author} {\bibfnamefont
			{J.~M.}\ \bibnamefont {Martinis}},\ }\bibfield  {title} {\enquote {\bibinfo
			{title} {Tunable coupler for superconducting xmon qubits: Perturbative
				nonlinear model},}\ }\href {\doibase 10.1103/PhysRevA.92.012320} {\bibfield
		{journal} {\bibinfo  {journal} {Phys. Rev. A}\ }\textbf {\bibinfo {volume}
			{92}},\ \bibinfo {pages} {012320} (\bibinfo {year} {2015})}\BibitemShut
	{NoStop}%
	\bibitem [{\citenamefont {Wulschner}\ \emph {et~al.}(2016)\citenamefont
		{Wulschner}, \citenamefont {Goetz}, \citenamefont {Koessel}, \citenamefont
		{Hoffmann}, \citenamefont {Baust}, \citenamefont {Eder}, \citenamefont
		{Fischer}, \citenamefont {Haeberlein}, \citenamefont {Schwarz}, \citenamefont
		{Pernpeintner}, \citenamefont {Xie}, \citenamefont {Zhong}, \citenamefont
		{Zollitsch}, \citenamefont {Peropadre}, \citenamefont {Garcia Ripoll},
		\citenamefont {Solano}, \citenamefont {Fedorov}, \citenamefont {Menzel},
		\citenamefont {Deppe}, \citenamefont {Marx},\ and\ \citenamefont
		{Gross}}]{Wulschner2016}%
	\BibitemOpen
	\bibfield  {author} {\bibinfo {author} {\bibfnamefont {F.}~\bibnamefont
			{Wulschner}}, \bibinfo {author} {\bibfnamefont {J.}~\bibnamefont {Goetz}},
		\bibinfo {author} {\bibfnamefont {F.~R.}\ \bibnamefont {Koessel}}, \bibinfo
		{author} {\bibfnamefont {E.}~\bibnamefont {Hoffmann}}, \bibinfo {author}
		{\bibfnamefont {A.}~\bibnamefont {Baust}}, \bibinfo {author} {\bibfnamefont
			{P.}~\bibnamefont {Eder}}, \bibinfo {author} {\bibfnamefont {M.}~\bibnamefont
			{Fischer}}, \bibinfo {author} {\bibfnamefont {M.}~\bibnamefont {Haeberlein}},
		\bibinfo {author} {\bibfnamefont {M.~J.}\ \bibnamefont {Schwarz}}, \bibinfo
		{author} {\bibfnamefont {M.}~\bibnamefont {Pernpeintner}}, \bibinfo {author}
		{\bibfnamefont {E.}~\bibnamefont {Xie}}, \bibinfo {author} {\bibfnamefont
			{L.}~\bibnamefont {Zhong}}, \bibinfo {author} {\bibfnamefont {C.~W.}\
			\bibnamefont {Zollitsch}}, \bibinfo {author} {\bibfnamefont {B.}~\bibnamefont
			{Peropadre}}, \bibinfo {author} {\bibfnamefont {J.-J.}\ \bibnamefont
			{Garcia Ripoll}}, \bibinfo {author} {\bibfnamefont {E.}~\bibnamefont
			{Solano}}, \bibinfo {author} {\bibfnamefont {K.~G.}\ \bibnamefont {Fedorov}},
		\bibinfo {author} {\bibfnamefont {E.~P.}\ \bibnamefont {Menzel}}, \bibinfo
		{author} {\bibfnamefont {F.}~\bibnamefont {Deppe}}, \bibinfo {author}
		{\bibfnamefont {A.}~\bibnamefont {Marx}}, \ and\ \bibinfo {author}
		{\bibfnamefont {R.}~\bibnamefont {Gross}},\ }\bibfield  {title} {\enquote
		{\bibinfo {title} {Tunable coupling of transmission-line microwave resonators
				mediated by an rf squid},}\ }\href {\doibase 10.1140/epjqt/s40507-016-0048-2}
	{\bibfield  {journal} {\bibinfo  {journal} {EPJ Quantum Technology}\ }\textbf
		{\bibinfo {volume} {3}},\ \bibinfo {pages} {10} (\bibinfo {year}
		{2016})}\BibitemShut {NoStop}%
	\bibitem [{\citenamefont {Kounalakis}\ \emph {et~al.}(2018)\citenamefont
		{Kounalakis}, \citenamefont {Dickel}, \citenamefont {Bruno}, \citenamefont
		{Langford},\ and\ \citenamefont {Steele}}]{Kounalakis2018}%
	\BibitemOpen
	\bibfield  {author} {\bibinfo {author} {\bibfnamefont {M.}~\bibnamefont
			{Kounalakis}}, \bibinfo {author} {\bibfnamefont {C.}~\bibnamefont {Dickel}},
		\bibinfo {author} {\bibfnamefont {A.}~\bibnamefont {Bruno}}, \bibinfo
		{author} {\bibfnamefont {N.~K.}\ \bibnamefont {Langford}}, \ and\ \bibinfo
		{author} {\bibfnamefont {G.~A.}\ \bibnamefont {Steele}},\ }\bibfield  {title}
	{\enquote {\bibinfo {title} {Tuneable hopping and nonlinear cross-kerr
				interactions in a high-coherence superconducting circuit},}\ }\href {\doibase
		10.1038/s41534-018-0088-9} {\bibfield  {journal} {\bibinfo  {journal} {npj
				Quantum Information}\ }\textbf {\bibinfo {volume} {4}},\ \bibinfo {pages}
		{38} (\bibinfo {year} {2018})}\BibitemShut {NoStop}%
	\bibitem [{\citenamefont {Koch}\ \emph {et~al.}(2010)\citenamefont {Koch},
		\citenamefont {Houck}, \citenamefont {Hur},\ and\ \citenamefont
		{Girvin}}]{Koch2011a}%
	\BibitemOpen
	\bibfield  {author} {\bibinfo {author} {\bibfnamefont {J.}~\bibnamefont
			{Koch}}, \bibinfo {author} {\bibfnamefont {A.~A.}\ \bibnamefont {Houck}},
		\bibinfo {author} {\bibfnamefont {K.~L.}\ \bibnamefont {Hur}}, \ and\
		\bibinfo {author} {\bibfnamefont {S.~M.}\ \bibnamefont {Girvin}},\ }\bibfield
	{title} {\enquote {\bibinfo {title} {Time-reversal-symmetry breaking in
				circuit-qed-based photon lattices},}\ }\href {\doibase
		10.1103/PhysRevA.82.043811} {\bibfield  {journal} {\bibinfo  {journal} {Phys.
				Rev. A}\ }\textbf {\bibinfo {volume} {82}},\ \bibinfo {pages} {043811}
		(\bibinfo {year} {2010})}\BibitemShut {NoStop}%
	\bibitem [{\citenamefont {Vermersch}\ \emph {et~al.}(2016)\citenamefont
		{Vermersch}, \citenamefont {Ramos}, \citenamefont {Hauke},\ and\
		\citenamefont {Zoller}}]{Vermersch2016}%
	\BibitemOpen
	\bibfield  {author} {\bibinfo {author} {\bibfnamefont {B.}~\bibnamefont
			{Vermersch}}, \bibinfo {author} {\bibfnamefont {T.}~\bibnamefont {Ramos}},
		\bibinfo {author} {\bibfnamefont {Philipp}\ \bibnamefont {Hauke}}, \ and\
		\bibinfo {author} {\bibfnamefont {Peter}\ \bibnamefont {Zoller}},\ }\bibfield
	{title} {\enquote {\bibinfo {title} {Implementation of chiral quantum optics
				with rydberg and trapped-ion setups},}\ }\href {\doibase
		10.1103/PhysRevA.93.063830} {\bibfield  {journal} {\bibinfo  {journal} {Phys.
				Rev. A}\ }\textbf {\bibinfo {volume} {93}},\ \bibinfo {pages} {063830}
		(\bibinfo {year} {2016})}\BibitemShut {NoStop}%
	\bibitem [{\citenamefont {Roushan}\ \emph {et~al.}(2017)\citenamefont
		{Roushan}, \citenamefont {Neill}, \citenamefont {Megrant}, \citenamefont
		{Chen}, \citenamefont {Babbush}, \citenamefont {Barends}, \citenamefont
		{Campbell}, \citenamefont {Chen}, \citenamefont {Chiaro}, \citenamefont
		{Dunsworth}, \citenamefont {Fowler}, \citenamefont {Jeffrey}, \citenamefont
		{Kelly}, \citenamefont {Lucero}, \citenamefont {Mutus}, \citenamefont
		{O’Malley}, \citenamefont {Neeley}, \citenamefont {Quintana}, \citenamefont
		{Sank}, \citenamefont {Vainsencher}, \citenamefont {Wenner}, \citenamefont
		{White}, \citenamefont {Kapit}, \citenamefont {Neven},\ and\ \citenamefont
		{Martinis}}]{Roushan2017}%
	\BibitemOpen
	\bibfield  {author} {\bibinfo {author} {\bibfnamefont {P.}~\bibnamefont
			{Roushan}}, \bibinfo {author} {\bibfnamefont {C.}~\bibnamefont {Neill}},
		\bibinfo {author} {\bibfnamefont {A.}~\bibnamefont {Megrant}}, \bibinfo
		{author} {\bibfnamefont {Y.}~\bibnamefont {Chen}}, \bibinfo {author}
		{\bibfnamefont {R.}~\bibnamefont {Babbush}}, \bibinfo {author} {\bibfnamefont
			{R.}~\bibnamefont {Barends}}, \bibinfo {author} {\bibfnamefont
			{B.}~\bibnamefont {Campbell}}, \bibinfo {author} {\bibfnamefont
			{Z.}~\bibnamefont {Chen}}, \bibinfo {author} {\bibfnamefont {B.}~\bibnamefont
			{Chiaro}}, \bibinfo {author} {\bibfnamefont {A.}~\bibnamefont {Dunsworth}},
		\bibinfo {author} {\bibfnamefont {A.}~\bibnamefont {Fowler}}, \bibinfo
		{author} {\bibfnamefont {E.}~\bibnamefont {Jeffrey}}, \bibinfo {author}
		{\bibfnamefont {J.}~\bibnamefont {Kelly}}, \bibinfo {author} {\bibfnamefont
			{E.}~\bibnamefont {Lucero}}, \bibinfo {author} {\bibfnamefont
			{J.}~\bibnamefont {Mutus}}, \bibinfo {author} {\bibfnamefont {P.~J.~J.}\
			\bibnamefont {O’Malley}}, \bibinfo {author} {\bibfnamefont
			{M.}~\bibnamefont {Neeley}}, \bibinfo {author} {\bibfnamefont
			{C.}~\bibnamefont {Quintana}}, \bibinfo {author} {\bibfnamefont
			{D.}~\bibnamefont {Sank}}, \bibinfo {author} {\bibfnamefont {A.}~\bibnamefont
			{Vainsencher}}, \bibinfo {author} {\bibfnamefont {J.}~\bibnamefont {Wenner}},
		\bibinfo {author} {\bibfnamefont {T.}~\bibnamefont {White}}, \bibinfo
		{author} {\bibfnamefont {E.}~\bibnamefont {Kapit}}, \bibinfo {author}
		{\bibfnamefont {H.}~\bibnamefont {Neven}}, \ and\ \bibinfo {author}
		{\bibfnamefont {J.}~\bibnamefont {Martinis}},\ }\bibfield  {title} {\enquote
		{\bibinfo {title} {Chiral ground-state currents of interacting photons in a
				synthetic magnetic field},}\ }\href {\doibase 10.1038/nphys3930} {\bibfield
		{journal} {\bibinfo  {journal} {Nat. Phys.}\ }\textbf {\bibinfo {volume}
			{13}},\ \bibinfo {pages} {146} (\bibinfo {year} {2017})}\BibitemShut
	{NoStop}%
	\bibitem [{\citenamefont {Wang}\ \emph {et~al.}(2020)\citenamefont {Wang},
		\citenamefont {Li},\ and\ \citenamefont {Li}}]{Wang2020X}%
	\BibitemOpen
	\bibfield  {author} {\bibinfo {author} {\bibfnamefont {X.}~\bibnamefont
			{Wang}}, \bibinfo {author} {\bibfnamefont {H.-R.}\ \bibnamefont {Li}}, \ and\
		\bibinfo {author} {\bibfnamefont {F.-L.}\ \bibnamefont {Li}},\ }\bibfield
	{title} {\enquote {\bibinfo {title} {Generating synthetic magnetism via
				floquet engineering auxiliary qubits in phonon-cavity-based lattice},}\
	}\href {https://iopscience.iop.org/article/10.1088/1367-2630/ab776e}
	{\bibfield  {journal} {\bibinfo  {journal} {New J. Phys.}\ }\textbf {\bibinfo
			{volume} {22}},\ \bibinfo {pages} {033037} (\bibinfo {year}
		{2020})}\BibitemShut {NoStop}%
	\bibitem [{\citenamefont {Cohen-Tannoudji}\ \emph {et~al.}(1998)\citenamefont
		{Cohen-Tannoudji}, \citenamefont {Dupont-Roc},\ and\ \citenamefont
		{Grynberg}}]{cohen1998atom}%
	\BibitemOpen
	\bibfield  {author} {\bibinfo {author} {\bibfnamefont {C.}~\bibnamefont
			{Cohen-Tannoudji}}, \bibinfo {author} {\bibfnamefont {J.}~\bibnamefont
			{Dupont-Roc}}, \ and\ \bibinfo {author} {\bibfnamefont {G.}~\bibnamefont
			{Grynberg}},\ }\href@noop {} {\emph {\bibinfo {title} {{Atom--Photon
					Interactions}}}}\ (\bibinfo  {publisher} {Wiley},\ \bibinfo {year}
	{1998})\BibitemShut {NoStop}%
	\bibitem [{\citenamefont {Scully}\ and\ \citenamefont
		{Zubairy}(1997)}]{Scully1997}%
	\BibitemOpen
	\bibfield  {author} {\bibinfo {author} {\bibfnamefont {M.~O.}\ \bibnamefont
			{Scully}}\ and\ \bibinfo {author} {\bibfnamefont {M.~S.}\ \bibnamefont
			{Zubairy}},\ }\href@noop {} {\emph {\bibinfo {title} {Quantum optics}}}\
	(\bibinfo  {publisher} {Cambridge University Press},\ \bibinfo {year}
	{1997})\BibitemShut {NoStop}%
	\bibitem [{\citenamefont {et~al}(2022)}]{Wang2022}%
	\BibitemOpen
	\bibfield  {author} {\bibinfo {author} {\bibfnamefont {C.-L.~Wang}\
			\bibnamefont {et~al}},\ }\bibfield  {title} {\enquote {\bibinfo {title}
			{Towards practical quantum computers: transmon qubit with a lifetime
				approaching 0.5 milliseconds},}\ }\href
	{https://doi.org/10.1038/s41534-021-00510-2} {\bibfield  {journal} {\bibinfo
			{journal} {npj Quantum Information}\ }\textbf {\bibinfo {volume} {8}},\
		\bibinfo {pages} {3} (\bibinfo {year} {2022})}\BibitemShut {NoStop}%
	\bibitem [{\citenamefont {Korotkov}(2011)}]{Korotkov2011}%
	\BibitemOpen
	\bibfield  {author} {\bibinfo {author} {\bibfnamefont {A.~N.}\ \bibnamefont
			{Korotkov}},\ }\bibfield  {title} {\enquote {\bibinfo {title} {Flying
				microwave qubits with nearly perfect transfer efficiency},}\ }\href {\doibase
		10.1103/PhysRevB.84.014510} {\bibfield  {journal} {\bibinfo  {journal} {Phys.
				Rev. B}\ }\textbf {\bibinfo {volume} {84}},\ \bibinfo {pages} {014510}
		(\bibinfo {year} {2011})}\BibitemShut {NoStop}%
	\bibitem [{\citenamefont {Johansson}\ \emph {et~al.}(2012)\citenamefont
		{Johansson}, \citenamefont {Nation},\ and\ \citenamefont
		{Nori}}]{Johansson12qutip}%
	\BibitemOpen
	\bibfield  {author} {\bibinfo {author} {\bibfnamefont {J.~R.}\ \bibnamefont
			{Johansson}}, \bibinfo {author} {\bibfnamefont {P.~D.}\ \bibnamefont
			{Nation}}, \ and\ \bibinfo {author} {\bibfnamefont {F.}~\bibnamefont
			{Nori}},\ }\bibfield  {title} {\enquote {\bibinfo {title} {Qutip: {A}n
				open-source {P}ython framework for the dynamics of open quantum systems},}\
	}\href {http://www.sciencedirect.com/science/article/pii/S0010465512000835}
	{\bibfield  {journal} {\bibinfo  {journal} {Comput. Phys. Commun.}\ }\textbf
		{\bibinfo {volume} {183}},\ \bibinfo {pages} {1760} (\bibinfo {year}
		{2012})}\BibitemShut {NoStop}%
	\bibitem [{\citenamefont {Johansson}\ \emph {et~al.}(2013)\citenamefont
		{Johansson}, \citenamefont {Nation},\ and\ \citenamefont
		{Nori}}]{Johansson13qutip}%
	\BibitemOpen
	\bibfield  {author} {\bibinfo {author} {\bibfnamefont {J.~R.}\ \bibnamefont
			{Johansson}}, \bibinfo {author} {\bibfnamefont {P.~D.}\ \bibnamefont
			{Nation}}, \ and\ \bibinfo {author} {\bibfnamefont {F.}~\bibnamefont
			{Nori}},\ }\bibfield  {title} {\enquote {\bibinfo {title} {Qutip 2: {A}
				{P}ython framework for the dynamics of open quantum systems},}\ }\href
	{http://www.sciencedirect.com/science/article/pii/S0010465512003955}
	{\bibfield  {journal} {\bibinfo  {journal} {Comput. Phys. Commun.}\ }\textbf
		{\bibinfo {volume} {184}},\ \bibinfo {pages} {1234} (\bibinfo {year}
		{2013})}\BibitemShut {NoStop}%
	\bibitem [{\citenamefont {Zhang}\ \emph {et~al.}(2021)\citenamefont {Zhang},
		\citenamefont {Carceller}, \citenamefont {Kjaergaard},\ and\ \citenamefont
		{S\o{}rensen}}]{Zhang2021}%
	\BibitemOpen
	\bibfield  {author} {\bibinfo {author} {\bibfnamefont {Y.-X.}\ \bibnamefont
			{Zhang}}, \bibinfo {author} {\bibfnamefont {C.}~\bibnamefont {Carceller}},
		\bibinfo {author} {\bibfnamefont {M.}~\bibnamefont {Kjaergaard}}, \ and\
		\bibinfo {author} {\bibfnamefont {A.~S.}\ \bibnamefont {S\o{}rensen}},\
	}\bibfield  {title} {\enquote {\bibinfo {title} {Charge-noise insensitive
				chiral photonic interface for waveguide circuit qed},}\ }\href {\doibase
		10.1103/PhysRevLett.127.233601} {\bibfield  {journal} {\bibinfo  {journal}
			{Phys. Rev. Lett.}\ }\textbf {\bibinfo {volume} {127}},\ \bibinfo {pages}
		{233601} (\bibinfo {year} {2021})}\BibitemShut {NoStop}%
	\bibitem [{\citenamefont {Koch}\ \emph {et~al.}(2007)\citenamefont {Koch},
		\citenamefont {Yu}, \citenamefont {Gambetta}, \citenamefont {Houck},
		\citenamefont {Schuster}, \citenamefont {Majer}, \citenamefont {Blais},
		\citenamefont {Devoret}, \citenamefont {Girvin},\ and\ \citenamefont
		{Schoelkopf}}]{Koch2017}%
	\BibitemOpen
	\bibfield  {author} {\bibinfo {author} {\bibfnamefont {J.}~\bibnamefont
			{Koch}}, \bibinfo {author} {\bibfnamefont {T.~M.}\ \bibnamefont {Yu}},
		\bibinfo {author} {\bibfnamefont {J.}~\bibnamefont {Gambetta}}, \bibinfo
		{author} {\bibfnamefont {A.~A.}\ \bibnamefont {Houck}}, \bibinfo {author}
		{\bibfnamefont {D.~I.}\ \bibnamefont {Schuster}}, \bibinfo {author}
		{\bibfnamefont {J.}~\bibnamefont {Majer}}, \bibinfo {author} {\bibfnamefont
			{A.}~\bibnamefont {Blais}}, \bibinfo {author} {\bibfnamefont {M.~H.}\
			\bibnamefont {Devoret}}, \bibinfo {author} {\bibfnamefont {S.~M.}\
			\bibnamefont {Girvin}}, \ and\ \bibinfo {author} {\bibfnamefont {R.~J.}\
			\bibnamefont {Schoelkopf}},\ }\bibfield  {title} {\enquote {\bibinfo {title}
			{Charge-insensitive qubit design derived from the cooper pair box},}\ }\href
	{https://journals.aps.org/pra/pdf/10.1103/PhysRevA.76.042319} {\bibfield
		{journal} {\bibinfo  {journal} {Phys. Rev. A}\ }\textbf {\bibinfo {volume}
			{76}} (\bibinfo {year} {2007})}\BibitemShut {NoStop}%
	\bibitem [{\citenamefont {Bello}\ \emph {et~al.}(2019)\citenamefont {Bello},
		\citenamefont {Platero}, \citenamefont {Cirac},\ and\ \citenamefont
		{Gonz{\'{a}}lez-Tudela}}]{Bello2019}%
	\BibitemOpen
	\bibfield  {author} {\bibinfo {author} {\bibfnamefont {M.}~\bibnamefont
			{Bello}}, \bibinfo {author} {\bibfnamefont {G.}~\bibnamefont {Platero}},
		\bibinfo {author} {\bibfnamefont {J.~I.}\ \bibnamefont {Cirac}}, \ and\
		\bibinfo {author} {\bibfnamefont {A.}~\bibnamefont {Gonz{\'{a}}lez-Tudela}},\
	}\bibfield  {title} {\enquote {\bibinfo {title} {Unconventional quantum
				optics in topological waveguide {QED}},}\ }\href {\doibase
		10.1126/sciadv.aaw0297} {\bibfield  {journal} {\bibinfo  {journal} {Sci.
				Adv.}\ }\textbf {\bibinfo {volume} {5}},\ \bibinfo {pages} {eaaw0297}
		(\bibinfo {year} {2019})}\BibitemShut {NoStop}%
	\bibitem [{\citenamefont {Gough}\ and\ \citenamefont
		{James}(2009)}]{Gough2009}%
	\BibitemOpen
	\bibfield  {author} {\bibinfo {author} {\bibfnamefont {J.}~\bibnamefont
			{Gough}}\ and\ \bibinfo {author} {\bibfnamefont {M.}~\bibnamefont {James}},\
	}\bibfield  {title} {\enquote {\bibinfo {title} {Quantum feedback networks:
				Hamiltonian formulation},}\ }\href {\doibase 10.1007/s00220-008-0698-8}
	{\bibfield  {journal} {\bibinfo  {journal} {Communications in Mathematical
				Physics}\ }\textbf {\bibinfo {volume} {287}},\ \bibinfo {pages} {1109}
		(\bibinfo {year} {2009})}\BibitemShut {NoStop}%
	\bibitem [{\citenamefont {Combes}\ \emph {et~al.}(2017)\citenamefont {Combes},
		\citenamefont {Kerckhoff},\ and\ \citenamefont {Sarovar}}]{Combes2017}%
	\BibitemOpen
	\bibfield  {author} {\bibinfo {author} {\bibfnamefont {J.}~\bibnamefont
			{Combes}}, \bibinfo {author} {\bibfnamefont {J.}~\bibnamefont {Kerckhoff}}, \
		and\ \bibinfo {author} {\bibfnamefont {M.}~\bibnamefont {Sarovar}},\
	}\bibfield  {title} {\enquote {\bibinfo {title} {The slh framework for
				modeling quantum input-output networks},}\ }\href {\doibase
		10.1080/23746149.2017.1343097} {\bibfield  {journal} {\bibinfo  {journal}
			{Advances in Physics-X}\ }\textbf {\bibinfo {volume} {2}},\ \bibinfo {pages}
		{784--888} (\bibinfo {year} {2017})}\BibitemShut {NoStop}%
\end{thebibliography}

\end{document}